\documentclass[prb,twocolumn,superscriptaddress,floatfix,letterpaper]{revtex4}
\pdfoutput=1

\usepackage{tikz}

\def\FBox{\draw [line width=0pt, white] (0,0)--(5,0)--(5,2)--(0,2)--(0,0);}

\def\iLnkaa{\draw [dotted] (0,0) -- (0.5,1); \draw [dotted] (3,0) -- (4,0.5);}
\def\jLnkaa{\draw [dotted] (0,2) -- (0.5,1); \draw [dotted] (3,2) -- (4,1.5);}
\def\kLnkaa{\draw [dotted] (2,2) -- (1.5,1); \draw [dotted] (5,2) -- (4,1.5);}
\def\lLnkaa{\draw [dotted] (2,0) -- (1.5,1); \draw [dotted] (5,0) -- (4,0.5);}
\def\mLnkaa{\draw [dotted] (0.5,1) -- (1.5,1);}
\def\nLnkaa{\draw [dotted] (4,0.5) -- (4,1.5);}

\def\iLnkbb{\draw  (0,0) -- (0.5,1); \draw  (3,0) -- (4,0.5);}
\def\jLnkbb{\draw  (0,2) -- (0.5,1); \draw  (3,2) -- (4,1.5);}
\def\kLnkbb{\draw  (2,2) -- (1.5,1); \draw  (5,2) -- (4,1.5);}
\def\lLnkbb{\draw  (2,0) -- (1.5,1); \draw  (5,0) -- (4,0.5);}
\def\mLnkbb{\draw  (0.5,1) -- (1.5,1);}
\def\nLnkbb{\draw  (4,0.5) -- (4,1.5);}

\def\iLnkbc{
\draw[->]  (0,0) -- (0.25,0.5);
\draw  (0.25,0.5) -- (0.5,1);
\draw[->]  (3,0) -- (3.5,0.25);
\draw (3.5,0.25) -- (4,0.5);
}
\def\jLnkbc{
\draw[->]  (0,2) -- (0.25,1.5);
\draw  (0.25,1.5) -- (0.5,1);
\draw[->]  (3,2) -- (3.5,1.75);
\draw  (3.5,1.75) -- (4,1.5);
}
\def\kLnkbc{
\draw[->]  (2,2) -- (1.75,1.5);
\draw  (1.75,1.5) -- (1.5,1);
\draw[->]  (5,2) -- (4.5,1.75);
\draw  (4.5,1.75) -- (4,1.5);
}
\def\lLnkbc{
\draw[->]  (1.5,1) -- (1.75,0.5) ;
\draw  (1.75,0.5) -- (2,0) ;
\draw[->]  (4,0.5) -- (4.5,0.25) ;
\draw  (4.5,0.25) -- (5,0) ;
}
\def\mLnkbc{
\draw[->]  (0.5,1) -- (1,1);
\draw  (1,1) -- (1.5,1);
}
\def\nLnkbc{
\draw[->]  (4,1.5) -- (4,1);
\draw  (4,1) -- (4,0.5);
}

\def\iLnkcb{
\draw  (0,0) -- (0.25,0.5);
\draw[<-]  (0.25,0.5) -- (0.5,1);
\draw  (3,0) -- (3.5,0.25);
\draw[<-] (3.5,0.25) -- (4,0.5);
}
\def\jLnkcb{
\draw  (0,2) -- (0.25,1.5);
\draw[<-]  (0.25,1.5) -- (0.5,1);
\draw  (3,2) -- (3.5,1.75);
\draw[<-]  (3.5,1.75) -- (4,1.5);
}
\def\kLnkcb{
\draw  (2,2) -- (1.75,1.5);
\draw[<-]  (1.75,1.5) -- (1.5,1);
\draw  (5,2) -- (4.5,1.75);
\draw[<-]  (4.5,1.75) -- (4,1.5);
}
\def\lLnkcb{
\draw  (1.5,1) -- (1.75,0.5) ;
\draw[<-]  (1.75,0.5) -- (2,0) ;
\draw  (4,0.5) -- (4.5,0.25) ;
\draw[<-]  (4.5,0.25) -- (5,0) ;
}
\def\mLnkcb{
\draw  (0.5,1) -- (1,1);
\draw[<-]  (1,1) -- (1.5,1);
}
\def\nLnkcb{
\draw  (4,1.5) -- (4,1);
\draw[<-]  (4,1) -- (4,0.5);
}

\newtheorem{theorem}{Theorem}

\begin{document}

\title{
Local unitary transformation, long-range quantum entanglement,\\
wave function renormalization, and topological order
}

\author{Xie Chen}
\affiliation{Department of Physics, Massachusetts Institute of Technology, Cambridge, Massachusetts 02139, USA}

\author{Zheng-Cheng Gu}
\affiliation{Kavli Institute for Theoretical Physics, University of California, Santa Barbara, CA 93106, USA}

\author{Xiao-Gang Wen}
\affiliation{Department of Physics, Massachusetts Institute of Technology, Cambridge, Massachusetts 02139, USA}

\date{Jul., 2010}
\begin{abstract}
Two gapped quantum ground states in the same phase are
connected by an adiabatic evolution which gives rise to a local
unitary transformation that maps between the states. On the
other hand, gapped ground states remain within the same phase
under local unitary transformations. Therefore, local
unitary transformations define an equivalence relation and
the equivalence classes are the universality classes that
define the different phases for gapped quantum systems.
Since local unitary transformations can remove local
entanglement, the above equivalence/universality classes
correspond to pattern of long range entanglement, which is
the essence of topological order.  The local unitary
transformation also allows us to define a wave function
renormalization scheme, under which a wave function can flow
to a simpler one within the same equivalence/universality
class. Using such a setup, we find conditions on the
possible fixed-point wave functions where the local unitary
transformations have \emph{finite} dimensions.  The
solutions of the conditions allow us to classify this type
of topological orders, which generalize the string-net
classification of topological orders.
We also describe an algorithm of wave function
renormalization induced by local unitary transformations.
The algorithm allows us to calculate the flow of
tensor-product wave functions which are not at the fixed
points.  This will allow us to calculate topological orders
as well as symmetry breaking  orders in a generic
tensor-product state.
\end{abstract}

\maketitle

{\small \setcounter{tocdepth}{2} \tableofcontents }

\section{Introduction -- new states beyond symmetry breaking}

According to the principle of emergence, the rich properties
and the many different forms of materials
originate from the different ways in which the
atoms are ordered in the materials.
Landau symmetry-breaking theory provides a general
understanding of those different orders and resulting
rich states of matter.\cite{L3726,GL5064} It points out that
different orders really correspond to different symmetries
in the organizations of the constituent atoms. As a material
changes from one order to another order (i.e., as the
material undergoes a phase transition), what happens is that
the symmetry of the organization of the atoms changes.

For a long time, we believed that Landau symmetry-breaking
theory describes all possible orders in materials, and all
possible (continuous) phase transitions.  However, in last
twenty years, it has become more and more clear that Landau
symmetry-breaking theory does not describe all possible
orders.  After the discovery of high $T_c$ superconductors
in 1986,\cite{BM8689} some theorists believed that quantum
spin liquids play a key role in understanding high $T_c$
superconductors\cite{A8796} and started to introduce various
spin liquids.\cite{BZA8773,AM8874,RK8876,AZH8845,DFM8826}
Despite the success of Landau symmetry-breaking theory in
describing all kinds of states, the theory cannot explain
and does not even allow the existence of spin liquids. This
leads many theorists to doubt the very existence of spin
liquids.  In 1987,  in an attempt to explain high
temperature superconductivity, an
infrared stable spin liquid -- chiral spin
state was discovered,\cite{KL8795,WWZcsp} which was shown to be
perturbatively stable and exist as quantum phase of matter
(at least in a large $N$ limit).  At first, not believing
Landau symmetry-breaking theory fails to describe spin
liquids, people still wanted to use symmetry-breaking to
describe the chiral spin state.  They identified the chiral
spin state as a state that breaks the time reversal and
parity symmetries, but not the spin rotation
symmetry.\cite{WWZcsp} However, it was quickly realized that
there are many different chiral spin states that have
exactly the same symmetry, so symmetry alone is not enough
to characterize different chiral spin states.  This means
that the chiral spin states contain a new kind of order that
is beyond symmetry description.\cite{Wtop} This new kind of
order was named\cite{Wrig} topological order.\footnote{The
name ``topological order'' is motivated by the low energy
effective theory of the chiral spin states, which is a
topological quantum field theory.\cite{W8951}.}

The key to identify (and define) new orders is to identify
new universal properties that are beyond the local order
parameters and long-range correlations used in the Landau
symmetry breaking theory.  Indeed, new quantum numbers, such
as ground state degeneracy\cite{Wtop}, the non-Abelian
Berry's phase of degenerate ground states\cite{Wrig} and
edge excitations\cite{Wedge}, were introduced to
characterize (and define) the different topological orders
in chiral spin states.  Recently, it was shown that
topological orders can also be characterized by topological
entanglement entropy.\cite{KP0604,LWtopent} More
importantly, those quantities were shown to be universal
(\ie robust against any local perturbation of the
Hamiltonian) for chiral spin states.\cite{Wrig}  The
existence of those universal properties establishes the
existence of topological order in chiral spin states.

Near the end of 1980's, the existence of chiral spin states
as a theoretical possibility, as well as their many amazing
properties, such as fractional
statistics,\cite{KL8795,WWZcsp} spin-charge
separation,\cite{KL8795,WWZcsp} chiral gapless edge
excitations,\cite{Wedge} were established reliably, at
least in the large $N$-limit introduced in \Ref{MA8938}.  Even
non-Abelian chiral spin states can be established reliably
in the large $N$-limit.\cite{Wnab} However, it took about 10
years to establish the existence of a chiral spin state
reliably without using large $N$-limit (based on an exactly
soluble model on honeycomb lattice).\cite{Kh}

Soon after the introduction of chiral spin states,
experiments indicated that high-temperature
superconductors do not break the time reversal and parity
symmetries. So chiral spin states do not describe
high-temperature superconductors. Thus the theory of
topological order became a theory with no experimental
realization. However, the similarity between chiral spin
states and fractional quantum Hall (FQH) states allows one
to use the theory of topological order to describe different
FQH states.\cite{WNtop} Just like chiral spin states,
different FQH states all have the same symmetry and are
beyond the Landau symmetry-breaking description.  Also like
chiral spin states, FQH states have ground state
degeneracies\cite{HR8529} that depend on the topology of the
space.\cite{WNtop,Wrig} Those ground state degeneracies are
shown to be robust against any perturbations.  Thus, the
different orders in different quantum Hall states can be
described by topological orders, and the topological order
does have experimental realizations.

The topology dependent ground state degeneracy, that signal
the presence of topological order, is an amazing phenomenon.
In FQH states, the correlation of any local operators are
short ranged. This seems to imply that FQH states are
``short sighted'' and they cannot know the topology of space
which is a global and long-distance property.  However, the
fact that ground state degeneracy does depend on the
topology of space implies that FQH states are not ``short
sighted'' and they do find a way to know the global and
long-distance structure of space.  So, despite the
short-range correlations of any local operators, the FQH
states must contain certain hidden long-range correlation.
But what is this hidden long-range correlation? This will be
one of the main topic of this paper.

Since high $T_c$ superconductors do not break the time
reversal and parity symmetries, nor any other lattice
symmetries, some people concentrated on finding spin liquids
that respect all those symmetries and hoping one of those
symmetric spin liquids hold the key to understand high $T_c$
superconductors.  Between 1987 and 1992, many symmetric spin
liquids were introduced and
studied.\cite{BZA8773,AM8874,KL8842,RK8876,AZH8845,DFM8826,RS9173,Wsrvb,LN9221}
The excitations in some of constructed spin liquids have a
finite energy gap, while in others there is no energy gap.
Those symmetric spin liquids do not break any symmetry and,
by definition, are beyond Landau symmetry-breaking
description.

By construction, topological order only describes the
organization of particles or spins in a gapped quantum
state.  So the theory of topological order only applies to
gapped spin liquids. Indeed, we find that the gapped spin
liquids do contain nontrivial topological orders\cite{Wsrvb}
(as signified by their topology dependent and robust ground
state degeneracies) and are described by topological quantum
field theory (such as $\mathbb{Z}_2$ gauge theory) at low
energies.  One of the simplest topologically ordered spin
liquids is the $\mathbb{Z}_2$ spin liquid which was first
introduced in 1991.\cite{RS9173,Wsrvb} The existence of
$\mathbb{Z}_2$ spin liquid as a theoretical possibility, as
well as its many amazing properties, such as spin-charge
separation,\cite{RS9173,Wsrvb} fractional mutual
statistics,\cite{Wsrvb} topologically protected ground state
degeneracy,\cite{Wsrvb} were established reliably (at least
in the large $N$-limit introduced in \Ref{MA8938}).
Later, Kitaev introduced the famous toric code model
which establishes the existence of the $\mathbb{Z}_2$ spin
liquid reliably without using large $N$-limit.\cite{K032}
The topologically protected degeneracy of the $\mathbb{Z}_2$
spin liquid was used to perform fault-tolerant quantum
computation.

The study of high $T_c$ superconductors also leads
to many gapless spin liquids. The stability and the
existence of those gapless spin liquid were in more doubts
than their gapped counterparts.  But careful analysis in
certain large $N$ limit do suggest that stable gapless spin
liquids can exist.\cite{MA8938,RWasl,HSF0437}
If we do believe in the existence
of gapless spin liquids, then the next question is how to
describe the orders (\ie the organizations of spins) in
those gapless spin liquids.  If gapped quantum state can
contain new type of orders that are beyond Landau's symmetry
breaking description, it is natural to expect that gapless
quantum states can also contain new type of orders.  But how
to show the existence of new orders in gapless states?

Just like topological order, the key to identify new orders
is to identify new universal properties that are beyond
Landau symmetry description.  Clearly we can no longer use
the ground state degeneracy to establish the existence of
new orders in gapless states.  To show the existence of new
orders in gapless states, a new universal quantity --
projective symmetry group (PSG) -- was
introduced.\cite{Wqoslpub} It was argued that (some) PSGs
are robust against any local perturbations of the  Hamiltonian
that do not change the symmetry of the
Hamiltonian.\cite{Wqoslpub,WZqoind,RWasl,HSF0437} So through
PSG, we establish the existence of new orders even in
gapless states. The new orders are called quantum order to
indicate that the new orders are related to patterns of
quantum entanglement in the many-body ground
state.\cite{W0275}

\section{Short-range and long-range quantum entanglement}

What is missed in Landau theory so that it fails to describe
those new orders? What is the new feature in the
organization of particles/spins so that the resulting order
cannot be described by symmetry?

To answer those questions, let us consider a simple
quantum system which can be described with Landau theory --
the transverse field Ising model in two dimensions:
$H= -B\sum X_{\v i} -J\sum  Z_{\v i} Z_{\v j}$,
where
$X_{\v i}$,
$Y_{\v i}$, and
$Z_{\v i}$ are the Pauli matrices on site $\v i$.
In $B\gg J$ limit, the ground state of the system is an equal-weight
superposition of
all possible spin-up and spin-down states:
$|\Phi^+\>= \sum_{\{\si_{\v i}\}} |\{\si_{\v i}\}\>$,
where $\{\si_{\v i}\}$ label a particular spin-up ($Z_{\v
i}=1$) and spin-down ($Z_{\v i}=-1$) configuration.
In the $J\gg B$ limit, the system has two degenerate ground states
$|\Phi^\up\> = |\up\up\cdots\up\>$ and
$|\Phi^\down\> = |\down\down\cdots\down\>$.

The transverse field Ising model has a $Z \to -Z$ symmetry.
The ground state $|\Phi^+\>$ respect such a symmetry while
the ground state $|\Phi^\up\>$ (or $|\Phi^\down\>$) break
the symmetry.  Thus the small $J$ ground state  $|\Phi^+\>$
and the small $B$ ground state $|\Phi^\up\>$ describe
different phases since they have different symmetries.

We note that $|\Phi^\up\>$ is the exact ground state of the
transverse field Ising model with $B=0$.  The state has no
quantum entanglement since $|\Phi^\up\>$ is a direct
product of local states:
$ |\Phi^\up\>=\otimes_{\v i} |\up\>_{\v i} $
where $|\up\>_{\v i}$ is an up-spin state at site $\v i$.
The state $|\Phi^+\>$ is the exact ground state of the
transverse field Ising model with $J=0$.  It is also a
state with no quantum entanglement:
$ |\Phi^+\>=\otimes_{\v i} (|\up\>_{\v i} + |\down\>_{\v i})
 \propto \otimes_{\v i} |+\>_{\v i}$
where $|+\>_{\v i}\equiv |\up\>_{\v i} + |\down\>_{\v i}$
is a state with spin in $x$-direction at
site $\v i$.

We see that the states (or phases) described by Landau
symmetry breaking theory has no quantum entanglement at
least in the $J=0$ or $B=0$ limits.  In the $J/B\ll 1$ and
$B/J\ll 1$ limits, the two ground states of the two limits
still represent two phases with different symmetries.
However, in this case, the ground states are not
unentangled.  On the other hand since the ground states in
the $J=0$ or $B=0$ limits have finite energy gaps and
short-range correlations, a small $J$ or a small $B$ can
only modify the states locally.  Thus we expect the ground
states have only short-range entanglement.

The above example, if generalized to other symmetry breaking
states, suggests the following conjecture:
\emph{If a gapped
quantum ground state is described by Landau symmetry
breaking theory, then it has short-range quantum
entanglement.}\footnote{The conjecture is for symmetry
breaking states with finite energy gaps.  However, it is
unclear whether it is possible to find a definition of
``short range quantum entanglement'' to make the conjecture
valid for gapless symmetry breaking states.}

The direct-product states and short-range entangled states
only represent a small subset of all possible quantum
many-body states. Thus, according to the point of view of
the above conjecture, we see that what is missed by Landau
symmetry breaking theory is long-range quantum
entanglement.  It is this long-range quantum entanglement
that makes a state to have nontrivial topological/quantum
order.

However, mathematically speaking, the above conjecture is
a null statement since the meaning of short-range quantum
entanglement is not defined.  In the following, we will try
to find a more precise description (or definition) of
short-range and long-range quantum entanglement.  We will
start with a careful discussion of quantum phases and
quantum phase transitions.

\section{Quantum phases and local unitary evolutions}
\label{QphUcl}

To give a precise definition of quantum phases, let us
consider a local quantum system whose Hamiltonian has a
smooth dependence on a parameter $g$: $H(g)$.  The ground state average
of a local operator $O$ of the system, $\<O\>(g)$, naturally
also depend on $g$.  If the function $\<O\>(g)$, in the
limit of infinite system size, has a singularity at $g_c$
for some local operators $O$, then the system described by
$H(g)$ has a quantum phase transition at $g_c$.  After
defining phase transition, we can define when two quantum
ground states belong to the same phase: Let $|\Phi(0)\>$ be
the ground state of $H(0)$ and $|\Phi(1)\>$ be the ground
state of $H(1)$.  If we can find a smooth path $H(g)$,
$0\leq g\leq 1$ that connect  the two Hamiltonian $H(0)$ and
$H(1)$ such that there is no phase transition along the
path, then the two quantum ground states $|\Phi(0)\>$ and
$|\Phi(1)\>$ belong to the same phase.  We note that
``connected by a smooth path'' define an equivalence relation
between quantum states.  A quantum phase is an equivalence
class of such equivalence relation. Such an  equivalence
class is called an universality class.

\begin{figure}
\begin{center}
\includegraphics[scale=0.5]{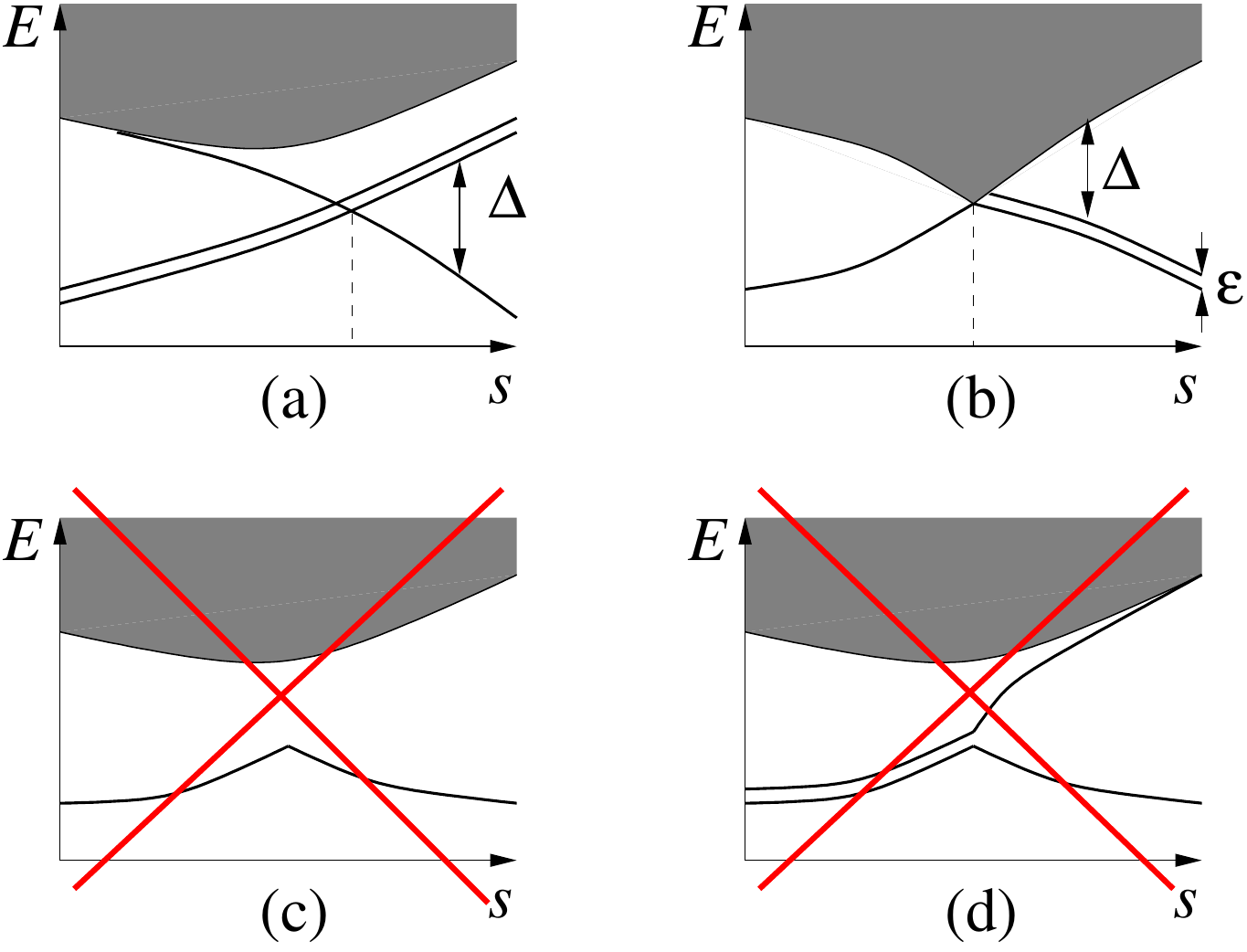}
\end{center}
\caption{
(Color online)
Energy spectrum of a gapped system as a function of
a parameter $s$ in the Hamiltonian.
(a,b) For gapped system, a quantum phase
transition can happen only when energy gap closes.
(a) describes a first order  quantum phase
transition (caused by level crossing).
(b) describes a continuous quantum phase
transition which has a continuum of gapless
excitations at the transition point.
(c) and (d) cannot happen for generic states.
A gapped system may have ground state degeneracy, where
the energy splitting between the ground states
vanishes when system size $L\to \infty$: $\lim_{L\to \infty}
\eps=0$. The energy gap $\Del$ between ground and excited states on the other hand
remains finite as $L\to \infty$.
}
\label{trans}
\end{figure}

If $|\Phi(0)\>$ is the ground state of $H(0)$ and in the
limit of infinite system size all excitations above
$|\Phi(0)\>$ have a gap, then for small enough $g$, we
believe that the systems described by $H(g)$ are also
gapped.\cite{BHM1044} In this case, we can show that, the
ground state of $H(g)$, $|\Phi(g)\>$, is in the same phase
as $|\Phi(0)\>$, for small enough $g$,  \ie the average
$\<O\>(g)$ is a smooth function of $g$ near $g=0$ for any
local operator $O$.\cite{HW0541} After scaling $g$ to $1$,
we find that:
\[\frm{\emph{If the energy gap for $H(g)$ is finite for all
$g$ in the segment $[0,1]$, then there is no phase
transition along the path $g$.
}}
\]
In other words, for gapped system, a quantum phase
transition can happen only when energy gap closes\cite{HW0541} (see Fig.
\ref{trans}).\footnote{
For each Hamiltonian, we can introduce an algebra of local
operators. If the energy gap for $H(g)$ is finite for all
$g$ in the segment $[0,1]$, \Ref{HW0541} show that the
algebra of local operators for different $g$'s are
isomorphic to each other at low energies, using a
quasi-adiabatic continuation of quantum states.}
Here, we would like to assume that the
reverse is also true: a closing of the energy gap for a
gapped state always induces a phase transition.  Or more
precisely
\[\frm{\emph{If two gapped states $|\Phi(0)\>$ and
$|\Phi(1)\>$ are in the same phase, then we can always
find a family
of Hamiltonian $H(g)$, such that
the energy gap for $H(g)$ are finite for all
$g$ in the segment $[0,1]$, and $|\Phi(0)\>$ and
$|\Phi(1)\>$ are ground states of $H(0)$ and $H(1)$
respectively.
}}
\]
The above two boxed statements imply that two gapped quantum
states are in the same phase $|\Phi(0)\> \sim |\Phi(1)\>$
if and only if they can be connected by an adiabatic
evolution that does not close the energy gap.

Given two states, $|\Phi(0)\>$ and $|\Phi(1)\>$, determining the existence of such a gapped adiabatic connection can be hard. We would like to have a more operationally practical equivalence relation between states in the same phase. Here we would like to show that \emph{two gapped states $|\Phi(0)\>$
and $|\Phi(1)\>$ are in the same phase, if and only if they are
related by a local unitary (LU) evolution. } We define a local unitary(LU) evolution as an unitary operation generated by time evolution of a local Hamiltonian for a finite time. That is,
\begin{align}
\label{LUdef}
 |\Phi(1)\> \sim |\Phi(0)\> \text{\ iff\ }
 |\Phi(1)\> =  \cT[e^{-i\int_0^1 dg\, \t H(g)}] |\Phi(0)\>
\end{align}
where $\cT$ is the path-ordering operator and $\t
H(g)=\sum_{\v i} O_{\v i}(g)$ is a sum of local Hermitian
operators. Note that $\t H(g)$ is in general different from
the adiabatic path $H(g)$ that connects the two states.

First, assume that two states $|\Phi(0)\>$ and $|\Phi(1)\>$
are in the same phase, therefore we can find a gapped
adiabatic path $H(g)$ between the states. The existence of a
gap prevents the system to be excited to higher energy
levels and leads to a local unitary evolution, the
Quasi-adiabatic Continuation as defined in \Ref{HW0541},
that maps from one state to the other. That is,
\begin{equation}
\label{lcltrn}
 |\Phi(1)\> = U |\Phi(0)\>, \ \ \ \ \ \
U=\cT[e^{-i\int_0^1 dg\, \t H(g)}]
\end{equation}
The exact form of $\t H(g)$ is given in \Ref{HW0541, BHM1044} and will be discussed in more detail in Appendix.

On the other hand, the reverse is also true: \emph{if two
gapped states $|\Phi(0)\>$ and $|\Phi(1)\>$ are related by a
local unitary evolution, then they are in the same
phase. } Since $|\Phi(0)\>$ and $|\Phi(1)\>$ are related by
a local unitary evolution, we have $ |\Phi(1)\>  =
\cT[e^{-i\int_0^1 dg\, \t H(g)}] |\Phi(0)\> $.  Let us
introduce
\begin{align}
|\Phi(s)\>  = U(s)
|\Phi(0)\> ,\ \ \ \ \ U(s)=
\cT[e^{-i\int_0^s dg\, \t H(g)}] .
\end{align}
Assume $|\Phi(0)\>$ is a ground state of $H(0)$, then
$|\Phi(s)\>$ is a ground state of $H(s)= U(s) H U^\dag(s)$. If $H(s)$ remains local and gapped for all $s \in [0,1]$, then we have found an adiabatic connection between $|\Phi(0)\>$ and $|\Phi(1)\>$.

First, let us show that $H(s)$ is  a local Hamiltonian.
Since $H$ is a local Hamiltonian, it has a form
$H=\sum_{\v i} O_{\v i}$ where $O_{\v i}$ only acts on a
cluster whose size is $\xi$.  $\xi$ is called the range of
interaction of $H$.  We see that $H(s)$ has a form
$H(s)=\sum_{\v i} O_{\v i}(s)$, where
$O_{\v i}(s) = U(s) O_{\v i} U^\dag(s)$.
To show that $O_{\v i}(s)$ only acts on a cluster of a
finite size, we note that for a local system
described by $\t H(g)$, the propagation velocities of its
excitations have a maximum value $v_{max}$.  Since $O_{\v
i}(s)$ can be viewed as the time evolution of $O_{\v i}$ by
$\t H(t)$ from $t=0$ to $t=s$, we find that $O_{\v i}(s)$
only acts on a cluster of size $\xi+\t \xi+s
v_{max}$,\cite{LR7251,HW0541}  where $\t\xi$ is the range of
interaction of $\t H$.  Thus $H(s)$ are indeed local
Hamiltonian.

If $H$ has a finite energy gap, then $H(s)$ also have a
finite energy gap for any $s$. As $s$ goes for $0$ to $1$,
the ground state of the local Hamiltonians, $H(s)$, goes
from $|\Phi(0)\>$ to $|\Phi(1)\>$.  Thus the two states
$|\Phi(0)\>$ and $|\Phi(1)\>$ belong to the same phase.
This completes our argument that states related by a local
unitary evolution belong to the same phase.

Thus through the above discussion, we show that
\[
\frm{
Two gapped ground states,\footnote{A state is called a gapped state
if there exist a local Hamiltonian such that the state is
the gapped ground state of the Hamiltonian.} $|\Phi(0)\>$ and $|\Phi(1)\>$, belong to the same
phase if and only if they are related by a local unitary evolution \eq{LUdef}.}
\]
A more detailed and more rigorous discussion of this
equivalence relation is given in Appendix A, where exact
bounds on locality and transformation error is given.

The relation \eq{LUdef} defines an equivalence relation between
$|\Phi(0)\>$ and $|\Phi(1)\>$. The equivalence classes
of such an equivalence relation represent different quantum phases.
So the above result implies that the equivalence classes
of the LU evolutions are the universality classes of
quantum phases for gapped states.

\section{Topological order is a pattern of long-range entanglement}
\label{TOLREn}

Using the LU evolution, we can obtain a more
precise description (or definition) of short-range entanglement:
\[
\frm{
\emph{A state has only short-range entanglement if and only if it can be
transformed into an unentangled state (\ie a direct-product state)
through a local unitary evolution}.
}
\]
If a state cannot be transformed into an unentangled state
through a LU evolution, then the state has long-range
entanglement. We also see that
\[
\frm{
\emph{All states with short-range entanglement can
transform into each other through local unitary
evolutions.
}
}
\]
Thus all states with short-range entanglement belong to the
same phase. The local unitary evolutions we consider here do
not have any symmetry. If we require certain symmetry of the
local unitary evolutions, states with short-range
entanglement may belong to different symmetry breaking
phases, which will be discussed in section \ref{symtop}.

Since a direct-product state is a state with trivial
topological order, we see that a state with a short-range
entanglement also has a trivial topological order.  This
leads us to conclude that a non-trivial topological order is
related to long-range entanglement.  Since two gapped states
related by a LU evolution belong to the same phase, thus two
gapped states related by a local unitary evolution have the
same topological order.  In other words,
\[
\frm{
\emph{Topological order describes the equivalent classes
defined by  local unitary evolutions}.
}
\]
Or more pictorially, topological order is a pattern of
long-range entanglement.  In \Ref{BHV}, it was shown that
the ``topologically non-trivial'' ground states, such as the
toric code,\cite{K032} cannot be changed into a
``topologically trivial'' state such as a product state by
any unitary locality-preserving operator.  In other words,
those ``topologically non-trivial'' ground states have
long-range entanglement.

\begin{figure}
\begin{center}
\includegraphics[scale=0.5]{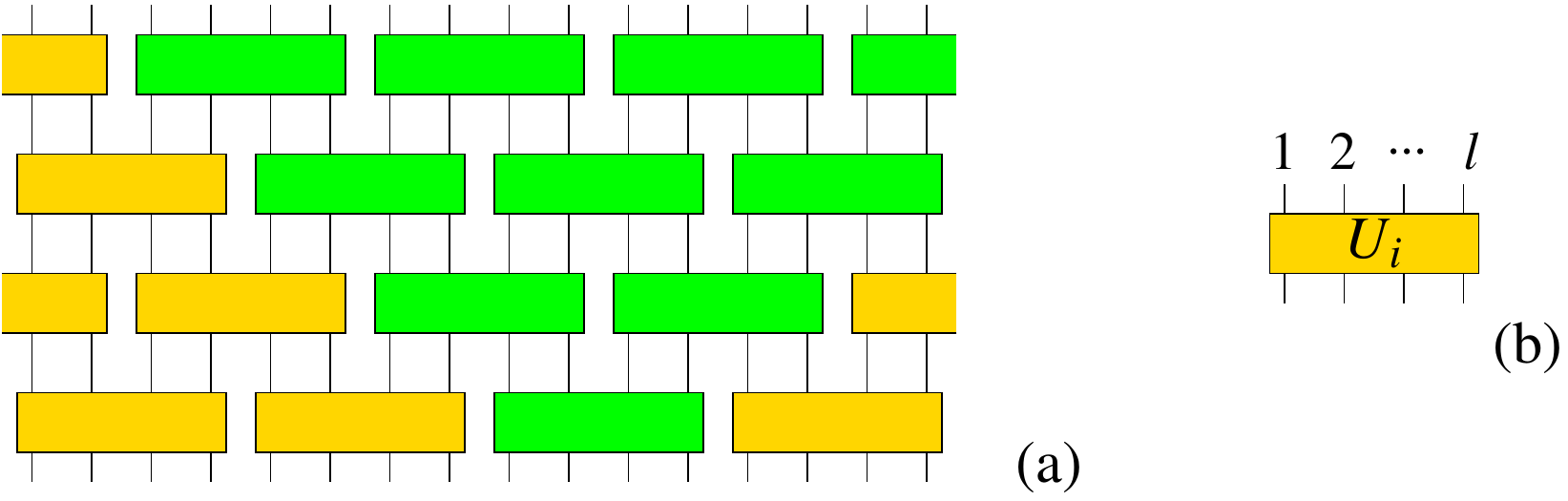}
\end{center}
\caption{
(Color online)
(a) A graphic representation of a
quantum circuit, which is formed by
(b) unitary operations on
patches of finite size $l$. The green shading
represents a causal structure.
}
\label{qc}
\end{figure}

\section{The LU evolutions and quantum circuits}

The LU evolutions introduced here is closely related to
\emph{quantum circuits with finite depth}.  To define
quantum circuits, let us introduce  piece-wise local unitary
operators.  A piece-wise local unitary operator has a form $
U_{pwl}= \prod_{i} U_i$ where $\{ U_i \}$ is a set of
unitary operators that act on non overlapping regions. The
size of each region is less than some finite number $l$. The
unitary operator $U_{pwl}$ defined in this way is called a
piece-wise local unitary operator with range $l$.  A quantum
circuit with depth $M$ is given by the product of $M$
piece-wise local unitary operators (see Fig. \ref{qc}):
$ U^M_{circ}= U_{pwl}^{(1)} U_{pwl}^{(2)} \cdots U_{pwl}^{(M)}$.
In quantum information theory, it is known that finite time
unitary evolution with local Hamiltonian (LU evolution
defined before) can be simulated with constant depth quantum
circuit and vice-verse. Therefore, the equivalence relation
\eqn{LUdef} can be equivalently stated in terms of constant
depth quantum circuits:
\begin{equation}
\label{PhiUcPhi}
|\Phi(1)\> \sim |\Phi(0)\> \text{ iff }
 |\Phi(1)\> = U^M_{circ} |\Phi(0)\>
\end{equation}
where $M$ is a constant independent of system size. Because
of their equivalence, we will use the term ``Local Unitary
Transformation'' to refer to both local unitary evolution and
constant depth quantum circuit in general.

The LU transformation defined through LU evolution
\eq{LUdef} is more general. It can be easily generalized to
study topological orders and quantum phases with symmetries
(see section \ref{symtop}).\cite{Wqoslpub,GW0931} The
quantum circuit has a more clear and simple causal
structure. However, the quantum circuit approach breaks the
translation symmetry.  So it is more suitable for studying
quantum phases that do not have translation symmetry.

In fact, people have been using quantum circuits to classify
many-body quantum states which correspond to quantum phases
of matter.  In \Ref{VCL}, the local unitary transformations
described by quantum circuits was used to define a
renormalization group transformations for states and
establish an equivalence relation in which states are
equivalent if they are connected by a local unitary
transformation.  Such an approach was used to classify 1D
matrix product states.  In \Ref{V0705}, the local unitary
transformations with disentanglers was used to perform a renormalization
group transformations for states, which give rise to the
multi-scale entanglement renormalization ansatz (MERA) in
one and higher dimensions.  The disentanglers and the
isometries in MERA can be used to study quantum phases and
quantum phase transitions in one and higher dimensions.
Later in this paper, we will use the quantum circuit
description of LU transformations to classify 2D topological
orders through classifying the fixed-point
LU transformations.

\section{Symmetry breaking orders and symmetry protected
topological orders}
\label{symtop}

In the above discussions, we have defined phases without any
symmetry consideration.  The $\t H(g)$ or $U_{pwl}$ in the
LU transformation does not need to have any symmetry and can
be sum/product of any local operators.  In this case, two
Hamiltonians with an adiabatic connection are in the same
phase even if they may have different symmetries.  Also, all
states with short-range entanglement belong to the same
phase (under the LU transformations that do not have any
symmetry).

On the other hand, we can consider only Hamiltonians $H$
with certain symmetries and define phases as the equivalent
classes of symmetric local unitary transformations:
\begin{equation}
\begin{array}{llll}
   |\Psi\>\sim &\cT\Big( e^{-\imth \int_0^1 d g\; \t H(g)} \Big)
|\Psi\> \
 \text{ or }\ |\Psi\> \sim U^M_{circ} |\Psi\> \nonumber
\end{array}
\end{equation}
where $\t H(g)$ or $U^M_{circ}$ has the same symmetries as
$H$. We note that the symmetric local unitary transformation
in the form $\cT\Big( e^{-\imth \int_0^1 d g\; \t H(g)}
\Big)$ always connect to the identity transformation
continuously.  This may not be the case for the
transformation in the form $U^M_{circ}$.  To rule out that
possibility, we define symmetric local unitary
transformations as those that connect to the identity
transformation continuously.

The equivalent classes of the symmetric LU transformations
have very different structures compared to those of LU
transformations without symmetry.  Each equivalent class of
the symmetric LU transformations is smaller and there are
more kinds of classes, in general.

\begin{figure}
\begin{center}
\includegraphics[scale=0.5]{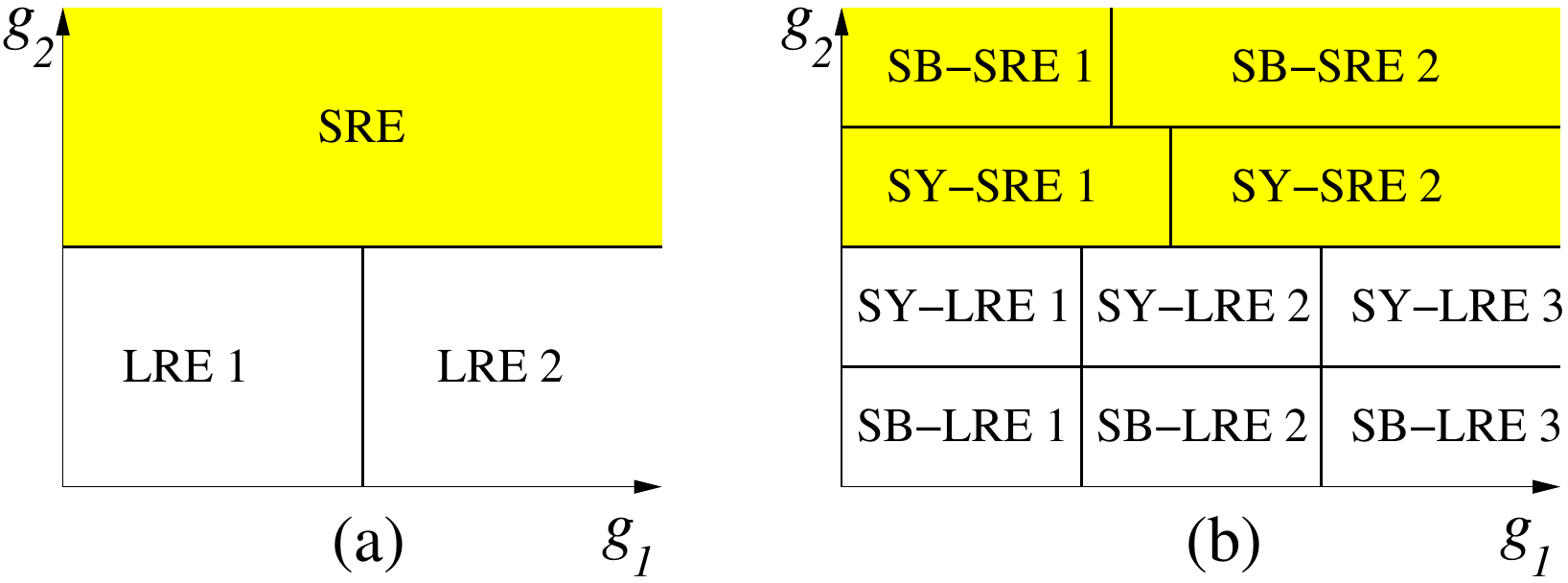}
\end{center}
\caption{
(Color online)
(a) The possible phases for a Hamiltonian $H(g_1,g_2)$ without
any symmetry.
(b) The possible phases for a Hamiltonian $H_\text{symm}(g_1,g_2)$ with
some symmetries.
The shaded regions in (a) and (b) represent the phases with
short range entanglement (\ie those ground states
can be transformed into a direct product state via a generic LU
transformations that do not have any symmetry.)
}
\label{topsymm}
\end{figure}

Fig. \ref{topsymm} compares the structure of phases for a
system without any symmetry and a system with some symmetry
in more detail.  For a system without any symmetry, all the
short-range-entangled (SRE) states (\ie those ground states
can be transformed into a direct product state via a generic
LU transformations that do not have any symmetry) are in the
same phase (SRE in Fig. \ref{topsymm}(a)).  On the other
hand, long range entanglement (LRE) can have many different
patterns that give rise to different topological phases (LRE
1 and LRE 2 in Fig.  \ref{topsymm}(a)).  The different
topological orders usually give rise to quasiparticles with
different fractional statistics and fractional charges.

For a system with some symmetries,
the phase structure can be much more complicated.
The short-range-entangled states no longer belong to
the same phase, since the equivalence relation is described
by more special symmetric LU transformations:\\
(A) States with short range entanglement belong to
different equivalent classes of the symmetric LU
transformations if they have different broken symmetries.
They correspond to the symmetry-breaking (SB)
short-range-entanglement phases SB-SRE 1 and SB-SRE 2 in
Fig.  \ref{topsymm}(b).  They are Landau's symmetry
breaking states.\\
(B) States with short range entanglement can belong to
different equivalent classes of the symmetric LU
transformations even if they do not break any symmetry of
the system. (In this case, they have the same symmetry.)  They correspond to
the symmetric (SY) short-range-entangled phases SY-SRE 1 and
SY-SRE 2 in Fig.  \ref{topsymm}(b).  We say those states
have symmetry protected topological orders.  Haldane
phase\cite{H8364} and $S_z=0$ phase of spin-1 chain are
examples of states with the same symmetry which belong to
two different equivalent classes of symmetric LU
transformations (with parity symmetry).\cite{GW0931,PBT0959}
Band and topological
insulators\cite{KM0501,BZ,KM0502,MB0706,FKM0703,QHZ0837} are
other examples of states that have the same symmetry and at
the same time belong to two different equivalent classes of
symmetric LU transformations (with time reversal symmetry).

Also, for a system with some symmetries,
the long-range-entangled states are
divided into more classes (more phases):\\
(C) Symmetry breaking and long range entanglement can appear
together in a state, such as SB-LRE 1, SB-LRE 2, \etc in
Fig.  \ref{topsymm}(b). The topological superconducting
states are examples of such phases.\cite{RG,KoLW}\\
(D) Long-range-entangled states that do not break any
symmetry can also belong to different phases such as the
symmetric long-range-entanglement phases SY-LRE 1, SY-LRE 2,
\etc in Fig.  \ref{topsymm}(b). The many different $\mathbb{Z}_2$
symmetric spin liquids with spin rotation, translation,
and time-reversal symmetries are examples of those
phases.\cite{Wqoslpub,KLW,KW}
Some time-reversal symmetric topological orders were also called
topological Mott-insulators or fractionalized
topological insulators.\cite{RQC,ZRV,PB,YK,MQK,SBM}

\section{
Local unitary transformation and wave function
renormalization
}
\label{wvrg}

After defining topological order as the equivalent classes
of many-body wave functions under LU transformations, we
like to ask: how to describe (or label) the different
equivalent classes (\ie the different topological orders or
patterns of long-range entanglement)?

One simple way to do so is to use the full wave function
which completely describe  the different topological orders.
But the full wave functions contain a lot of non-universal
short-range entanglement.  As a result, such a labeling
scheme is a very inefficient many-to-one labeling scheme of
topological orders.  To find a more efficient or even
one-to-one labeling scheme, we need to remove the
non-universal short-range entanglement.

\begin{figure}
\begin{center}
\includegraphics[scale=0.4]{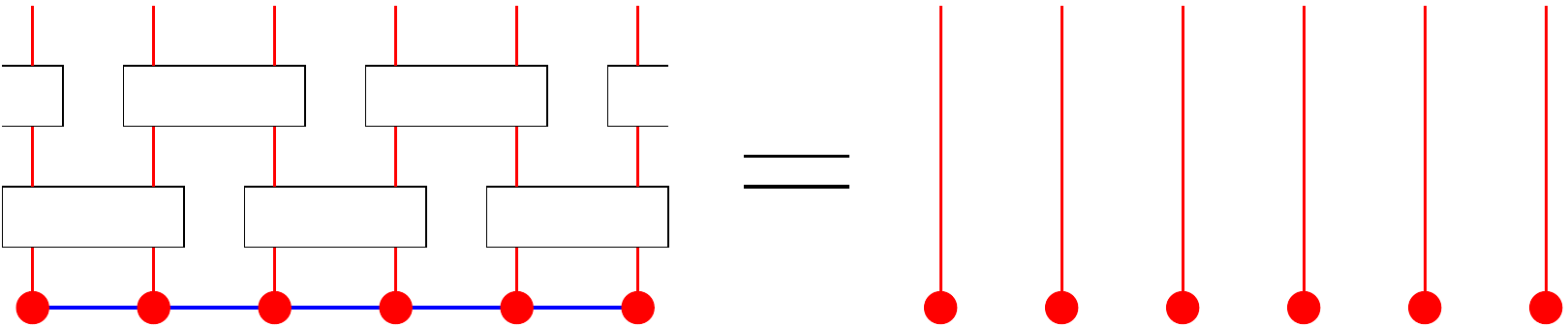}
\end{center}
\caption{
(Color online)
A finite depth quantum circuit can transform a
state $|\Phi\>$ into a direct-product state, if and only if
the state $|\Phi\>$ has no long-range quantum entanglement.
Here, a dot represents a site with physical degrees of
freedom. A vertical line carries an index that
label the different physical states on a site.
The presence of horizontal lines between dots
represents quantum entanglement.
}
\label{transSE}
\end{figure}

\begin{figure} \begin{center}
\includegraphics[scale=0.4]{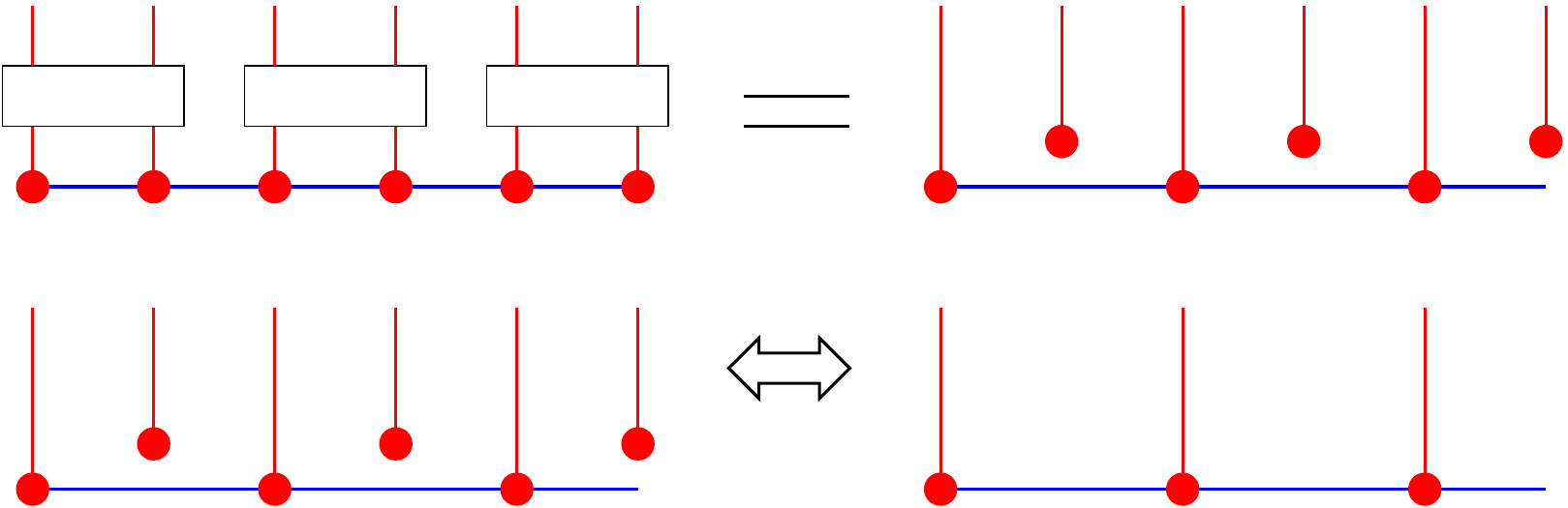}
\end{center}
\caption{
(Color online)
A piece-wise local unitary transformation can transform
some degrees of freedom in a state $\Phi\>$ into a direct
product.  Removing/adding the degrees of freedom in the form
of direct product defines an additional equivalence relation
that defines the topological order (or classes of long-range
entanglement).
}
\label{entre}
\end{figure}

As the first application of the notion of LU transformation,
we would like to describe a wave function renormalization
group flow introduced in \Ref{LWstrnet},\onlinecite{V0705}.  The wave
function renormalization can remove the short-range
entanglement and simplify the wave function.  In \Ref{LWstrnet},
the wave function renormalization for string-net states is generated by the
following two basic moves
\begin{align}
\label{Omv}
 \Phi
\bpm \includegraphics[height=0.3in]{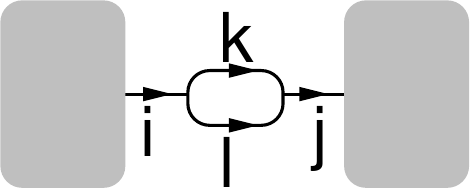} \epm  =&
\delta_{ij}
\Phi
\bpm \includegraphics[height=0.3in]{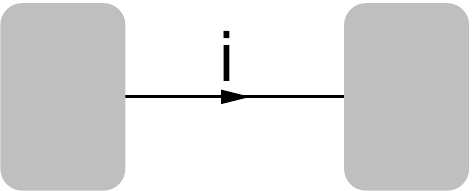} \epm
\\
\label{Fmv}
 \Phi
\bpm \includegraphics[height=0.3in]{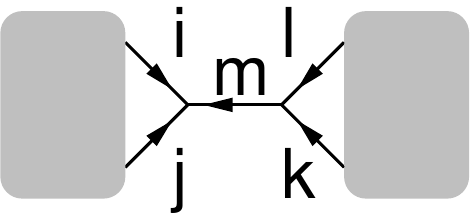} \epm  =&
\sum_{n}
F^{jim^*}_{lk^*n^*}
\Phi
\bpm \includegraphics[height=0.3in]{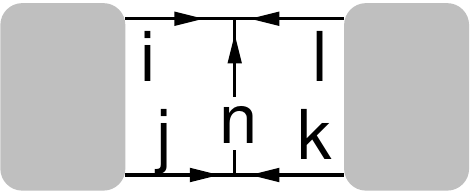} \epm
\end{align}
(Note that the definition of the F-tensor in \Ref{LWstrnet}
is slightly different from the definition in this paper.)
The two basic moves can generate a generic wave function
renormalization which can reduce the string-net wave
functions to very simple forms.\cite{LWstrnet,LWtopent}
Later in \Ref{V0705}, the  wave function renormalization for
generic states was discussed in a more general setting, and
was called MERA.  The two basic string-net moves \eq{Omv}
and \eq{Fmv} correspond to the isometry and the disentangler
in MERA respectively.  In the MERA approach, the isometries
and the disentanglers are applied in a layered fashion,
while in the string-net approach, the two basic moves can be
applied arbitrarily.  In this section, we will follow the
MERA setup to describe the wave function renormalization.
Later in this paper, we will follow the string-net setup to
study the fixed-point wave functions.

Note that we can use a LU transformation $U$ to transform some
degrees of freedom in a state into direct product (see Fig.
\ref{entre}). We then remove those degrees of freedom in the
form of direct product. Such a procedure does not change
the topological order. The reverse process of adding
degrees of freedom in the form of direct product also does not
change the topological order.  We call the local
transformation in Fig. \ref{entre} that changes the degrees
of freedom a generalized local unitary (gLU) transformation.
It is clear that a generalized local unitary transformation
inside a region $A$ does not change the reduced density
matrix $\rho_A$ for the region $A$.  This is the reason why
we say that (generalized) local unitary transformations
cannot change long-range entanglement and topological order.
Similarly, the addition or removal of decoupled
degrees of freedom to or from the Hamiltonian, $H
\leftrightarrow H\otimes H_{dp}$, will not change the phase
of the Hamiltonian (\ie the ground states of $H$ and
$H\otimes H_{dp}$ are in the same phase), if those degrees
of freedom form a direct product state (\ie the ground state
of $H_{dp}$ is a  direct product state).

\begin{figure}
\begin{center}
\includegraphics[scale=0.45]{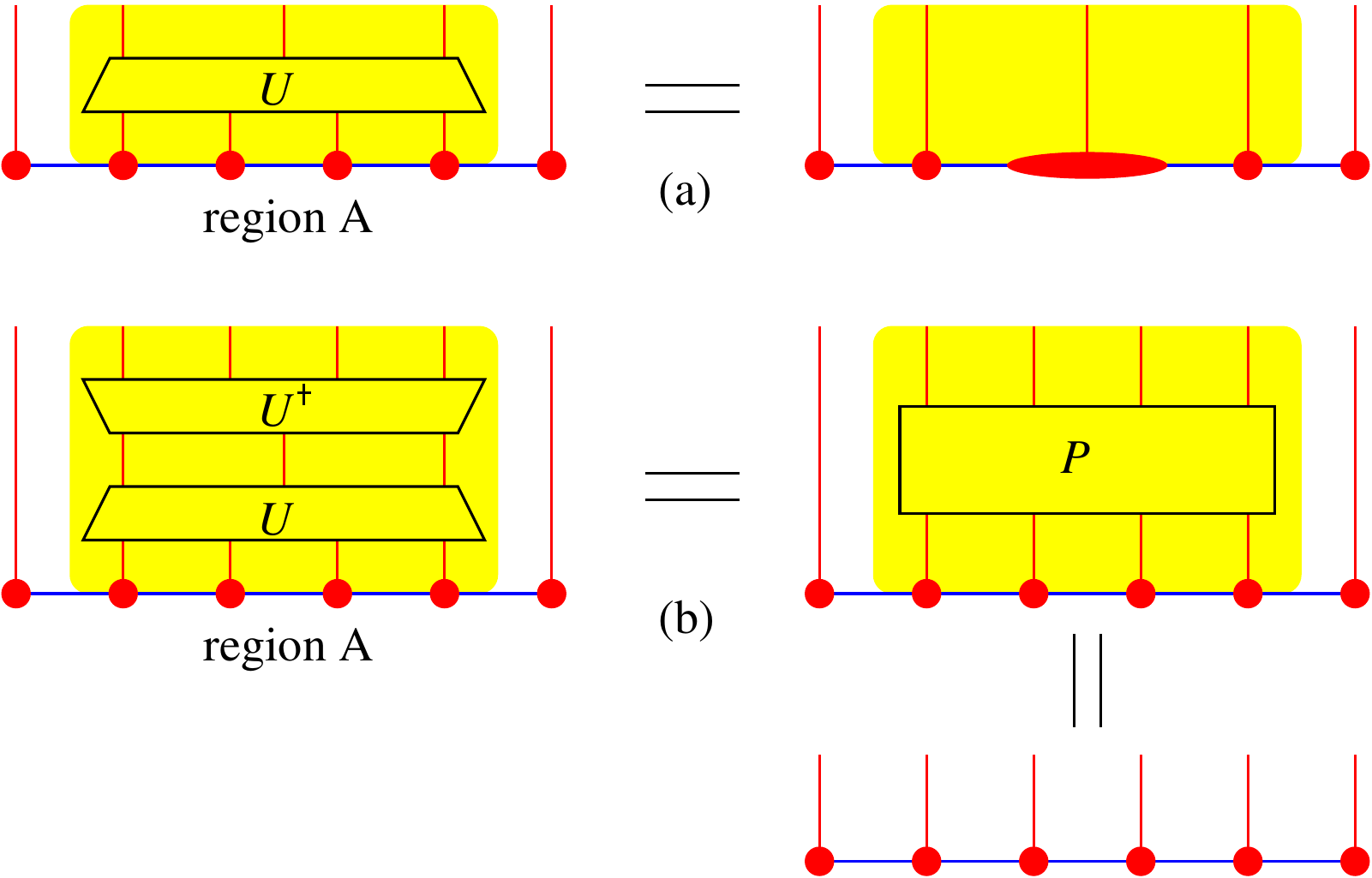}
\end{center}
\caption{
(Color online)
(a) A gLU transformation $U$ acts in
region A of a state $|\Phi\>$, which reduces the
degree freedom in region A to those contained only in the
support space of $|\Phi\>$ in region A.
(b) $U^\dag U=P$ is a projector that does not change the
state $|\Phi\>$.
}
\label{gLUT}
\end{figure}

Let us define the gLU transformation $U$ more carefully
and in a more general setting.
Consider a state $|\Phi\>$.  Let $\rho_A$ be the reduced
density matrix of $|\Phi\>$ in region $A$.  $\rho_A$ may act
in a subspace of the total Hilbert space $V_A$ in region A,
which is called the support space $V^{sp}_A$ of region $A$.
The dimension $D^{sp}_A$ of $V^{sp}_A$ is called support
dimension of region $A$.  Now the Hilbert space $V_A$ in
region A can be written as $V_A=V_A^{sp}\oplus \bar
V_A^{sp}$. Let $|\t \psi_{i}\>$, $i=1,...,D^{sp}_A$ be a
basis of this support space $V^{sp}_A$, $|\t \psi_{i}\>$,
$i=D^{sp}_A+1,...,D_A$ be a basis of $\bar V^{sp}_A$, where
$D_A$ is the dimension of $V_A$, and $|\psi_{i}\>$,
$i=1,...,D_A$ be a basis of $V_A$.  We can introduce a LU
transformation $U^{full}$ which rotates the basis
$|\psi_{i}\>$ to $|\t \psi_{i}\>$.  We note that in the new
basis, the wave function only has non-zero amplitudes on the
first $D^{sp}_A$ basis vectors.  Thus, in the new basis
$|\t \psi_{i}\>$, we can reduce the range of the label $i$
from $[1,D_A]$ to $[1,D^{sp}_A]$ without losing any
information.  This motivates us to introduce the gLU
transformation as a rotation from $|\psi_{i}\>$,
$i=1,...,D_A$ to
$|\t\psi_{i}\>$, $i=1,...,D^{sp}_A$. The rectangular matrix $U$ is given by
$U_{ij}=\<\t\psi_{i}|\psi_{j}\>$.  We also regard the
inverse of $U$, $U^\dag$, as a gLU transformation.  A LU
transformation is viewed as a special case of gLU
transformation where the degrees of freedom are not changed.
Clearly $U^\dag U=P$ and $U U^\dag=P'$ are two projectors.
The action of $P$ does not change the state $|\Phi\>$ (see
Fig. \ref{gLUT}(b)).

We note that
despite the reduction of the
degrees of freedom, a gLU transformation defines an
equivalent relation.  Two states related by a gLU
transformation belong to the same phase.  The
renormalization flow induced by the gLU transformations
always flows within the same phase.

\section{
Wave function renormalization
and a classification of topological order
}

As an application of the wave function renormalization, in
this section, we will study the structure of fixed-point
wave functions under the wave function renormalization,
which will lead to a classification of topological order
(without any symmetry).

We note that as wave functions flow to a fixed point, the
gLU transformations in each step of the renormalization also
flow to a fixed point.  So instead of studying fixed-point wave
functions, here, we will study the fixed-point gLU
transformations.  For this purpose, we need to fix the
renormalization scheme.  In the following, we will discuss a
renormalization scheme motivated by the string-net wave
function.\cite{LWstrnet} After we specify a proper wave
function renormalization scheme, then the fixed-point wave
function is simply the wave function whose ``form'' does not
change under the wave function renormalization.

Since those fixed-point gLU transformations do not change
the fixed-point wave function, their actions on the
fixed-point wave function do not depend on the order of the
actions. This allows us to obtain many conditions that gLU
transformations must satisfy.  From those conditions, we can
determine the forms of allowed fixed-point gLU
transformations.  This leads to a general description and a
classification of topological orders and their corresponding
fixed-point wave functions.

The renormalization scheme that we will discuss was first
used in \Ref{LWstrnet} to characterize the scale invariant
string-net wave function.  It is also used in \Ref{LWtopent}
to simplify the string-net state in a region, which allows
us to calculate the entanglement entropy of the string-net
state exactly.  A similar approach was used in
\Ref{KRV0923} to show quantum-double/string-net wave function to be a
fixed-point wave function and its connection to 2D
MERA.\cite{V0705}  In the
following, we will generalize those discussions by not
starting with string-net wave functions.  We just try to
construct local unitary transformations at a fixed point. We
will see that the fixed-point conditions on the gLU
transformations lead to a mathematical structure that is
similar to the tensor category theory --
the mathematical framework behind the string-net states.

\subsection{Quantum states on a graph}

\begin{figure}[t]
\begin{center}
\includegraphics[scale=0.45]{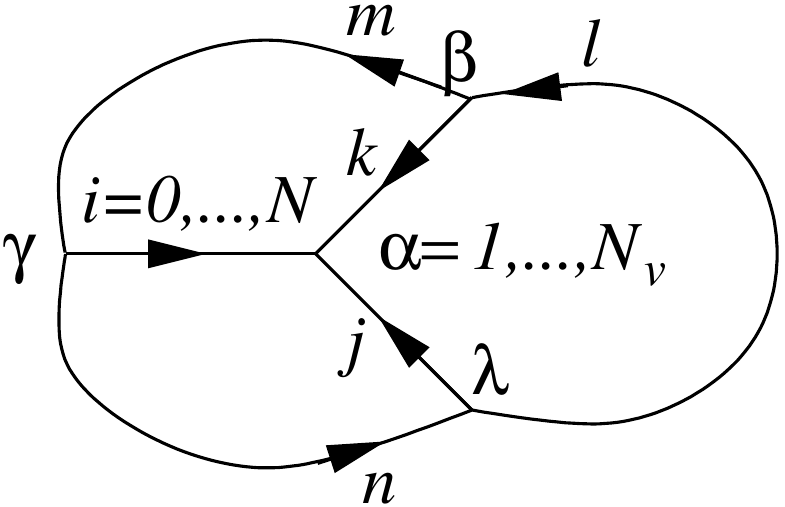}
\end{center}
\caption{
A quantum state on a graph $G$.  There are $N+1$ states on
each edge which are labeled by $l=0,...,N$.   There are
$N_v$ states on each vertex which are labeled by $\al=1,...,
N_v$.
}
\label{NijkV}
\end{figure}

\begin{figure}[t]
\begin{center}
\includegraphics[scale=0.5]{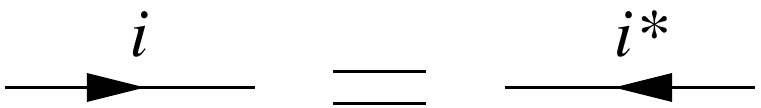}
\end{center}
\caption{
The mapping $i\to i^*$ corresponds to the reverse
of the orientation of the edge
}
\label{llstar}
\end{figure}

Since the wave function renormalization may change the
lattice structure, we will consider quantum states defined on
a generic trivalence graph $G$: Each edge has $N+1$ states,
labeled by $i=0,...,N$ (see Fig.  \ref{NijkV}).  We assume
that the index $i$ on the edge admits an one-to-one mapping
$*$: $i\to i^*$ that satisfies $(i^*)^*=i$.  As a result,
the edges of the graph are oriented.  The mapping $i\to i^*$
corresponds to the reverse of the orientation of the edge
(see Fig.  \ref{llstar}).
Each vertex also has physical states, labeled by $\al=1,...,
N_v$ (see Fig. \ref{NijkV}).

Each labeled graph (see Fig. \ref{NijkV}) corresponds to a
state and all the labeled graphs form an orthonormal basis.
Our fixed-point state is a superposition of those basis
states: $|\Phi_\text{fix}\>=\sum_\text{all conf.}
\Phi_\text{fix}\left( \bmm
\includegraphics[scale=0.18]{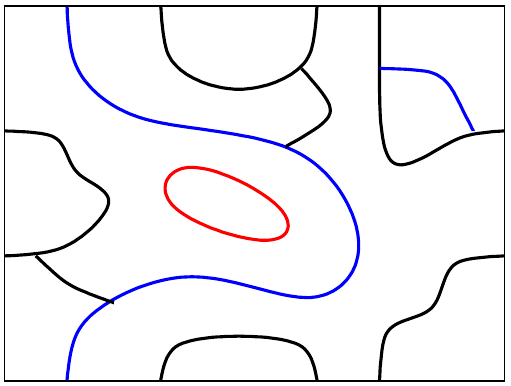}\emm \right) \left |\bmm
\includegraphics[scale=0.2]{strnet}\emm\right \>$.

Here we will make an important assumption about the
fixed-point wave function.  We will assume that the
fixed-point wave function is ``topological'': two labeled
graphs have the same amplitude if the two labeled graphs can
be deformed into each other continuously on the plane
without the vertices crossing the links.  For example $
\psi_\text{fix}\bpm \includegraphics[scale=0.25]{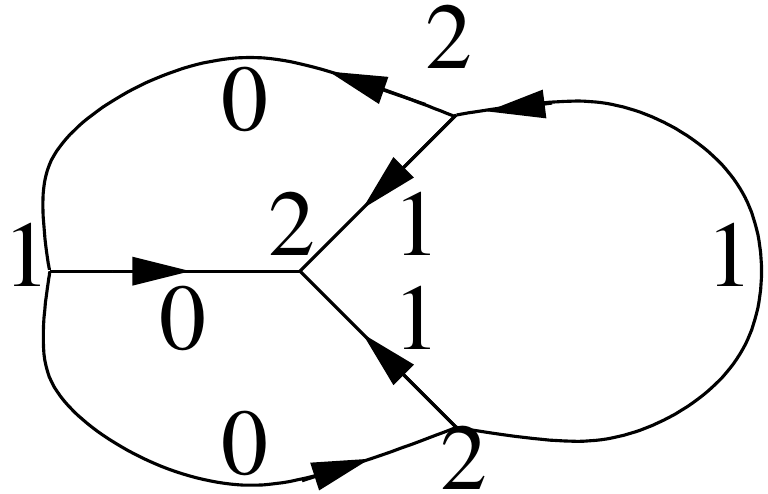}\epm
=\psi_\text{fix}\bpm \includegraphics[scale=0.25]{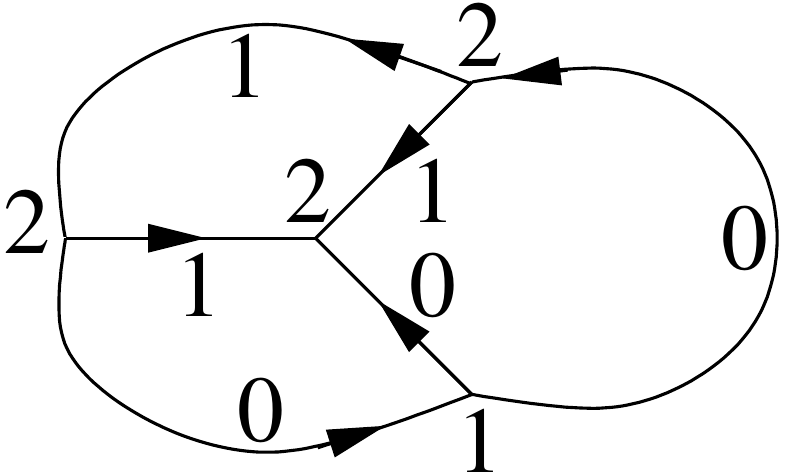}\epm
$.  Due to such an assumption, the topological orders studied
in this paper may not be most general.

We also assume that all the fixed-point states on each
different graphs to have the same ``form''.  This assumption
is motivated by the fact that during wave function
renormalization, we transform a state on one graph to a
state on a different graph.  The ``fixed-point'' means that
the wave functions on those different graphs are all
determined by the same collection of the rules, which
defines the meaning of having the same ``form''.  However,
the wave function for a given graph can have different total
phases if the wave function is calculated by applying the
rules in different orders.  Those rules are noting but the
fixed-point gLU transformations.

\subsection{The structure of entanglement in a fixed-point
wave function} \label{entstru}

\begin{figure}[t]
\begin{center}
\includegraphics[scale=0.45]{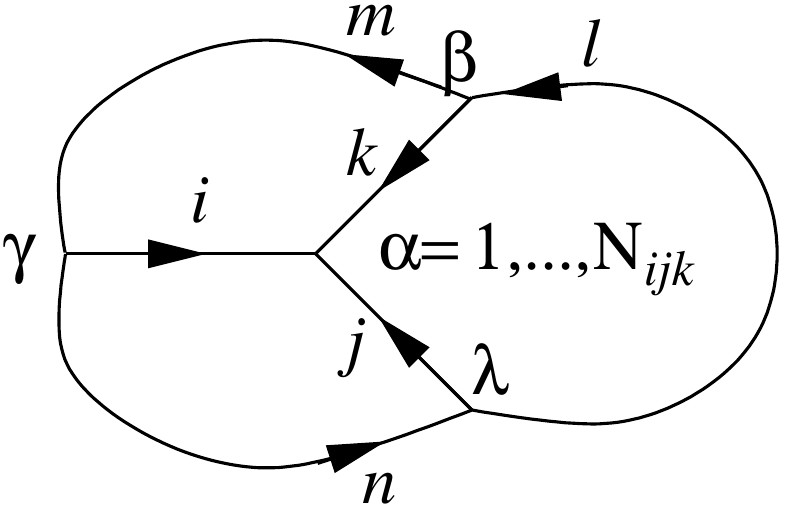}
\end{center}
\caption{
A quantum state on a graph $G$.
If the three edges of a vertex are in the states $i$, $j$,
and $k$ respectively, then the vertex has $N_{ijk}$ states,
labeled by $\al=1,..., N_{ijk}$. Note the orientation of the
edges are point towards to vertex.  Also note that $i\to
j\to k$ runs anti-clockwise.
}
\label{Nijk}
\end{figure}

Before describing the wave function renormalization, let us
examine the structure of entanglement of a fixed-point
wave function.  First, let us consider a fixed-point wave
function $\Phi_\text{fix}$ on a graph.
We examine the wave function on a patch of the graph,
for example, $\bmm\includegraphics[scale=.35]{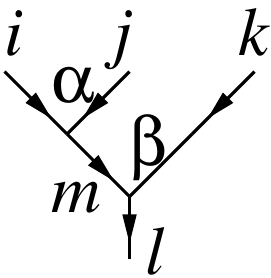}\emm$.
The fixed-point wave function
$\Phi_\text{fix}\bpm\includegraphics[scale=.35]{F1g}\epm$
(only the relevant part of the graph is drawn)  can be
viewed as a function of $\al,\bt,m$:
$\psi_{ijkl,\Ga}(\al,\bt,m)=\Phi_\text{fix} \bpm
\includegraphics[scale=.35]{F1g} \epm$ if we fix $i,j,k,l$
and the indices on other part of the graph.  (Here the
indices on other part of the graph is summarized by $\Ga$.)
As we vary the indices $\Ga$ on other part of graph (still
keeping $i$, $j$, $k$, and $l$ fixed), the wave function of
$\al,\bt,m$, $\psi_{ijkl,\Ga}(\al,\bt,m)$, may change.  All
those $\psi_{ijkl,\Ga}(\al,\bt,m)$ form a linear space of
dimension $D_{ijkl^*}$.  $D_{ijkl}$ is an important concept
that will appear later.  We note that the two vertices $\al$
and $\bt$ and the link $m$ form a region surrounded by the
links $i,j,k,l$.  So we will call the dimension-$D_{ijkl^*}$
space the support space $V_{ijkl^*}$ and $D_{ijkl^*}$ the support
dimension for the state $\Phi_\text{fix}$ on the region
surrounded by the fixed boundary state $i,j,k,l$.

Similarly, we can define $D_{ijk}$ as the support dimension
of the
$\Phi_\text{fix}\bpm\includegraphics[scale=.35]{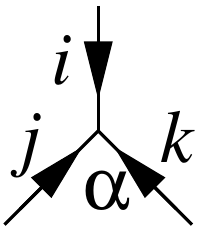}\epm$
on a region bounded by links $i,j,k$.  Since the region
contains only a single vertex $\al$ with $N_v$ physics
states, we have $D_{ijk}\leq N_v$.  We can use a local
unitary transformation on the vertex to reduce the range of
$\al$ to $1,...,N_{ijk}$ where $N_{ijk}=D_{ijk}$.  In the
rest of this paper, we will implement such a reduction.  So,
the number of physical states on a vertex depend on the
physical states of the edges that connect to the vertex.  If
the three edges of a vertex are in the states $i$, $j$, and
$k$ respectively, then the vertex has $N_{ijk}$ states,
labeled by $\al=1,..., N_{ijk}$ (see Fig. \ref{Nijk}).  Here
we assume that
\begin{align}
 N_{ijk}=N_{jki} .
\end{align}

We note that in the fixed-point wave function
$\Phi_\text{fix}\bpm\includegraphics[scale=.35]{F1g}\epm$,
the number of choices of $\al,\bt,m$ is
$N_{ijkl^*}=\sum_{m=0}^N N_{jim^*} N_{kml^*}$.  Thus the
support dimension $D_{ijkl^*}$ satisfies $D_{ijkl^*}\leq
N_{ijkl^*}$.  Here we will make an important assumption --
the saturation assumption: \emph{The fixed-point wave
function saturate the inequality:}
\begin{align}
\label{NijklD}
D_{ijkl^*}= N_{ijkl^*}\equiv \sum_{m=0}^N N_{jim^*} N_{kml^*} .
\end{align}
We will see that the entanglement structure described by
such a saturation assumption is invariant under the wave
function renormalization.

\subsection{The first type of wave function renormalization }

Our wave function renormalization scheme contains two types
of renormalization.  The first type of renormalization does
not change the degrees of freedom and corresponds to a local
unitary transformation.
It corresponds to locally deform the graph $\bmm
\includegraphics[scale=.35]{F1g} \emm$ to $\bmm
\includegraphics[scale=.35]{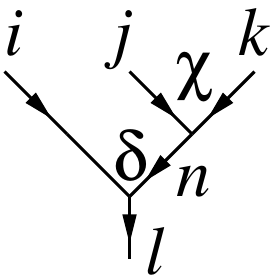} \emm$.  (The parts that are
not drawn are the same.) The fixed-point wave function on
the new graph is given by
$\Phi_\text{fix}\bpm\includegraphics[scale=.35]{F2g}\epm$.
Again, such a wave function can be viewed as a function of
$\chi,\del,n$:
$\t\psi_{ijkl,\Ga}(\chi,\del,n)=\Phi_\text{fix} \bpm
\includegraphics[scale=.35]{F2g} \epm$ if we fix $i,j,k,l$
and the indices on other part of the graph.  The support
dimension of the state
$\Phi_\text{fix}\bpm\includegraphics[scale=.35]{F2g}\epm$ on
the region surrounded by $i,j,k,l$ is  $\t D_{ijkl^*}$.
Again
$\t D_{ijkl^*} \leq \t N_{ijkl^*}$, where
$\t N_{ijkl^*}\equiv \sum_{n=0}^N N_{kjn^*} N_{l^*ni}$ is
number of choices of $\chi,\del,n$.
The saturation assumption implies that $ \t
N_{ijkl^*}=\t D_{ijkl^*} $.

The two fixed-point
wave functions
$\Phi_\text{fix}\bpm\includegraphics[scale=.35]{F1g}\epm$
and
$\Phi_\text{fix}\bpm\includegraphics[scale=.35]{F2g}\epm$
are related via a local unitary transformation.
Thus $ D_{ijkl^*}=\t D_{ijkl^*}$, which implies
\begin{align}
\label{NNNN}
\sum_{m=0}^N N_{jim^*} N_{kml^*}
=\sum_{n=0}^N N_{kjn^*} N_{l^*ni} .
\end{align}
We express such an unitary transformation
as
\begin{align}
\label{IHwavepsi}
\psi_{ijkl,\Ga}(\al,\bt,m) \simeq
\sum_{n=0}^N
\sum_{\chi=1}^{N_{kjn^*}}
\sum_{\del=1}^{N_{nil^*}}
 F^{ijm,\al\bt}_{kln,\chi\del}
\t\psi_{ijkl,\Ga}(\chi,\del,n)
\end{align}
or graphically as
\begin{align}
\label{IHwave}
 \Phi_\text{fix}
\bpm \includegraphics[scale=.40]{F1g} \epm   \simeq
\sum_{n=0}^N
\sum_{\chi=1}^{N_{kjn^*}}
\sum_{\del=1}^{N_{nil^*}}
 F^{ijm,\al\bt}_{kln,\chi\del}
\Phi_\text{fix}
\bpm \includegraphics[scale=.40]{F2g} \epm .
\end{align}
where $\simeq$ means equal up to a constant phase factor.
(Note that the total phase of the wave function is
unphysical.) We will call such a wave function
renormalization step a F-move.

We would like to remark that \eqn{IHwave}
relates two wave functions on two graphs $G_1$ and $G_2$
which only differ by a local reconnection.
We can choose the phase of $F^{ijm,\al\bt}_{kln,\chi\del}$
to make $\simeq$ into $=$:
\begin{align*}
 \Phi_\text{fix}
\bpm \includegraphics[scale=.40]{F1g} \epm   =
\sum_{n=0}^N
\sum_{\chi=1}^{N_{kjn^*}}
\sum_{\del=1}^{N_{nil^*}}
 F^{ijm,\al\bt}_{kln,\chi\del}
\Phi_\text{fix}
\bpm \includegraphics[scale=.40]{F2g} \epm .
\end{align*}
But such choice of phase only works for a particular pair of
graphs $G_1$ and $G_2$.  To use
$F^{ijm,\al\bt}_{kln,\chi\del}$ to relate all pair of states
that only differ by a local reconnection, in general, we may
have a phase ambiguity, with the value of phase depend on the
pair of graphs.  So \eqn{IHwave} can only be a relation up to
a total phase factor.

For fixed $i$, $j$, $k$, and $l$, the matrix $F^{ij}_{kl}$
with matrix elements $(F^{ij}_{kl})^{m,\al\bt}_{n,\chi\del}
= F^{ijm,\al\bt}_{kln,\chi\del} $ is a
$N_{ijkl^*}$ dimensional matrix (see \eq{NNNN}).  The
mapping $ \t\psi_{ijkl,\Ga}(\chi,\del,n) \to
\psi_{ijkl,\Ga}(\al,\bt,m)$ generated by the matrix
$F^{ij}_{kl}$ is unitary.  Since, as we change $\Ga$,
$\t\psi_{ijkl,\Ga}(\chi,\del,n)$ and
$\psi_{ijkl,\Ga}(\al,\bt,m)$ span two $N_{ijkl^*}$ dimensional
spaces.  Thus $F^{ij}_{kl}$ is a $N_{ijkl^*}\times N_{ijkl^*}$
unitary matrix
\begin{align}
\label{2FFstar}
\sum_{n,\chi,\del}
F^{ijm',\al'\bt'}_{kln,\chi\del}
(F^{ijm,\al\bt}_{kln,\chi\del})^*
=\del_{m\al\bt,m'\al'\bt'},
\end{align}
where $\del_{m\al\bt,m'\al'\bt'} =1$ when $m=m'$,
$\al=\al'$, $\bt=\bt'$, and $\del_{m\al\bt,m'\al'\bt'} =0$
otherwise.

We can deform
$\bmm \includegraphics[scale=.35]{F1g} \emm$
to
$\bmm \includegraphics[scale=.35]{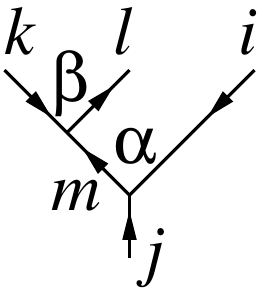} \emm$,
and
$\bmm \includegraphics[scale=.35]{F2g} \emm$
to
$\bmm \includegraphics[scale=.35]{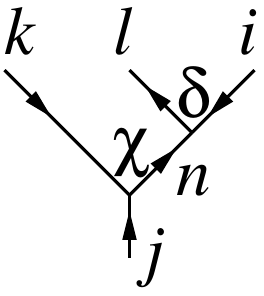} \emm$.
We see that
\begin{align}
\label{IHwaveA}
 \Phi_\text{fix}
\bpm \includegraphics[scale=.40]{F1gA} \epm  \simeq
\sum_{n=0}^N
\sum_{\chi,\del}
 F^{kl^*m^*,\bt\al}_{ij^*n^*,\del\chi}
\Phi_\text{fix}
\bpm \includegraphics[scale=.40]{F2gA} \epm .
\end{align}
Eqn. (\ref{IHwave}) and eqn. (\ref{IHwaveA}) relate the same
pair of graphs, and thus
\begin{align}
\label{1FF}
 F^{ijm,\al\bt}_{kln,\chi\del} \simeq F^{kl^*m^*,\bt\al}_{ij^*n^*,\del\chi}
\end{align}
(where we have used the condition $N_{ijkl}=D_{ijkl}$.)

Using the relation \eq{2FFstar}, we can rewrite
\eqn{IHwave} as
\begin{align}
\label{HIwavestar}
\Phi_\text{fix}
\bpm \includegraphics[scale=.40]{F2g} \epm \simeq
\sum_{m=0}^N
\sum_{\al,\bt}
 (F^{ijm,\al\bt}_{kln,\chi\del})^*
\Phi_\text{fix}
\bpm \includegraphics[scale=.40]{F1g} \epm  .
\end{align}
We can also express
$\Phi_\text{fix}\bpm \includegraphics[scale=.35]{F2g} \epm $
as
\begin{align}
\label{HIwave}
\Phi_\text{fix}
\bpm \includegraphics[scale=.40]{F2g} \epm \simeq
\sum_{m=0}^N
\sum_{\al,\bt}
 F^{jkn,\chi\del}_{l^*i^*m^*,\bt\al}
\Phi_\text{fix}
\bpm \includegraphics[scale=.40]{F1g} \epm
\end{align}
using the relabeled \eqn{IHwave}.
So we have
\begin{align}
\label{1FstarFap}
 (F^{ijm,\al\bt}_{kln,\chi\del})^*
\simeq
 F^{jkn,\chi\del}_{l^*i^*m^*,\bt\al} .
\end{align}
Since the total phase of the wave function is unphysical,
the  total phase of $F^{ijm,\al\bt}_{kln,\chi\del}$ can be
chosen arbitrarily.  We can choose the total phase of
$F^{ijm,\al\bt}_{kln,\chi\del}$ to make
\begin{align}
\label{1FstarF}
 (F^{ijm,\al\bt}_{kln,\chi\del})^*
=
 F^{jkn,\chi\del}_{l^*i^*m^*,\bt\al} .
\end{align}
If we apply \eqn{1FstarF} twice, we reproduce \eqn{1FF}.
Thus \eqn{1FF} is not independent and can be dropped.

{}From the graphic representation \eq{IHwave}, We note that
\begin{align}
\label{Feq0}
& F^{ijm,\al\bt}_{kln,\chi\del} = 0 \text{ when}
\\
&
N_{jim^*}<1 \text{ or }
N_{kml^*}<1 \text{ or }
N_{kjn^*}<1 \text{ or }
N_{nil^*}<1
.
\nonumber
\end{align}
When $N_{jim^*}<1$ or $ N_{kml^*}<1$, the left-hand-side of
\eqn{IHwave} is always zero.  Thus
$F^{ijm,\al\bt}_{kln,\chi\del} = 0$ when $N_{jim^*}<1$ or $
N_{kml^*}<1$.  When $N_{kjn^*}<1$ or  $N_{nil^*}<1$, wave
function on the right-hand-side of \eqn{IHwave} is always
zero.  So we can choose $F^{ijm,\al\bt}_{kln,\chi\del} = 0$
when $N_{kjn^*}<1$ or  $N_{nil^*}<1$.

The F-move \eq{IHwave} maps the wave
functions on two different graphs through a
local unitary transformation.  Since we can locally transform one graph to
another graph through different paths, the F-move
\eq{IHwave} must satisfy certain self consistent condition.
For example the graph $ \bmm
\includegraphics[scale=.35]{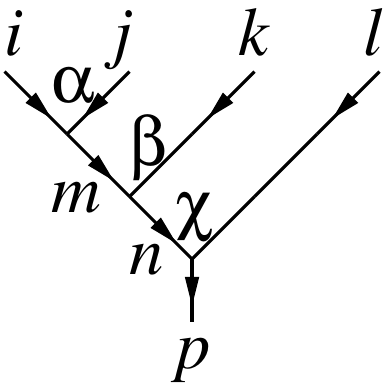} \emm $ can be transformed
to $ \bmm \includegraphics[scale=.35]{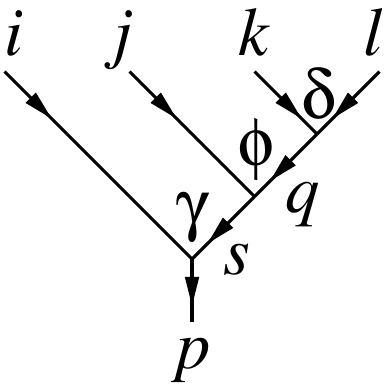} \emm $ through
two different paths; one contains two steps of local
transformations and the other contains three steps of local
transformations as described by \eqn{IHwave}.
The two paths lead to the following relations between the
wave functions:
\begin{widetext}
\begin{align}
\label{FFrel}
\Phi_\text{fix} \bpm \includegraphics[scale=.40]{pent1g} \epm
& \simeq \sum_{q,\del,\eps}
F^{mkn,\bt\chi}_{lpq,\del\eps}
\Phi_\text{fix} \bpm \includegraphics[scale=.40]{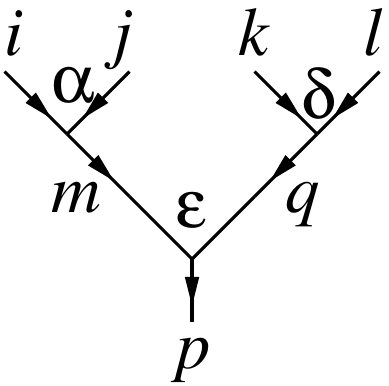} \epm
 \simeq \sum_{q,\del,\eps;s,\phi,\ga}
F^{mkn,\bt\chi}_{lpq,\del\eps}
F^{ijm,\al\eps}_{qps,\phi\ga}
\Phi_\text{fix} \bpm \includegraphics[scale=.40]{pent3g} \epm ,
\end{align}
\begin{align}
\label{FFFrel}
\Phi_\text{fix} \bpm \includegraphics[scale=.40]{pent1g} \epm
& \simeq \sum_{t,\eta,\vphi}
F^{ijm,\al\bt}_{knt,\eta\vphi}
\Phi_\text{fix} \bpm \includegraphics[scale=.40]{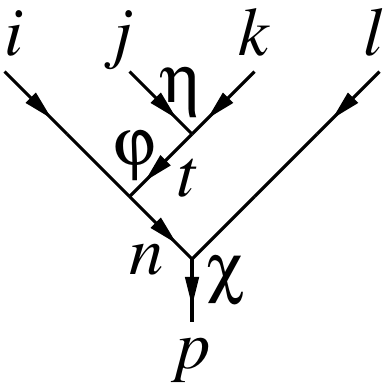} \epm
 \simeq \sum_{t,\eta,\vphi;s,\ka,\ga}
F^{ijm,\al\bt}_{knt,\eta\vphi}
F^{itn,\vphi\chi}_{lps,\ka\ga}
\Phi_\text{fix} \bpm \includegraphics[scale=.40]{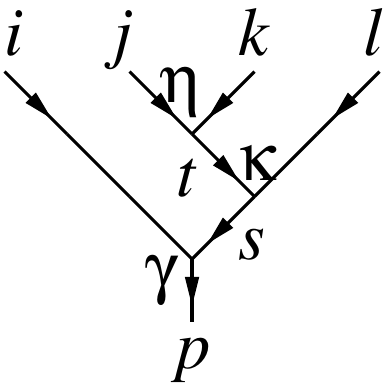} \epm
\nonumber\\
& \simeq \sum_{t,\eta,\ka;\vphi;s,\ka,\ga;q,\del,\phi}
F^{ijm,\al\bt}_{knt,\eta\vphi}
F^{itn,\vphi\chi}_{lps,\ka\ga}
F^{jkt,\eta\ka}_{lsq,\del\phi}
\Phi_\text{fix} \bpm \includegraphics[scale=.40]{pent3g} \epm .
\end{align}
\end{widetext}
The consistence of the above two relations leads
a condition on the $F$ tensor.

To obtain such a condition, let us fix $i$, $j$, $k$, $l$,
$p$, and view $\Phi_\text{fix} \bpm \includegraphics[scale=.35]{pent1g}
\epm$ as a function of $\al,\bt,\chi,m,n$:
$\psi(\al,\bt,\chi,m,n)=\Phi_\text{fix} \bpm
\includegraphics[scale=.35]{pent1g} \epm$.  As we vary
indices on other part of graph, we obtain different wave
functions $\psi(\al,\bt,\chi,m,n)$ which form a dimension
$D_{ijklp^*}$ space.  In other words, $D_{ijklp^*}$ is the
support dimension of the state $\Phi_\text{fix}$ on the region
$\al,\bt,\chi,m,n$ with boundary state $i,j,k,l,p$ fixed (see the
discussion in section \ref{entstru}).  Since the number of
choices of $\al,\bt,\chi,m,n$ is $N_{ijklp^*}=\sum_{m,n}
N_{jim^*} N_{kmn^*}N_{lnp^*}$, we have $D_{ijklp^*} \leq
N_{ijklp^*}$.  Here we require a similar saturation condition as
in \eq{NijklD}:
\begin{align}
\label{NijklpD}
N_{ijklp^*} =  D_{ijklp^*}
\end{align}
Similarly, the number of choices of $\del,\phi,\ga,q,s$ in
$\Phi_\text{fix} \bpm \includegraphics[scale=.35]{pent3g} \epm $ is
also $N_{ijklp^*}$. Here we again
assume $\t D_{ijklp^*}=N_{ijklp^*}$, where
$\t D_{ijklp^*}$ is the support dimension of
$\Phi_\text{fix} \bpm \includegraphics[scale=.35]{pent3g} \epm $
on the region bounded by $i,j,k,l,p$.

So the two relations \eq{FFrel} and \eq{FFFrel}
can be viewed as two relations between a pair of vectors
in the two $D_{ijklp^*}$ dimensional vector spaces.
As we vary indices on other part of graph
(still keeping $i,j,k,l,p$ fixed),
each vector in the pair can span the full
$D_{ijklp^*}$ dimensional vector space.
So the validity of the two relations \eq{FFrel} and \eq{FFFrel}
implies that
\begin{align}
\label{penid}
&\ \ \ \
\sum_{t}
\sum_{\eta=1}^{N_{kjt^*}}
\sum_{\vphi=1}^{N_{tin^*}}
\sum_{\ka=1}^{N_{lts^*}}
F^{ijm,\al\bt}_{knt,\eta\vphi}
F^{itn,\vphi\chi}_{lps,\ka\ga}
F^{jkt,\eta\ka}_{lsq,\del\phi}
\nonumber\\
& = \e^{\imth \th_F}
\sum_{\eps=1}^{N_{qmp^*}}
F^{mkn,\bt\chi}_{lpq,\del\eps}
F^{ijm,\al\eps}_{qps,\phi\ga}
.
\end{align}
which is a generalization of the famous pentagon identity
(due to the extra constant phase factor $\e^{\imth \th_F}$).  We
will call such a relation projective pentagon identity.  The
projective pentagon identity is a set of nonlinear equations
satisfied by the rank-10 tensor
$F^{ijm,\al\bt}_{kln,\chi\del}$ and $\th_F$.  The above
consistency relation is equivalent to the requirement that
the local unitary transformations described by \eqn{IHwave}
on different paths all commute with each other up to a total
phase factor.

\subsection{The second type of wave function renormalization}

The second type of wave function renormalization does change the
degrees of freedom and corresponds to a generalized local
unitary transformation.  One way to implement the second
type renormalization is to reduce $ \bmm
\includegraphics[scale=.35]{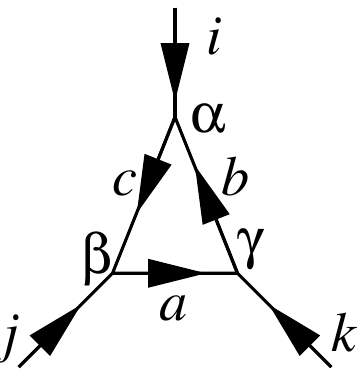} \emm $ to
$\bmm \includegraphics[scale=.35]{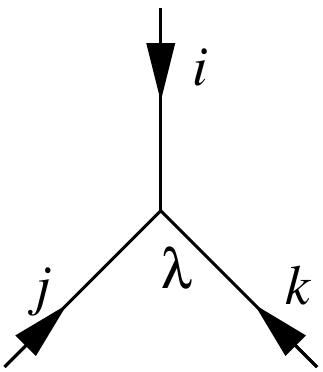} \emm
$ (the part of the graph that is not drawn is
unchanged):
\begin{align}
\label{4to1wave}
\Phi_\text{fix}\bpm \includegraphics[scale=.40]{4to1a} \epm
\simeq \sum_{\la=1}^{N_{ijk}}
F ^{abc,\al\bt\ga} _{ijk,\la}
\Phi_\text{fix} \bpm \includegraphics[scale=.40]{4to1b} \epm .
\end{align}

\begin{figure}[t]
\begin{center}
\includegraphics[scale=0.5]{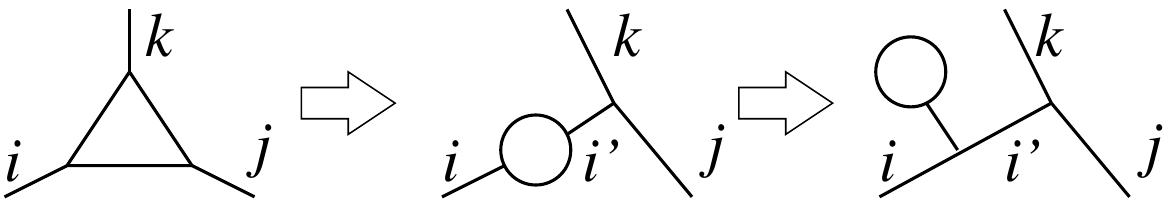}
\end{center}
\caption{
A ``triangle'' graph can be transformed into a ``tadpole''
via two steps of the first type of wave function renormalization
(\ie two steps of local unitary transformations).
}
\label{tritpol}
\end{figure}

But we can define a simpler second type renormalization, by noting that $\bmm
\includegraphics[scale=.35]{4to1a} \emm$ can be reduced to $\bmm
\includegraphics[scale=.35]{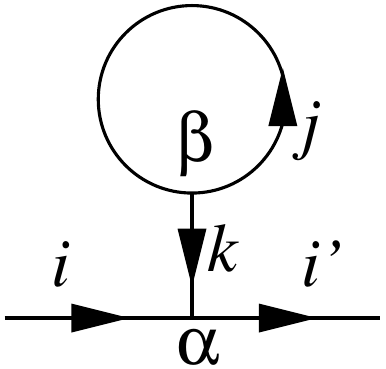} \emm$ via the first type of renormalization
steps (see Fig. \ref{tritpol}), which are local unitary transformations.  In
the simplified second type renormalization, we want to reduce $\bmm
\includegraphics[scale=.35]{tpolij} \emm$ to $\bmm
\includegraphics[scale=.35]{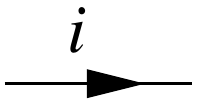} \emm$, so that we still have a trivalence
graph.  This requires that
the support dimension $D_{ii'^*}$ of
the fixed-point wave function $\Phi_\text{fix}
\bpm
\includegraphics[scale=.35]{tpolij} \epm $
is given by
\begin{align}
 D_{ii'^*}=\del_{ii'}.
\end{align}
This implies that
\begin{align}
\Phi_\text{fix} \bpm \includegraphics[scale=.40]{tpolij} \epm
\simeq \del_{ii'}
\Phi_\text{fix} \bpm \includegraphics[scale=.40]{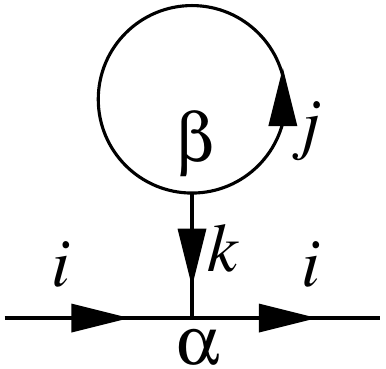} \epm  .
\end{align}
The simplified second type renormalization can now be written as
(since $D_{ii^*}=1$)
\begin{align}
\label{PhiP}
\Phi_\text{fix} \bpm \includegraphics[scale=.40]{tpol} \epm
\simeq  P_i^{kj,\al\bt}
\Phi_\text{fix} \bpm \includegraphics[scale=.40]{iline} \epm  .
\end{align}
We will call such a wave function renormalization step
a P-move.\cite{KRV0923}  Here $P_i^{kj,\al\bt} $ satisfies
\begin{align}
\label{Pnorm}
\sum_{k,j}
\sum_{\al=1}^{N_{kii^*}}
\sum_{\bt=1}^{N_{j^*jk^*}}
P_i^{kj,\al\bt} (P_i^{kj,\al\bt})^*=1
\end{align}
and
\begin{align}
 P_i^{kj,\al\bt}=0, \text{ if }
N_{kii^*}<1 \text{ or }
N_{j^*jk^*} <1 .
\end{align}
The condition \eq{Pnorm} ensures that the two wave functions
on the two sides of \eqn{PhiP} have the same normalization.
We note that the number of choices for the four indices
$(j,k,\al,\bt)$ in $P_i^{kj,\al\bt}$ must be equal or
greater than $1$:
\begin{align}
\label{Di}
 D_i=\sum_{j,k} N_{ii^*k} N_{jk^*j^*} \geq 1 .
\end{align}

Notice that
\begin{align}
\label{tpoltpol}
&\ \ \ \
\Phi_\text{fix} \bpm \includegraphics[scale=.40]{tpol} \epm
\simeq \sum_{m,\la,\ga}
F^{jj^*k,\bt\al}_{i^*i^*m^*,\la\ga}
\Phi_\text{fix} \bpm \includegraphics[scale=.40]{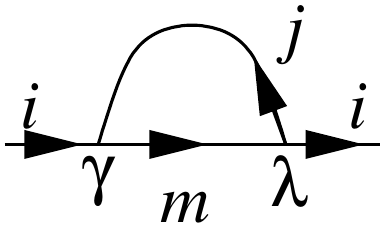} \epm
\nonumber\\
&\simeq
\sum_{m,\la,\ga,l,\nu,\mu}
F^{jj^*k,\bt\al}_{i^*i^*m^*,\la\ga}
F^{i^*mj,\la\ga}_{m^*i^*l,\nu\mu}
\Phi_\text{fix} \bpm \includegraphics[scale=.40]{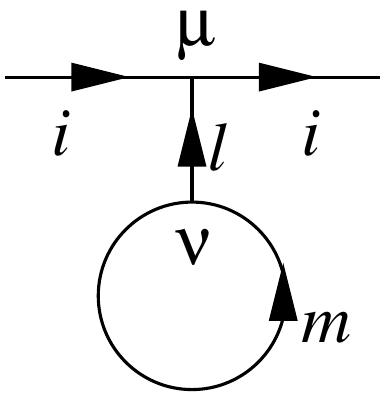} \epm
\end{align}
Using \eqn{PhiP} and its variation
\begin{align}
\label{PhiP1}
\Phi_\text{fix} \bpm \includegraphics[scale=.40]{tpolud} \epm
\simeq P_{i^*}^{lm,\mu\nu}
\Phi_\text{fix} \bpm \includegraphics[scale=.40]{iline} \epm  .
\end{align}
we can rewrite \eqn{tpoltpol} as
\begin{align}
\label{PFFP}
\e^{\imth\th_{P1}} P_i^{kj,\al\bt}=
\sum_{m,\la,\ga,l,\nu,\mu}
F^{jj^*k,\bt\al}_{i^*i^*m^*,\la\ga}
F^{i^*mj,\la\ga}_{m^*i^*l,\nu\mu}
P_{i^*}^{lm,\mu\nu}
\end{align}
which is a condition on $P_i^{kj,\al\bt}$.

More conditions on $F^{ijm,\al\bt}_{kln,\chi\del}$
and $P_i^{kj,\al\bt}$ can be obtained by noticing that
\begin{align}
\Phi_\text{fix} \bpm \includegraphics[scale=.40]{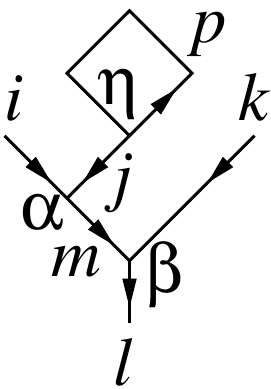} \epm
 \simeq
\sum_{n=0}^N
\sum_{\chi,\del}
 F^{ijm,\al\bt}_{kln,\chi\del}
\Phi_\text{fix} \bpm \includegraphics[scale=.40]{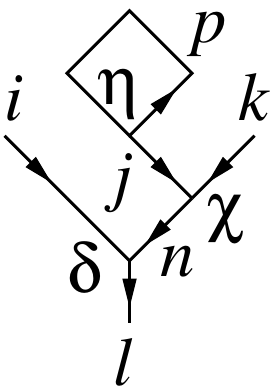} \epm ,
\end{align}
which implies that
\begin{align}
&\ \ \ \
P^{jp,\al\eta}_i \del_{im}\Phi_\text{fix} \bpm \includegraphics[scale=.40]{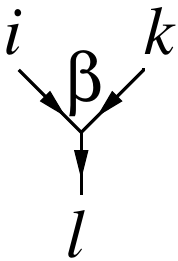} \epm
\nonumber\\
& \simeq
\sum_{n,\chi,\del}
 F^{ijm,\al\bt}_{kln,\chi\del}
P^{jp,\chi\eta}_{k^*} \del_{kn}
\Phi_\text{fix} \bpm \includegraphics[scale=.40]{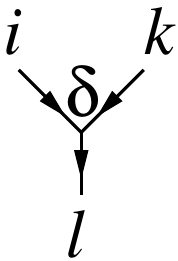} \epm .
\end{align}
We find
\begin{align}
\label{PFP}
&
\e^{\imth\th_{P2}} P^{jp,\al\eta}_i \del_{im} \del_{\bt\del}
=
\sum_{\chi=1}^{N_{kjk^*}} F^{ijm,\al\bt}_{klk,\chi\del} P^{jp,\chi\eta}_{k^*}
\nonumber\\
&\text{for all } k,i,l \text{ satisfying }
N_{kil^*}>0
.
\end{align}

\subsection{The fixed-point wave functions from the fixed-point gLU
transformations}

In the last section, we discussed the conditions that a
fixed-point gLU transformation
$(F^{ijm,\al\bt}_{kln,\ga\la},P_i^{kj,\al\bt})$ must
satisfy.  After finding a fixed-point gLU transformation
$(F^{ijm,\al\bt}_{kln,\ga\la},P_i^{kj,\al\bt})$, in this
section, we are going to discuss how to calculate the
corresponding fixed-point wave function $\Phi_\text{fix}$
from the solved fixed-point gLU transformation
$(F^{ijm,\al\bt}_{kln,\ga\la},P_i^{kj,\al\bt})$.

First we note that, using the two types of wave function
renormalization introduced above, we can reduce any graph to
$\bmm \includegraphics[scale=.35]{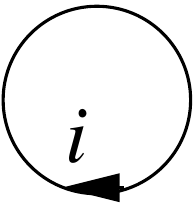} \emm$.  So, once we
know $\Phi_\text{fix} \bpm \includegraphics[scale=.35]{oi}
\epm$, we can reconstruct the full fixed-point wave function
$\Phi_\text{fix}$ on any connected graph.

Let us assume that
\begin{align}
\Phi_\text{fix} \bpm \includegraphics[scale=.40]{oi} \epm
=A^i=A^{i^*}
\end{align}
Here $A^i$ satisfy
\begin{align}
\label{AiAistar}
A^i= A^{i^*},\ \ \ \ \sum_i A^i (A^i)^*=1.
\end{align}
The condition $\sum_i A^i (A^i)^*=1$ is simply the
normalization condition of the wave function.  The condition
$A^i = A^{i^*}$ come from the fact that the graph $\bmm
\includegraphics[scale=.35]{oi} \emm$ can be deformed into
the graph $\bmm \includegraphics[scale=.35]{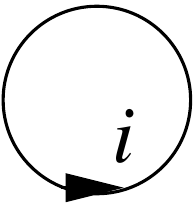} \emm$ on a
sphere.

To find the conditions that determine $A^i$, let us first
consider the fixed-point wave function where the index on a
link is $i$: $\Phi_\text{fix}(i,\Ga)= \Phi_\text{fix}\bpm
\includegraphics[scale=.35]{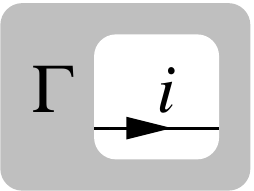} \epm$, where $\Ga$ are
indices on other part of graph.
We note that the graph $\bmm
\includegraphics[scale=.35]{iGa1} \emm$ can be deformed into
the graph $\bmm \includegraphics[scale=.35]{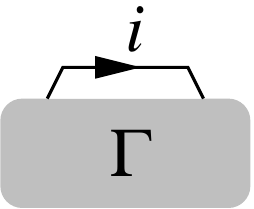} \emm$ on a
sphere.
Thus  $\Phi_\text{fix}\bpm \includegraphics[scale=.35]{iGa1} \epm=
\Phi_\text{fix}\bpm \includegraphics[scale=.35]{iGa2} \epm$.
Using the F-moves and the P-moves, we
can reduce $\bmm \includegraphics[scale=.35]{iGa2} \emm$ to
$\bmm \includegraphics[scale=.35]{oi} \emm$:
\begin{align}
 \Phi_\text{fix}\bpm \includegraphics[scale=.40]{iGa1} \epm=
\Phi_\text{fix}\bpm \includegraphics[scale=.40]{iGa2} \epm
\simeq f(i,\Ga) \Phi_\text{fix} \bpm \includegraphics[scale=.40]{oi} \epm
\end{align}
We see that
\begin{align}
\label{Anonz}
\Phi_\text{fix} \bpm \includegraphics[scale=.35]{oi} \epm=A^i\neq 0
\end{align}
for all $i$. Otherwise, any wave function with $i$-link
will be zero.

To find more conditions on $A^i$, we note that
\begin{align}
\Phi_\text{fix} \bpm \includegraphics[scale=.40]{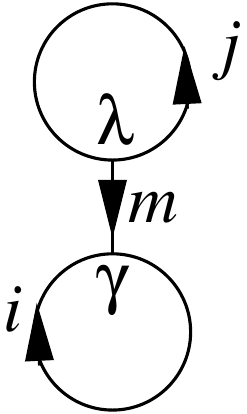} \epm
\simeq  P^{mj,\ga\la}_i
\Phi_\text{fix} \bpm \includegraphics[scale=.40]{oi} \epm
=  P^{mj,\ga\la}_iA^i .
\end{align}
By rotating $\bmm \includegraphics[scale=.35]{ooijm} \emm$
by 180$^\circ$, we can show that
$P^{mj,\ga\la}_i A^i
\simeq P^{m^*i^*,\la\ga}_{j^*} A^j$ or
\begin{align}
\label{PAPA}
P^{mj,\ga\la}_i A^i
=\e^{\imth \th_{A1}} P^{m^*i^*,\la\ga}_{j^*} A^j .
\end{align}
We also note that
\begin{align}
\Phi_\text{fix} \bpm \includegraphics[scale=.40]{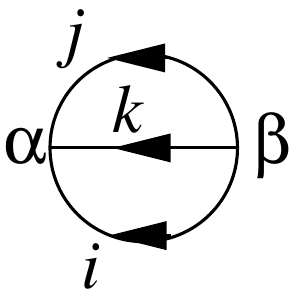} \epm
& \simeq
\sum_{m,\la,\ga}  F^{ijk^*,\al\bt}_{j^*im,\la\ga}
\Phi_\text{fix} \bpm \includegraphics[scale=.40]{ooijm} \epm
.
\end{align}
This allows us to show
\begin{align}
\label{PhiPA}
 \Phi^\th_{ikj,\al\bt}
=\e^{\imth \th'}
\sum_{m,\la,\ga}
F^{ijk^*,\al\bt}_{j^*im,\la\ga}
P^{mj,\ga\la}_i A^i
\end{align}
where
\begin{align}
\label{Phiikj}
&
\Phi^\th_{ikj,\al\bt},
\equiv
\Phi_\text{fix} \bpm \includegraphics[scale=.40]{thetaijk} \epm
\nonumber\\
&  \Phi^\th_{ikj,\al\bt}
= \e^{\imth \th_{A2}} \Phi^\th_{kji,\al\bt},
\nonumber\\
& \Phi^\th_{ikj,\al\bt}=0, \text{ if } N_{ikj}=0.
\end{align}
The condition $\Phi^\th_{ikj,\al\bt}
\simeq  \Phi^\th_{kji,\al\bt}$ comes from the
fact that
the graph
$\bmm \includegraphics[scale=.35]{thetaijk} \emm$
and the graph
$\bmm \includegraphics[scale=.35]{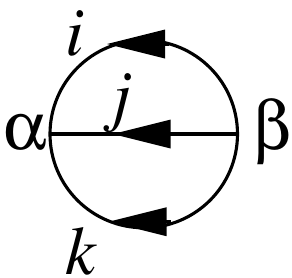} \emm$
can be deformed into each other on a sphere.

Also, for any given $i,j,k,\al$ that satisfy $N_{kji}>0$,
the wave function $\Phi_\text{fix} \bpm
\includegraphics[scale=.35]{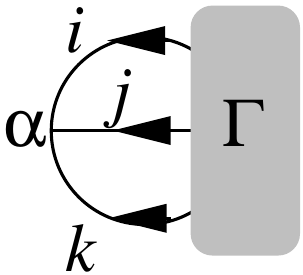} \epm$ must be non-zero
for some $\Ga$, where $\Ga$ represents indices on other part
of the graph.  Then after some F-moves and P-moves, we can
reduce $\bmm \includegraphics[scale=.35]{ijkGa} \emm$ to
$\bmm \includegraphics[scale=.35]{thetakij} \emm$.  So, for
any given $i,j,k,\al$ that satisfy $N_{kji}>0$,
$\Phi_\text{fix} \bpm \includegraphics[scale=.35]{thetakij}
\epm$ is non-zero for some $\bt$.  Since such a statement is
true for any choices of basis on the vertex $\al$, we  find
that for any given $i,j,k,\al$ that satisfy $N_{kji}>0$ and
for any non-zero vector $v_\al$, $\sum_\al v_\al
\Phi_\text{fix} \bpm \includegraphics[scale=.35]{thetakij}
\epm$ is non-zero for some $\bt$.  This means that the
matrix $M_{kji}$ is invertible, where $M_{kji}$ is a matrix
whose elements are given by $ (M_{kji})_{\al\bt} =
\Phi_\text{fix} \bpm \includegraphics[scale=.35]{thetakij}
\epm $. Let us define
$\det\Big[ \Phi_\text{fix} \bpm \includegraphics[scale=.35]{thetakij}
\epm \Big] = \det(M_{kji})$, we find that
\begin{align}
\label{Phinonz}
 \det\Big[ \Phi_\text{fix} \bpm \includegraphics[scale=.40]{thetakij}
\epm \Big]
 =\det[ \Phi^\th_{kji,\al\bt} ]
\neq 0.
\end{align}
The above also implies that
\begin{align}
\label{NNstar}
 N_{kji}=N_{i^*j^*k^*} .
\end{align}
The conditions
eqns. (\ref{AiAistar}, \ref{PAPA}, \ref{PhiPA}, \ref{Phiikj})
allow us to determine $A^i$ (and $\Phi^\th_{ikj,\al\bt}$).

{}From \eqn{Phiikj}, we see that relation
$\Phi_\text{fix}\bpm \includegraphics[scale=.35]{thetaijk} \epm =
\Phi_\text{fix}\bpm \includegraphics[scale=.35]{thetakij} \epm$
leads to some equations for
$F^{ijm,\al\bt}_{kln,\ga\la}$, $P_i^{kj,\al\bt}$, and $A^i$.
More equations for
$F^{ijm,\al\bt}_{kln,\ga\la}$, $P_i^{kj,\al\bt}$, and $A^i$,
can be obtained by using the relations
\begin{align}
\label{tetra}
\Phi_\text{fix} \bpm \includegraphics[scale=.40]{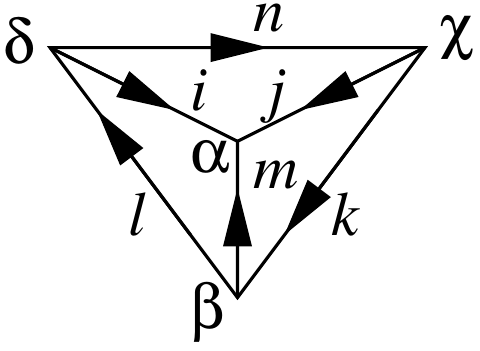} \epm &=
\Phi_\text{fix} \bpm \includegraphics[scale=.40]{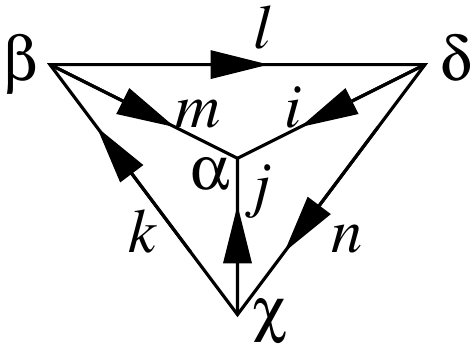} \epm
\nonumber\\
&=
\Phi_\text{fix} \bpm \includegraphics[scale=.40]{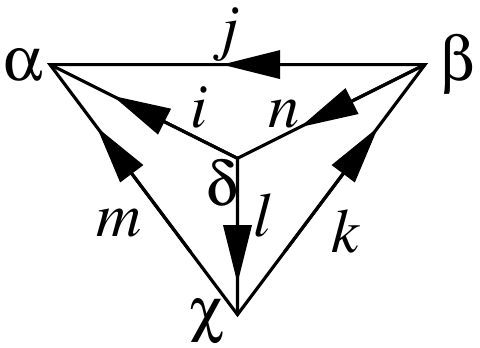} \epm
,
\end{align}
from the tetrahedron rotation symmetry\cite{LWstrnet,Hong} and
\begin{align}
\label{tetraR}
&
\Phi_\text{fix} \bpm \includegraphics[scale=.40]{tetra1} \epm
 \simeq
\sum_{\ga\la} F^{ijm^*,\al\bt}_{kln,\ga\la}
\Phi_\text{fix} \bpm \includegraphics[scale=.40]{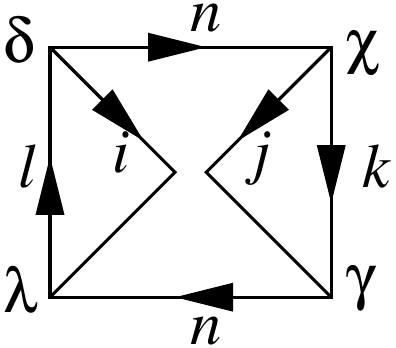} \epm
\nonumber\\
& \simeq
\sum_{\ga\la,p\si\eps}
F^{ijm^*,\al\bt}_{kln,\ga\la}
F^{ln^*i^*,\del\la}_{nlp^*,\si\eps}
\Phi_\text{fix} \bpm \includegraphics[scale=.40]{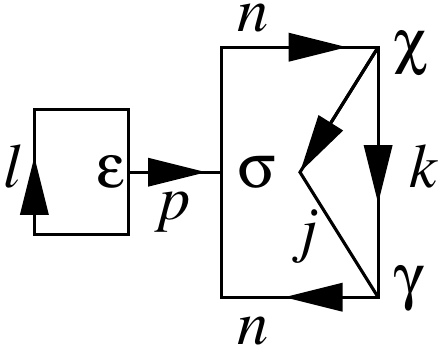} \epm
\nonumber\\
& \simeq
\sum_{\ga\la,p\si\eps}
F^{ijm^*,\al\bt}_{kln,\ga\la}
F^{ln^*i^*,\del\la}_{nlp^*,\si\eps}
P^{pl^*,\si\eps}_{n}
\Phi_\text{fix} \bpm \includegraphics[scale=.40]{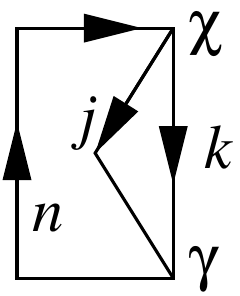} \epm
\nonumber\\
& =
\sum_{\ga\la,p\si\eps}
F^{ijm^*,\al\bt}_{kln,\ga\la}
F^{ln^*i^*,\del\la}_{nlp^*,\si\eps}
P^{pl^*,\si\eps}_{n}
\Phi^\th_{kjn^*,\ga\chi}
.
\end{align}
It is not clear if \eqn{tetra} and \eqn{tetraR} will lead to
new independent equations or not.  In the following
discussions, we will not include \eqn{tetra} and
\eqn{tetraR}. We find that, at least for simple cases, the
equations without \eqn{tetra} and \eqn{tetraR} are enough to
completely determine the solutions.

To summarize, the conditions (\ref{NNNN}, \ref{Di}, \ref{NNstar},
\ref{2FFstar}, \ref{1FstarF}, \ref{penid},
\ref{PFFP}, \ref{PFP}, \ref{Anonz}, \ref{AiAistar},
\ref{PAPA}, \ref{PhiPA}, \ref{Phiikj}, \ref{Phinonz}) form a
set of non-linear equations whose variables are $N_{ijk}$,
$F^{ijm,\al\bt}_{kln,\ga\la}$, $P_i^{kj,\al\bt}$, $A^i$, and
$(\th_F,\th_{P1},\th_{P2})$.
Finding $N_{ijk}$,
$F^{ijm,\al\bt}_{kln,\ga\la}$, $P_i^{kj,\al\bt}$, and $A^i$
that satisfy such a set of non-linear equations corresponds
to finding a fixed-point gLU transformation that has a
non-trivial fixed-point wave function.  So the solutions
$(N_{ijk}, F^{ijm,\al\bt}_{kln,\ga\la}, P_i^{kj,\al\bt},
A^i)$ give us a characterization of topological orders.
This may lead to a classification of topological order from
the local unitary transformation point of view.

\section{Simple solutions of the fixed-point conditions}

In this section, let us find some simple solutions of the
fixed-point conditions (\ref{NNNN}, \ref{Di}, \ref{NNstar},
\ref{2FFstar}, \ref{1FstarF}, \ref{penid},
\ref{PFFP}, \ref{PFP}, \ref{Anonz}, \ref{AiAistar},
\ref{PAPA}, \ref{PhiPA}, \ref{Phiikj}, \ref{Phinonz}) for
the fixed-point gLU transformations
$(N_{ijk},F^{ijm,\al\bt}_{kln,\ga\la},P_i^{kj,\al\bt})$ and
the fixed-point wave function $A^i$.

\subsection{Unimportant phase factors in the solutions}

Formally, the solutions of the fixed-point conditions are
not isolated. They are parameterized by several continuous
phase factors.  In this section, we will discuss the origin
of those phase factors.  We will see that those different
phase factors do not correspond to different states of
matter (\ie different equivalence classes of gLU
transformations).  So after removing those unimportant phase
factors, the solutions of the fixed-point conditions are
isolated (at least for the simple examples studied here).

We notice that, apart from two normalization conditions,
all of the fixed-point conditions are
linear in $P_i^{kj,\al\bt}$ and $A^i$.  Thus if
$(F^{ijm,\al\bt}_{kln,\ga\la},P_i^{kj,\al\bt},A^i)$ is a
solution, then $(F^{ijm,\al\bt}_{kln,\ga\la}, \e^{\imth
\phi_1} P_i^{kj,\al\bt}, \e^{\imth \phi_2} A^i)$ is also a
solution. However, the two phase factors $\e^{\imth
\phi_{1,2}}$ do not lead to different fixed-point wave
functions, since they only affect the
total phase of the wave function and are unphysical.  Thus
the total phases of $P^{kj}_i$ and $A^i$ can be adjusted. We
can use this degree of freedom to set, say, $P^{00,11}_0
\geq 0$ and  $A^0 > 0$.

Similarly the total phase of $F^{ijm,\al\bt}_{kln,\ga\la}$
is also unphysical and can be adjusted. We have used this
degree of freedom to reduce \eqn{1FstarFap} to
\eqn{1FstarF}.  But this does not totally fix the total
phase of $F^{ijm,\al\bt}_{kln,\ga\la}$. The transformation
$F^{ijm,\al\bt}_{kln,\ga\la}\to
-F^{ijm,\al\bt}_{kln,\ga\la}$ does not affect \eqn{1FstarF}.
We can use such a transformation to set the real part of a
non-zero component of $F^{ijm,\al\bt}_{kln,\ga\la}$ to be
positive.

The above three phase factors are unphysical.  However, the
fixed-point solutions may also contain phase factors that do
correspond to different fixed-point wave functions.  For
example, the local unitary transformation $\e^{\imth
\th_{l_0} \hat M_{l_0}}$ does not affect the fusion rule
$N_{ijk}$, where $\hat M_{l_0}$ is the number of links with
$|l_0\>$-state and $|l_0^*\>$-state.  Such a local unitary
transformation changes
$(F^{ijm,\al\bt}_{kln,\ga\la},P_i^{kj,\al\bt},A^i)$ and
generates a continuous family of the fixed-point wave
functions parameterized by $\th_{l_0}$.  Those wave functions
are related by local unitary transformations that
continuously connect to identity.  Thus, those fixed-point
wave functions all belong to the same phase.

Similarly, we can consider the following local unitary
transformation $|\al\> \to \sum_\bt U^{(i_0j_0k_0)}_{\al\bt}|\bt\>$ that
acts on each vertex with states $|i_0\>,|j_0\>,|k_0\>$ on
the three edges connecting to the vertex.  Such a local
unitary transformation also does not affect the fusion rule
$N_{ijk}$.  The new local unitary transformation
changes $(F^{ijm,\al\bt}_{kln,\ga\la},P_i^{kj,\al\bt},A^i)$
and generates a continuous family of the fixed-point wave
functions parameterized by the unitary matrix $U^{(i_0j_0k_0)}_{\al\bt}$.
Again, those fixed-point wave functions all belong to the
same phase.

In the following, we will study some  simple solutions of
the fixed-point conditions.  We find that, for those
examples, the solutions have no addition continuous parameter
apart from the phase factors discussed above.  This
suggests that the solutions of the fixed-point conditions
correspond to isolated zero-temperature phases.

\subsection{$N=1$ loop state}
\label{fixphase}

Let us first consider a system where there are only two
states $|0\>$ and $|1\>$ on each link of the graph.  We
choose $i^*=i$ and the simplest fusion rule that satisfies
\eqn{NNNN}, \eqn{Di}, \eqn{NNstar} is
\begin{align}
& N_{000}= N_{110}= N_{101}= N_{011}=1,
\nonumber\\
&
 \text{other } N_{ijk}=0.
\end{align}
Since $N_{ijk}\leq 1$, there is no states on the vertices.
So the indices $\al,\bt,...$ labeling the states on a vertex
can be suppressed.

The above fusion rule corresponds to the fusion rule for the
$N=1$ loop state discussed in \Ref{LWstrnet}.
So we will call the corresponding graphic state
$N=1$ loop state.

Due to the relation \eq{1FstarF}, the
different components of the tensor $F^{ijm}_{kln}$ are not
independent.  There are only four independent potentially non-zero
components which are denoted as $f_0$,...,$f_3$:
\begin{align}
\label{Floop}
F^{000}_{000}
\bmm\begin{tikzpicture}[scale=0.27]
\FBox \iLnkaa \jLnkaa \kLnkaa \lLnkaa \mLnkaa \nLnkaa
\end{tikzpicture}\emm
&=f_{0}
\nonumber\\
F^{000}_{111}
\bmm\begin{tikzpicture}[scale=0.26]
\FBox \iLnkaa \jLnkaa \kLnkbb \lLnkbb \mLnkaa \nLnkbb
\end{tikzpicture}\emm
&=(F^{011}_{100}
\bmm\begin{tikzpicture}[scale=0.26]
\FBox \iLnkaa \jLnkbb \kLnkbb \lLnkaa \mLnkbb \nLnkaa
\end{tikzpicture}\emm
)^*=(F^{101}_{010}
\bmm\begin{tikzpicture}[scale=0.26]
\FBox \iLnkbb \jLnkaa \kLnkaa \lLnkbb \mLnkbb \nLnkaa
\end{tikzpicture}\emm
)^*
\nonumber\\
&=F^{110}_{001}
\bmm\begin{tikzpicture}[scale=0.26]
\FBox \iLnkbb \jLnkbb \kLnkaa \lLnkaa \mLnkaa \nLnkbb
\end{tikzpicture}\emm
=f_{1}
\nonumber\\
F^{011}_{011}
\bmm\begin{tikzpicture}[scale=0.26]
\FBox \iLnkaa \jLnkbb \kLnkaa \lLnkbb \mLnkbb \nLnkbb
\end{tikzpicture}\emm
&=(F^{101}_{101}
\bmm\begin{tikzpicture}[scale=0.26]
\FBox \iLnkbb \jLnkaa \kLnkbb \lLnkaa \mLnkbb \nLnkbb
\end{tikzpicture}\emm
)^*=f_{2}
\nonumber\\
F^{110}_{110}
\bmm\begin{tikzpicture}[scale=0.26]
\FBox \iLnkbb \jLnkbb \kLnkbb \lLnkbb \mLnkaa \nLnkaa
\end{tikzpicture}\emm
&=f_{3}
\end{align}
We note that $F^{ijm}_{kln}$ in \eqn{IHwave} relates wave
functions on two graphs. In the above we have drawn the
two related graphs after the $F$ tensor, where the first
graph following $F$ corresponds to the graph on the left-hand
side of \eqn{IHwave} and the second graph corresponds to the
graph on the right-hand side of \eqn{IHwave}.  The doted
line corresponds to the $|0\>$-state on the link and the
solid line corresponds to the $|1\>$-state on the link.
There are four potentially non-zero components
in $P^{kj}_i$, which are denoted by
$p_0$,...,$p_3$:
\begin{align}
P^{00}_{0}=p_{0} ,\ \ \
P^{01}_{0}=p_{1} ,\ \ \
P^{00}_{1}=p_{2} ,\ \ \
P^{01}_{1}=p_{3}.
\end{align}

We can adjust the total phases of $p_i$ and $A^i$ to make
$p_0\geq 0$ and $A^0\geq 0$.  We can also use the local unitary
transformation $\e^{\imth \th_{l_0} \hat M_{l_0}}$ with
$l_0=1$ to make $f_1\geq 0$, since the $F$'s described by
$f_1$ in \eqn{Floop} are the only $F$'s that change the
number of $|1\>$-links.

The fixed-point conditions (\ref{NNNN}, \ref{Di},
\ref{NNstar}, \ref{2FFstar}, \ref{1FstarF},
\ref{penid}, \ref{PFFP}, \ref{PFP}, \ref{Anonz},
\ref{AiAistar}, \ref{PAPA}, \ref{PhiPA}, \ref{Phiikj},
\ref{Phinonz}) form a set of non-linear equations on the ten
variables $f_i$, $p_i$, and $A^i$.  Many of the non-linear
equations are dependent or even equivalent.  Using a
computer algebraic system, we simplify the set of non-linear
equations.  The simplified equations are
(after making the phase choice described above)
\begin{align}
&
f_0= f_1 = f_2 = 1, \ \
f_3 = \eta,
\nonumber\\
&
p_1 =p_3=\eta p_0, \ \
p_2=p_0,
\nonumber\\
&
p_0^{2}+|p_1|^2 = 1,\ \
|p_2|^2+|p_3|^2=1,
\\
&
p_1 A^0=p_2 A^1, \ \
\eta p_3 A^1=p_1 A^0 ,\ \
|A^0|^2+|A^1|^2=1
\nonumber
\end{align}
where $\eta=\pm 1$.  The above simplified
equations can be solved exactly. We find two solutions
parameterized by $\eta=\pm 1$:
\begin{align}
&
f_0=f_1 = f_2 = 1, \ \
f_3 = \eta,
\nonumber\\
&
p_0=p_2=\frac{1}{\sqrt 2}, \ \
p_1=p_3=\frac{\eta}{\sqrt 2},
\nonumber\\
&
A^0 = \frac{1}{\sqrt 2}, \ \
A^1 = \frac{\eta}{\sqrt 2} .
\end{align}
We also find
\begin{align}
 \e^{\imth \th_F}=
 \e^{\imth \th_{P1}}=
 \e^{\imth \th_{P2}}=
 \e^{\imth \th_{A1}}=
 \e^{\imth \th_{A2}}=
1.
\end{align}

The $\eta=1$ fixed-point state corresponds to the $\mathbb{Z}_2$ loop
condensed state whose low energy effective field theory is
the $\mathbb{Z}_2$ gauge theory.\cite{FNS0428,LWstrnet}
We call such a state, simply, the $\mathbb{Z}_2$ state.
The $\eta=-1$ fixed-point state corresponds to the double-semion
state whose low energy effective field theory is
the $U(1)\times U(1)$ Chern-Simons gauge theory\cite{FNS0428,LWstrnet}
\begin{align}
\cL=\frac {1}{4\pi}\Big(
2 a_{1\mu}\prt_{\nu}a_{1\la}\eps^{\mu\nu\la}
-2 a_{2\mu}\prt_{\nu}a_{2\la}\eps^{\mu\nu\la}
\Big) .
\end{align}

\subsection{$N=1$ string-net state}

To obtain another class of simple solutions, we modify the fusion
rule to
\begin{align}
& N_{000}= N_{110}= N_{101}= N_{011}= N_{111} =1,
\nonumber\\
& \text{other } N_{ijk}=0.
\end{align}
while keeping everything the same.
The above $N_{ijk}$ also
satisfies \eqn{NNNN}, \eqn{Di}, and \eqn{NNstar}.

The new fusion rule corresponds to the fusion rule for the
$N=1$ string-net state discussed in \Ref{LWstrnet}.
So we will call the corresponding graphic state
$N=1$ string-net state.

Again, due to the relation \eq{1FstarF}, the
different components of the tensor $F^{ijm}_{kln}$ are not
independent.  Now there are seven independent potentially non-zero
components which are denoted as $f_0$,...,$f_6$:
\begin{align}
F^{000}_{000}
\bmm\begin{tikzpicture}[scale=0.26]
\FBox \iLnkaa \jLnkaa \kLnkaa \lLnkaa \mLnkaa \nLnkaa
\end{tikzpicture}\emm
&=f_{0}
\nonumber\\
F^{000}_{111}
\bmm\begin{tikzpicture}[scale=0.26]
\FBox \iLnkaa \jLnkaa \kLnkbb \lLnkbb \mLnkaa \nLnkbb
\end{tikzpicture}\emm
&=(F^{011}_{100}
\bmm\begin{tikzpicture}[scale=0.26]
\FBox \iLnkaa \jLnkbb \kLnkbb \lLnkaa \mLnkbb \nLnkaa
\end{tikzpicture}\emm
)^*=(F^{101}_{010}
\bmm\begin{tikzpicture}[scale=0.26]
\FBox \iLnkbb \jLnkaa \kLnkaa \lLnkbb \mLnkbb \nLnkaa
\end{tikzpicture}\emm
)^*
\nonumber\\
&=F^{110}_{001}
\bmm\begin{tikzpicture}[scale=0.26]
\FBox \iLnkbb \jLnkbb \kLnkaa \lLnkaa \mLnkaa \nLnkbb
\end{tikzpicture}\emm
=f_{1}
\nonumber\\
F^{011}_{011}
\bmm\begin{tikzpicture}[scale=0.26]
\FBox \iLnkaa \jLnkbb \kLnkaa \lLnkbb \mLnkbb \nLnkbb
\end{tikzpicture}\emm
&=(F^{101}_{101}
\bmm\begin{tikzpicture}[scale=0.26]
\FBox \iLnkbb \jLnkaa \kLnkbb \lLnkaa \mLnkbb \nLnkbb
\end{tikzpicture}\emm
)^*=f_{2}
\nonumber\\
F^{011}_{111}
\bmm\begin{tikzpicture}[scale=0.26]
\FBox \iLnkaa \jLnkbb \kLnkbb \lLnkbb \mLnkbb \nLnkbb
\end{tikzpicture}\emm
&=(F^{101}_{111}
\bmm\begin{tikzpicture}[scale=0.26]
\FBox \iLnkbb \jLnkaa \kLnkbb \lLnkbb \mLnkbb \nLnkbb
\end{tikzpicture}\emm
)^*=F^{111}_{011}
\bmm\begin{tikzpicture}[scale=0.26]
\FBox \iLnkbb \jLnkbb \kLnkaa \lLnkbb \mLnkbb \nLnkbb
\end{tikzpicture}\emm
\nonumber\\
&=(F^{111}_{101}
\bmm\begin{tikzpicture}[scale=0.26]
\FBox \iLnkbb \jLnkbb \kLnkbb \lLnkaa \mLnkbb \nLnkbb
\end{tikzpicture}\emm
)^*=f_{3}
\nonumber\\
F^{110}_{110}
\bmm\begin{tikzpicture}[scale=0.26]
\FBox \iLnkbb \jLnkbb \kLnkbb \lLnkbb \mLnkaa \nLnkaa
\end{tikzpicture}\emm
&=f_{4}
\nonumber\\
F^{110}_{111}
\bmm\begin{tikzpicture}[scale=0.26]
\FBox \iLnkbb \jLnkbb \kLnkbb \lLnkbb \mLnkaa \nLnkbb
\end{tikzpicture}\emm
&=(F^{111}_{110}
\bmm\begin{tikzpicture}[scale=0.26]
\FBox \iLnkbb \jLnkbb \kLnkbb \lLnkbb \mLnkbb \nLnkaa
\end{tikzpicture}\emm
)^*=f_{5}
\nonumber\\
F^{111}_{111}
\bmm\begin{tikzpicture}[scale=0.26]
\FBox \iLnkbb \jLnkbb \kLnkbb \lLnkbb \mLnkbb \nLnkbb
\end{tikzpicture}\emm
&=f_{6}
\end{align}
Note that $F$'s described by $f_1$ and $f_5$ are the only
$F$'s that change the number of $|1\>$-links and the number
of $|1\>|1\>|1\>$-vertices.  So we can use the local unitary
transformation $\e^{\imth (\th \hat M_1+\phi \hat M_{111})}$
to make $f_1$ and $f_5$ to be positive real numbers. (Here
$\hat M_1$ is the total number of $|1\>$-links and  $\hat
M_{111}$ is the total number of $|1\>|1\>|1\>$-vertices.) We
also use the freedom of adjusting the total sign of
$F^{ijm}_{kln}$ to make Re$(f_0)\geq 0$.

There are five potentially non-zero components
in $P^{kj}_i$, which are denoted by
$p_0$,...,$p_4$:
\begin{align}
P^{00}_{0} &=p_{0} ,
&
P^{01}_{0} &=p_{1} ,
&
P^{00}_{1} &=p_{2} ,
\nonumber\\
P^{01}_{1} &=p_{3} ,
&
P^{11}_{1} &=p_{4} .
\end{align}
We use the freedom of adjusting the total phase of
$P^{kj}_i$ to make $p_0$ to be a positive number.
We can also use the freedom of adjusting the total phase of
$A^i$ to make $A^0$ to be a positive number.

The fixed-point conditions (\ref{2FFstar},
\ref{1FstarF}, \ref{penid}, \ref{PFFP}, \ref{PFP},
\ref{Anonz}, \ref{AiAistar}, \ref{PAPA}, \ref{PhiPA},
\ref{Phiikj}, \ref{Phinonz}) form a set of non-linear
equation on the variables $f_i$, $p_i$, and $A^i$, which can
be simplified.  The simplified equations have the following
form
\begin{align}
& f_0= f_1= f_2= f_3=1,\ \
f_4=f_5^2=-f_6>0,
\nonumber\\
&
p_1^2 f_4^2+p_1^2=1,\ \
p_0=f_4 p_1, \ \
p_2=p_0, \ \
p_3=p_1, \ \
p_4=0
\nonumber\\
&
A^0=f_4 A^1 ,\ \
(A^0)^2+(A^1)^2=1, \ \
f_4^2 + f_4 -1=0.
\end{align}
Let $\ga$ be the positive solution
of $\ga^2+\ga=1$:
$\ga=\frac{\sqrt{5}-1}{2}$.
We see that $f_5=\sqrt{\ga}$. The above can be written as
\begin{align}
& f_0= f_1= f_2= f_3=1,\ \ \
 f_4=-f_6=\ga, \ \ \
f_5=\sqrt{\ga},
\nonumber\\
&
p_0=p_2=\frac{\ga}{\ga^2+1} , \ \ \
p_1 =p_3= \frac{1}{\ga^2+1}, \ \ \
p_4=0,
\nonumber\\
&
 A^0=\frac{\ga}{\ga^2+1},\ \ \ \
 A^1=\frac{1}{\ga^2+1} .
\end{align}
We also find
\begin{align}
 \e^{\imth \th_F}=
 \e^{\imth \th_{P1}}=
 \e^{\imth \th_{P2}}=
 \e^{\imth \th_{A1}}=
 \e^{\imth \th_{A2}}=
1.
\end{align}

The fixed-point state corresponds to the $N=1$ string-net
condensed state\cite{LWstrnet} whose low energy effective
field theory is the doubled $SO(3)$ Chern-Simons gauge
theory.\cite{FNS0428}

\subsection{An $N=2$ string-net state -- the $\mathbb{Z}_3$ state}

The above simple examples correspond to non-orientable
string-net states. Here we will give an example of orientable
string-net state.  We choose $N=2$, $0^*=0$, $1^*=2$,
$2^*=1$, and
\begin{align}
N_{000}&=
N_{012}=
N_{120}=
N_{201}=
N_{021}=
N_{102}=
N_{210}
\nonumber\\
&=
N_{111}=
N_{222}=1.
\end{align}
The above $N_{ijk}$
satisfies \eqn{NNNN}, \eqn{Di}, and \eqn{NNstar}.

Due to the relation \eq{1FstarF}, the
different components of the tensor $F^{ijm}_{kln}$ are not
independent.  There are eight independent potentially non-zero
components which are denoted as $f_0$,...,$f_7$:
\begin{align}
F^{000}_{000}
\bmm\begin{tikzpicture}[scale=0.26]
\FBox \iLnkaa \jLnkaa \kLnkaa \lLnkaa \mLnkaa \nLnkaa
\end{tikzpicture}\emm
&=f_{0}
\nonumber\\
F^{000}_{111}
\bmm\begin{tikzpicture}[scale=0.26]
\FBox \iLnkaa \jLnkaa \kLnkbc \lLnkbc \mLnkaa \nLnkbc
\end{tikzpicture}\emm
&=(F^{011}_{200}
\bmm\begin{tikzpicture}[scale=0.26]
\FBox \iLnkaa \jLnkbc \kLnkcb \lLnkaa \mLnkbc \nLnkaa
\end{tikzpicture}\emm
)^*=F^{120}_{002}
\bmm\begin{tikzpicture}[scale=0.26]
\FBox \iLnkbc \jLnkcb \kLnkaa \lLnkaa \mLnkaa \nLnkcb
\end{tikzpicture}\emm
\nonumber\\
&=(F^{202}_{020}
\bmm\begin{tikzpicture}[scale=0.26]
\FBox \iLnkcb \jLnkaa \kLnkaa \lLnkcb \mLnkcb \nLnkaa
\end{tikzpicture}\emm
)^*=f_{1}
\nonumber\\
F^{000}_{222}
\bmm\begin{tikzpicture}[scale=0.26]
\FBox \iLnkaa \jLnkaa \kLnkcb \lLnkcb \mLnkaa \nLnkcb
\end{tikzpicture}\emm
&=(F^{022}_{100}
\bmm\begin{tikzpicture}[scale=0.26]
\FBox \iLnkaa \jLnkcb \kLnkbc \lLnkaa \mLnkcb \nLnkaa
\end{tikzpicture}\emm
)^*=(F^{101}_{010}
\bmm\begin{tikzpicture}[scale=0.26]
\FBox \iLnkbc \jLnkaa \kLnkaa \lLnkbc \mLnkbc \nLnkaa
\end{tikzpicture}\emm
)^*\nonumber\\
&=F^{210}_{001}
\bmm\begin{tikzpicture}[scale=0.26]
\FBox \iLnkcb \jLnkbc \kLnkaa \lLnkaa \mLnkaa \nLnkbc
\end{tikzpicture}\emm
=f_{2}
\nonumber\\
F^{011}_{011}
\bmm\begin{tikzpicture}[scale=0.26]
\FBox \iLnkaa \jLnkbc \kLnkaa \lLnkbc \mLnkbc \nLnkbc
\end{tikzpicture}\emm
&=F^{022}_{022}
\bmm\begin{tikzpicture}[scale=0.26]
\FBox \iLnkaa \jLnkcb \kLnkaa \lLnkcb \mLnkcb \nLnkcb
\end{tikzpicture}\emm
=(F^{101}_{202}
\bmm\begin{tikzpicture}[scale=0.26]
\FBox \iLnkbc \jLnkaa \kLnkcb \lLnkaa \mLnkbc \nLnkcb
\end{tikzpicture}\emm
)^*\nonumber\\
&=(F^{202}_{101}
\bmm\begin{tikzpicture}[scale=0.26]
\FBox \iLnkcb \jLnkaa \kLnkbc \lLnkaa \mLnkcb \nLnkbc
\end{tikzpicture}\emm
)^*=f_{3}
\nonumber\\
F^{011}_{122}
\bmm\begin{tikzpicture}[scale=0.26]
\FBox \iLnkaa \jLnkbc \kLnkbc \lLnkcb \mLnkbc \nLnkcb
\end{tikzpicture}\emm
&=(F^{101}_{121}
\bmm\begin{tikzpicture}[scale=0.26]
\FBox \iLnkbc \jLnkaa \kLnkbc \lLnkcb \mLnkbc \nLnkbc
\end{tikzpicture}\emm
)^*=F^{112}_{021}
\bmm\begin{tikzpicture}[scale=0.26]
\FBox \iLnkbc \jLnkbc \kLnkaa \lLnkcb \mLnkcb \nLnkbc
\end{tikzpicture}\emm
\nonumber\\
&=(F^{112}_{102}
\bmm\begin{tikzpicture}[scale=0.26]
\FBox \iLnkbc \jLnkbc \kLnkbc \lLnkaa \mLnkcb \nLnkcb
\end{tikzpicture}\emm
)^*=f_{4}
\nonumber\\
F^{022}_{211}
\bmm\begin{tikzpicture}[scale=0.26]
\FBox \iLnkaa \jLnkcb \kLnkcb \lLnkbc \mLnkcb \nLnkbc
\end{tikzpicture}\emm
&=(F^{202}_{212}
\bmm\begin{tikzpicture}[scale=0.26]
\FBox \iLnkcb \jLnkaa \kLnkcb \lLnkbc \mLnkcb \nLnkcb
\end{tikzpicture}\emm
)^*=F^{221}_{012}
\bmm\begin{tikzpicture}[scale=0.26]
\FBox \iLnkcb \jLnkcb \kLnkaa \lLnkbc \mLnkbc \nLnkcb
\end{tikzpicture}\emm
\nonumber\\
&=(F^{221}_{201}
\bmm\begin{tikzpicture}[scale=0.26]
\FBox \iLnkcb \jLnkcb \kLnkcb \lLnkaa \mLnkbc \nLnkbc
\end{tikzpicture}\emm
)^*=f_{5}
\nonumber\\
F^{112}_{210}
\bmm\begin{tikzpicture}[scale=0.26]
\FBox \iLnkbc \jLnkbc \kLnkcb \lLnkbc \mLnkcb \nLnkaa
\end{tikzpicture}\emm
&=(F^{120}_{221}
\bmm\begin{tikzpicture}[scale=0.26]
\FBox \iLnkbc \jLnkcb \kLnkcb \lLnkcb \mLnkaa \nLnkbc
\end{tikzpicture}\emm
)^*=(F^{210}_{112}
\bmm\begin{tikzpicture}[scale=0.26]
\FBox \iLnkcb \jLnkbc \kLnkbc \lLnkbc \mLnkaa \nLnkcb
\end{tikzpicture}\emm
)^*
\nonumber\\
&=F^{221}_{120}
\bmm\begin{tikzpicture}[scale=0.26]
\FBox \iLnkcb \jLnkcb \kLnkbc \lLnkcb \mLnkbc \nLnkaa
\end{tikzpicture}\emm
=f_{6}
\nonumber\\
F^{120}_{110}
\bmm\begin{tikzpicture}[scale=0.26]
\FBox \iLnkbc \jLnkcb \kLnkbc \lLnkbc \mLnkaa \nLnkaa
\end{tikzpicture}\emm
&=(F^{210}_{220}
\bmm\begin{tikzpicture}[scale=0.26]
\FBox \iLnkcb \jLnkbc \kLnkcb \lLnkcb \mLnkaa \nLnkaa
\end{tikzpicture}\emm
)^*=f_{7}
\end{align}
There are nine potentially non-zero components
in $P^{kj}_i$, which are denoted by
$p_0$,...,$p_8$:
\begin{align}
&
P^{00}_{0}=p_{0},\ \
P^{01}_{0}=p_{1},\ \
P^{02}_{0}=p_{2},\ \
P^{00}_{1}=p_{3},\ \
P^{01}_{1}=p_{4},
\nonumber\\
&
P^{02}_{1}=p_{5},\ \
P^{00}_{2}=p_{6},\ \
P^{01}_{2}=p_{7},\ \
P^{02}_{2}=p_{8}.
\end{align}
Using the transformations discussed in section
\ref{fixphase}, we can fix the phases of $f_1$, $f_2$, $f_6$,
and $p_0$ to make them positive.

The fixed-point conditions (\ref{2FFstar},
\ref{1FstarF}, \ref{penid}, \ref{PFFP}, \ref{PFP},
\ref{Anonz}, \ref{AiAistar}, \ref{PAPA}, \ref{PhiPA},
\ref{Phiikj}, \ref{Phinonz}) form a set of non-linear
equation on the variables $f_i$, $p_i$, and $A^i$, which can
be solved exactly. After fixing the phases using the
transformations discussed in section \ref{fixphase}, we find
only one solution
\begin{align}
&
f_i=1,\ \ i=0,1,...,7,
\nonumber\\
&
p_i=\frac{1}{\sqrt 3},\ \ i=0,1,...,8,
\nonumber\\
&
A^0=A^1=A^2=\frac{1}{\sqrt 3}.
\end{align}
We also find
\begin{align}
 \e^{\imth \th_F}=
 \e^{\imth \th_{P1}}=
 \e^{\imth \th_{P2}}=
 \e^{\imth \th_{A1}}=
 \e^{\imth \th_{A2}}=
1.
\end{align}

The fixed-point state corresponds to the $\mathbb{Z}_3$
string-net condensed state\cite{LWstrnet} whose low energy
effective field theory is the $U(1)\times U(1)$ Chern-Simons
gauge theory\cite{LWstrnet,KLW}
\begin{align}
\cL=\frac {1}{4\pi}\Big(
3 a_{1\mu}\prt_{\nu}a_{2\la}\eps^{\mu\nu\la}
+3 a_{2\mu}\prt_{\nu}a_{1\la}\eps^{\mu\nu\la}
\Big) ,
\end{align}
which is the $\mathbb{Z}_3$ gauge theory.

We note that all the above simple solutions also satisfy the
standard pentagon identity, although we solved the weaker
projective pentagon identity.  It is not clear if we can
find non-trivial solutions that do not satisfy the standard
pentagon identity.

\section{A classification of time reversal invariant
topological orders}

There are several ways to define time reversal operation
for the graphic states.
The simplest one is given by
\begin{align}
 \hat T: \Phi(\Ga) \to \Phi^*(\Ga)
\end{align}
where $\Ga$ represents the labels on the vertices and links
which are not changed under $\hat T$.
(This corresponds to the situation where the different
states on the links and the vertices are realized by
different occupations of scalar bosons.)
For such a time reversal transformation,
$\hat T^2=1$ and
the real solutions of the
fixed-point conditions (\ref{NNNN}, \ref{Di}, \ref{NNstar},
\ref{2FFstar}, \ref{1FstarF}, \ref{penid},
\ref{PFFP}, \ref{PFP}, \ref{Anonz}, \ref{AiAistar},
\ref{PAPA}, \ref{PhiPA}, \ref{Phiikj}, \ref{Phinonz}) give
us a classification of time reversal invariant topological
orders in local spin systems.  Note that the time reversal
invariant topological orders  are equivalent class of local
orthogonal transformations that connect to the identity
transformation continuously.

Different real solutions
$(N_{ijk},F^{ijm,\al\bt}_{kln,\ga\la},P_i^{kj,\al\bt},A^i)$
of the fixed-point conditions do not always correspond
to different time reversal invariant topological orders.  The
solutions differ by some unimportant phase factors (which
are $\pm 1$ signs) correspond to the same topological
order.

To understand the above result, we notice that,
from the structure of the fixed-point conditions, if
$(F^{ijm,\al\bt}_{kln,\ga\la},P_i^{kj,\al\bt},A^i)$ is a
solution, then $(\eta_F F^{ijm,\al\bt}_{kln,\ga\la}, \eta_P
P_i^{kj,\al\bt}, \eta_A A^i)$ is also a solution, where
$\eta_F=\pm 1$, $\eta_P=\pm 1$, and $\eta_A=\pm 1$. However,
the three phase factors $\eta_F$, $\eta_P$, and $\eta_A$ do
not lead to different fixed-point wave functions, since they
only affect the total phase of the wave function and are
unphysical.

On the other hand, the fixed-point solutions may also
contain phase factors that do correspond to different
fixed-point wave functions.  For example, the local orthogonal
transformation $\e^{\imth \pi \hat M_{l_0}}$ does not affect
the fusion rule $N_{ijk}$, where $\hat M_{l_0}$ is the
number of links with $|l_0\>$-state and $|l_0^*\>$-state.
Such a local orthogonal transformation changes
$(F^{ijm,\al\bt}_{kln,\ga\la},P_i^{kj,\al\bt},A^i)$ and
generates a discrete family of the fixed-point wave
functions.

Similarly, we can consider the following local orthogonal
transformation $|\al\> \to \sum_\bt
O^{(i_0j_0k_0)}_{\al\bt}|\bt\>$ that acts on each vertex
with states $|i_0\>,|j_0\>,|k_0\>$ on the three edges
connecting to the vertex.  Such a local orthogonal
transformation also does not affect the fusion rule
$N_{ijk}$.  The new local orthogonal transformation changes
$(F^{ijm,\al\bt}_{kln,\ga\la},P_i^{kj,\al\bt},A^i)$ and
generates a family of the fixed-point wave functions
parameterized by the orthogonal matrix
$O^{(i_0j_0k_0)}_{\al\bt}$.

Now the question is that do those solutions related by local
orthogonal transformations have the same time reversal
invariant topological order or not.  We know that two gapped
wave functions have the same time reversal invariant
topological order if and only if they can be connected by
local orthogonal transformation that \emph{connects to
identity continuously}.  It is well known that an orthogonal
matrix whose determinant is $-1$ does not connect to
identity.  Thus it appears that local orthogonal
transformations some times can generate different time
reversal invariant topological orders.

However, when we use the equivalent classes of local
orthogonal transformations to define  time reversal
invariant topological orders, we not only assume the local
orthogonal transformations to connect to identity
continuously, we also assume that we can expand the local
Hilbert spaces (say by increasing the range of the indices
$i$ and $\al$ that label the states on the edges and the
vertices). The local orthogonal transformations can act
on those enlarged Hilbert spaces and can connect to identity
in those enlarged Hilbert spaces.  Even when a local
orthogonal transformation cannot be deformed into identity
in the original Hilbert space, it can always be deformed
into identity continuously in an enlarged Hilbert space.
Thus two real wave functions related by a local orthogonal
transformation always have the same time reversal invariant
topological order.

For example, an orthogonal matrix $\bpm 1 &0\\ 0&-1\\ \epm$
that acts on states $|0\>$ and $|1\>$ does not connect to identity
within the space of two by two orthogonal matrices.
However, we can embed the above orthogonal matrix into a
three by three orthogonal matrix that acts on $|0\>$, $|1\>$,
and $|2\>$: $\bpm 1& 0& 0\\ 0&-1& 0\\ 0& 0 &-1\\ \epm$.
Such a three by three orthogonal matrix does connect to
identity within the space of three by three orthogonal
matrices.  This completes our argument that all local
orthogonal transformations can connect to identity
continuously at least in an enlarged Hilbert space.

So, after factoring out the unimportant phase factors
discussed above, the real solutions of the fixed-point
conditions may uniquely correspond to time reversal
invariant topological orders. The four types of real
solutions discussed in the last section are examples of four
different time reversal invariant topological orders.

For the $N=1$ loop states and the $N=1$ string-net state, we
only have two states on each link. In this case, we can
treat the two states as the two states of an electron spin.
The time reversal transformation now becomes
\begin{align}
 \hat T:
c_0|0\>+c_1|1\>
\to
-c_1^*|0\>+c_0^*|1\>
\end{align}
on each link.  For such a time reversal transformation $\hat
T^2=-1$. The two $N=1$ loop states and the $N=1$ string-net
state are not invariant under such an time-reversal
transformation.

\section{Wave function renormalization for tensor product
states}

\subsection{Motivation}

Once the fixed-point states have been identified and the
labeling of topological orders has being found, we then face
the next important issue: given a generic ground state wave
function of a system, how to identify the topological orders
in the state? In other words, how to calculate the data
$(N_{ijk},F^{ijm,\al\bt}_{kln,\ga\la},P_i^{kj,\al\bt}, A^i)$
that characterize the topological orders from a generic wave
function?

One way to address the above issue is to  have a general
renormalization procedure which flows other states in the
same phase to the simple fixed-point state so that we can
identify topological order from the resulting fixed-point
state.  That is, we want to find a local unitary
transformation which removes local entanglement and gets rid
of unnecessary degrees of freedom from the state.  How to
find the appropriate unitary transformation for a specific
state is then the central problem in this renormalization
procedure. Such a procedure for one dimensional tensor product states(TPS) (also called matrix product states) has been given in \Ref{VCL}. Here we will propose a method to renormalize two dimensional TPS, where nontrivial topological orders emerge. The basic idea is to use the gLU transformation
discussed in section \ref{wvrg}.  Note that, through the gLU
transformation, we can reduce the number of labels in a
region A to the minimal value without loosing any quantum
information (see Fig.  \ref{gLUT}). This is because the gLU
transformation is a lossless projection into the support
space of the state in the region A.  By performing such gLU
transformations on \emph{overlapping} regions repeatedly (see
Fig. \ref{qc}a), we can reduce a generic wave function to
the simple fixed-point form discussed above. It should be noted that any state reducible in this way can be represented as a MERA \cite{V0705}.

In the following, we will present this renormalization procedure for two dimensional TPS where we find a method to
calculate the proper gLU transformations. The tensor
product states are many-body entangled quantum states
described with local tensors. By making use of the
entanglement information contained in the local tensors, we
are able to come up with an efficient algorithm to
renormalize two dimensional TPS.

This algorithm can be very useful in the study of quantum
phases. Due to the efficiency in representation, TPS has
found wide application as variational ansatz states in the
studies of quantum many-body systems
\cite{GMN0391,NMG0415,M0460,VC0466,VCM0843,GLW0809,JWX0803}.  Suppose that
in a variational study we have found a set of tensors which
describe the ground state of a two dimensional many-body Hamiltonian and
want to determine the phase this state belongs to.  We can
apply our renormalization algorithm to this tensor product
state, which removes local entanglement and flows the state
to its fixed point. By identifying the kind of order present
in the fixed-point state, we can obtain the phase
information for the original state.

In this section, we will give a detailed description of the
algorithm and in the next section we will present its
application to some simple (but nontrivial) cases.  The
states we are concerned with have translational symmetry and
can be described with a translational invariant tensor
network. To be specific, we discuss states on a hexagonal
lattice. Generalization to other regular lattices is
straight-forward.

\subsection{Tensor product states}

Consider a two-dimensional spin model on a hexagonal lattice with
one spin (or one qudit) living at each vertex. The Hilbert space of each
spin is $D$-dimensional. The state can be represented by
assigning to every vertex a set of tensors
$T^{i}_{\alpha\beta\gamma}$, where $i$ labels the local physical
dimension and takes value from $1$ to $D$. $\alpha,\beta,\gamma$ are
inner indices along the three directions in the hexagonal lattice
respectively. The dimension of the inner indices is $d$. Note that
the figures in this note are all sideviews with inner indices in the
horizontal plane and the physical indices pointing in the vertical
direction, if not specified otherwise.
\begin{figure}[tb] \centering
\includegraphics[width=3.5in]{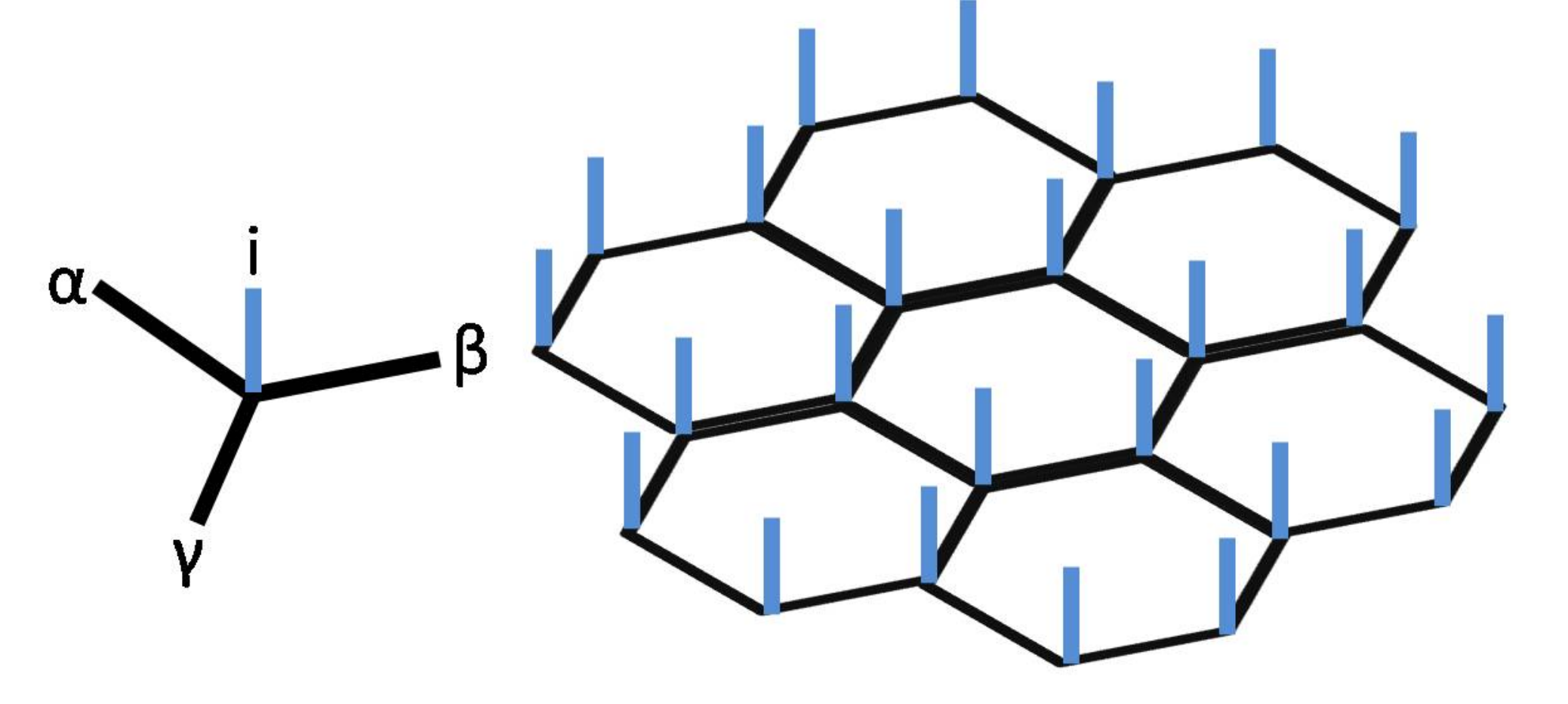}
\caption{(Color online) Left: Tensor $T$ representing a
2D quantum state on hexagonal lattice. $i$ is the physical index,
$\alpha,\beta,\gamma$ are inner indices. Right: a tensor product
state where each vertex is associated with a tensor. The inner
indices of the neighboring tensors connect according to the
underlying hexagonal lattice.} \label{fig:TPS}
\end{figure}
The wave function is given in terms of these tensors by
\begin{equation}
|\psi\rangle=\sum_{i_1,i_2,...i_m...}\text{tTr}(T^{i_1}T^{i_2}...T^{i_m}...)|i_1
i_2 ... i_m...\rangle \label{TPS}
\end{equation}
where $\text{tTr}$ denotes tensor contraction of all the connected
inner indices on the links of the hexagonal lattice.

A renormalization procedure of quantum states is composed of local
unitary transformations and isometry maps such that the state flows
along the path $|\psi^{(0)}\rangle$, $|\psi^{(1)}\rangle$,
$|\psi^{(2)}\rangle$,...and finally towards a fixed point
$|\psi^{\infty}\rangle$. With the tensor product representation,
flow of states corresponds to a flow of tensors $T^{(0)}$,
$T^{(1)}$, $T^{(2)}$... We will give the detailed procedure of how
the tensors are mapped from one step to the next in the following
section.

\subsection{Renormalization algorithm}

In one round of renormalization, we start from tensor $T^{(n)}$, do
some operation to it which corresponds to local unitary
transformations on the state, and map $T^{(n)}$ to $T^{(n+1)}$. The
whole procedure can be broken into two parts: the F-move and the P
move, in accordance with the two steps introduced in the previous
section.

\subsubsection{Step 1: F-move}

In the F-move, we take a
$\bmm\includegraphics[scale=.13]{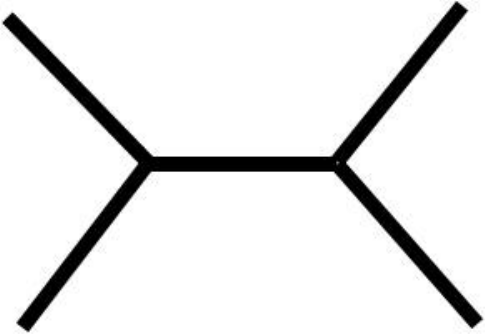}\emm$ configuration
in the tensor network and map it to a $\bmm
\includegraphics[scale=.13]{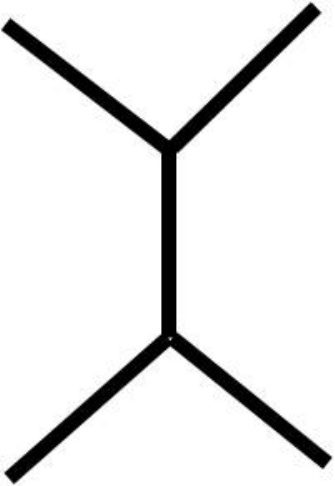}\emm$ configuration by
doing a local unitary operation. We will see that the tensor
product representation of a state leads to a natural way of
choosing an appropriate unitary operation for the
renormalization of the state.

In order to do so, first we define the double tensor
$\mathbb{T}$ of tensor $T$ as
\begin{equation}
\mathbb{T}_{\alpha'\beta'\gamma',\alpha\beta\gamma} =\sum_i
(T^{i}_{\alpha'\beta'\gamma'})^* \times T^{i}_{\alpha\beta\gamma}
\end{equation}
Graphically the double tensor $\mathbb{T}$ is represented by
two layers of tensor $T$ with the physical indices
connected.
\begin{figure}[tb] \centering
\includegraphics[width=1.5in]{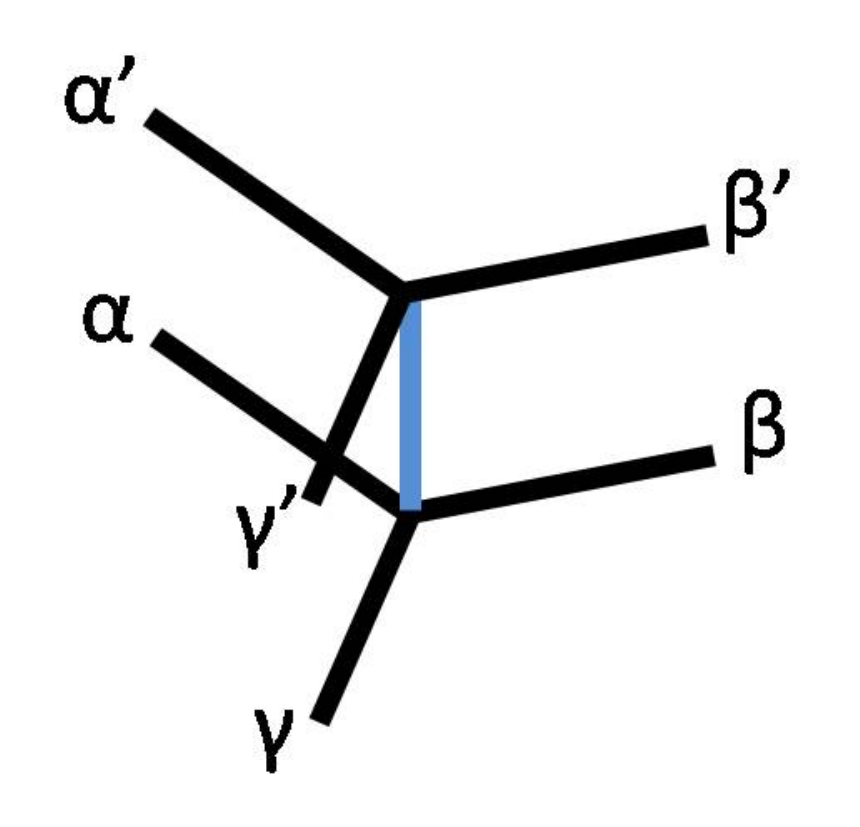}
\caption{
(Color online) Double tensor $\mathbb{T}$ represented
as two layers of tensor $T$ with the physical indices contracted.
The gray layer is the lower layer.} \label{fig:DT}
\end{figure}
The tensor $T$ giving rise to the same double tensor $\mathbb{T}$ is
not unique. Any tensor $T'$ which differs from $T$ by an unitary
transformation $U$ on physical index $i$ gives the same
$\mathbb{T}$ as $U$ and $U^{\dagger}$ cancels out in the
contraction of $i$. On the other hand, an unitary transformation on
$i$ is the only degree of freedom possible, i.e. any $T'$ which
gives rise to the same $\mathbb{T}$ as $T$ differs from $T$ by a
unitary on $i$. Therefore, in the process of turning a tensor $T$ into a double tensor $\mathbb{T}$ and then split it again into a different tensor $T'$, we apply a non-trivial local unitary operation on the corresponding state. A well designed way of splitting the double tensor will give us the appropriate unitary transformation we need, as we show below.

\begin{figure}[tb] \centering
\includegraphics[width=3.5in]{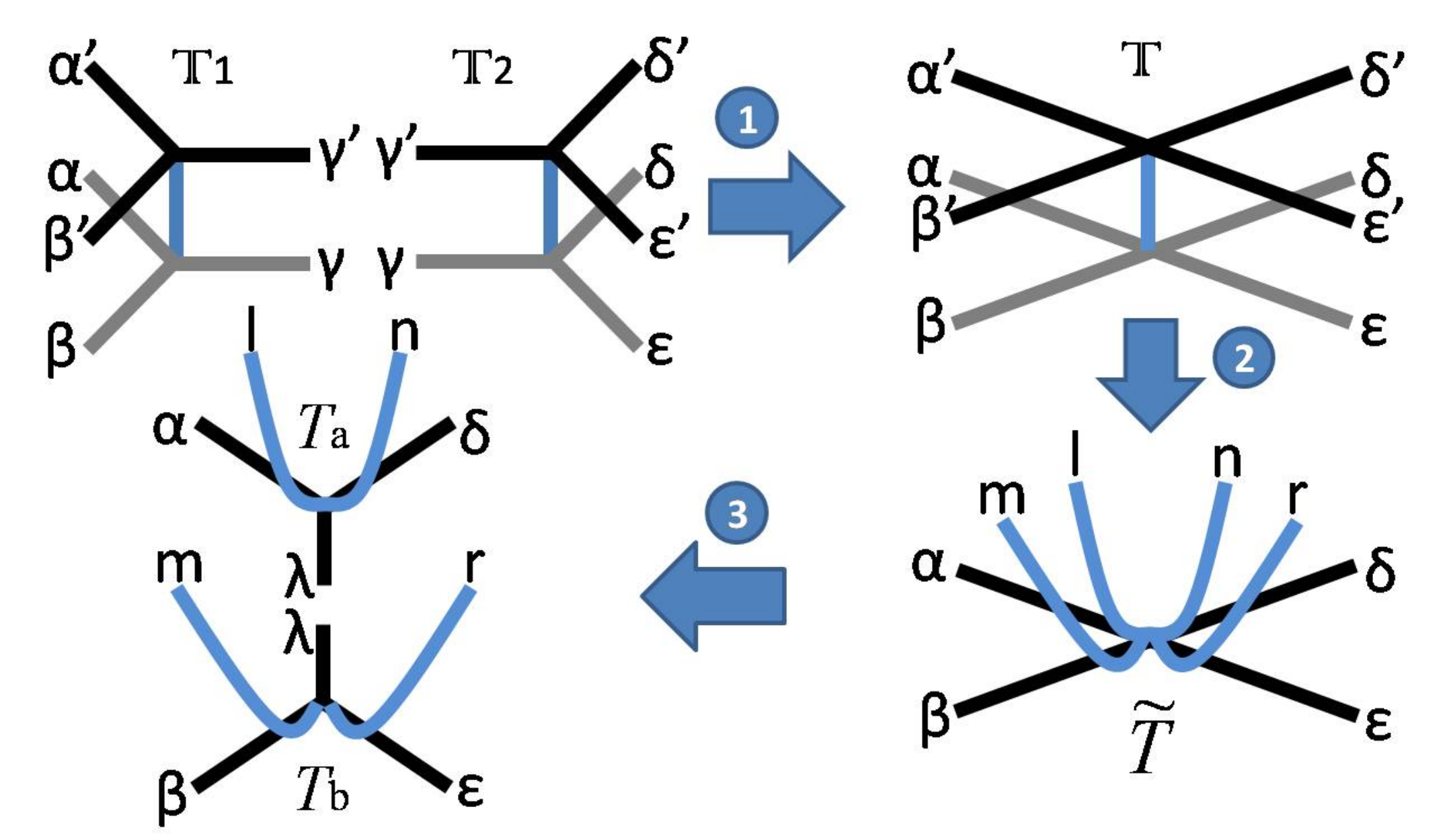}
\caption{
(Color online) F-move in the renormalization procedure: (1) combining double tensors $\mathbb{T}_1$ and $\mathbb{T}_2$ on neighboring sites into a single double tensor $\mathbb{T}$ (2)Splitting double tensor $\mathbb{T}$ into tensor $\tilde{T}$ (3) SVD decomposition of tensor $\tilde{T}$ into tensors $T_a$ and $T_b$.} \label{fig:F-move}
\end{figure}

F-move has the following steps. First, construct double tensors for two neighboring sites on the lattice and combine them into a single double tensor with $8$ inner indices.
\begin{equation}
\mathbb{T}_{\alpha'\beta'\delta'\epsilon',\alpha\beta\delta\epsilon}=\sum_{\gamma',\gamma} \mathbb{T}_{1,\alpha'\beta'\gamma',\alpha\beta\gamma}\times \mathbb{T}_{2,\delta'\epsilon'\gamma',\delta\epsilon\gamma}
\end{equation}

Note that with respect to the bipartition of indices $\alpha'\beta'\delta'\epsilon'$ and $\alpha\beta\delta\epsilon$, $\mathbb{T}$ is Hermitian
\begin{equation}
\mathbb{T}_{\alpha'\beta'\delta'\epsilon',\alpha\beta\delta\epsilon}=\left(\mathbb{T}_{\alpha\beta\delta\epsilon,\alpha'\beta'\delta'\epsilon'}\right)^*
\end{equation}
and positive semidefinite.
Therefore it has a spectral decomposition with positive eigenvalues $\{\lambda_j \ge 0\}$. The corresponding eigenvectors are $\{\hat{T}^j\}$
\begin{equation}
\mathbb{T}_{\alpha'\beta'\delta'\epsilon',\alpha\beta\delta\epsilon}= \sum_j \lambda_j \left(\hat{T}^j_{\alpha'\beta'\delta'\epsilon'}\right)^* \times \hat{T}^j_{\alpha\beta\delta\epsilon}
\end{equation}
This spectral decomposition lead to a special way of decomposing
double tensor $\mathbb{T}$ into tensors. Define a rank $8$ tensor
$\tilde{T}$(as shown in Fig. \ref{fig:F-move} after step 2) as follows:
\begin{equation}
\tilde{T}^{lmnr}_{\alpha\beta\delta\epsilon}=\sum_j \sqrt{\lambda_j}
\left(\hat{T}^j_{lmnr}\right)^* \times
\hat{T}^j_{\alpha\beta\delta\epsilon}
\end{equation}
$\tilde{T}$ has four inner indices
$\alpha,\beta,\delta,\epsilon$ of dimension $d$ and four
physical indices $l,m,n,r$ also of dimension $d$ which are
in the direction of $\alpha,\beta,\delta,\epsilon$
respectively.  As $\{\hat{T}^j\}$ form an orthonormal set,
it is easy to check that $\tilde{T}$ gives rise to double
tensor $\mathbb{T}$. Going from $T_1$ and $T_2$ to
$\tilde{T}$, we have implemented a local unitary
transformation on the physical degrees of freedom on the two
sites, so that in $\tilde{T}$ the physical indices and the
inner indices represent the same configuration. In some
sense, we are keeping only the physical degrees of freedom
necessary for entanglement with the rest of the system while
getting rid of those that are only entangled within this
local region. Now we do a singular value decomposition of
tensor $\tilde{T}$ in the direction orthogonal to the link
between $T_1$ and $T_2$ and $\tilde{T}$ is decomposed into
tensors $T_a$ and $T_b$.
\begin{equation}
\tilde{T}^{lmnr}_{\alpha\beta\delta\epsilon}=\sum_{\lambda} T^{ln}_{a,\alpha\delta\lambda}\times T^{mr}_{b,\beta\epsilon\lambda}
\label{SVD}
\end{equation}
This step completes the F-move. Ideally, this step should
be done exactly so we are only applying local unitary
operations to the state. Numerically, we keep some large but
finite cutoff dimension for the SVD step, so this step is
approximate.

\begin{figure}[tb] \centering
\includegraphics[width=1.4in]{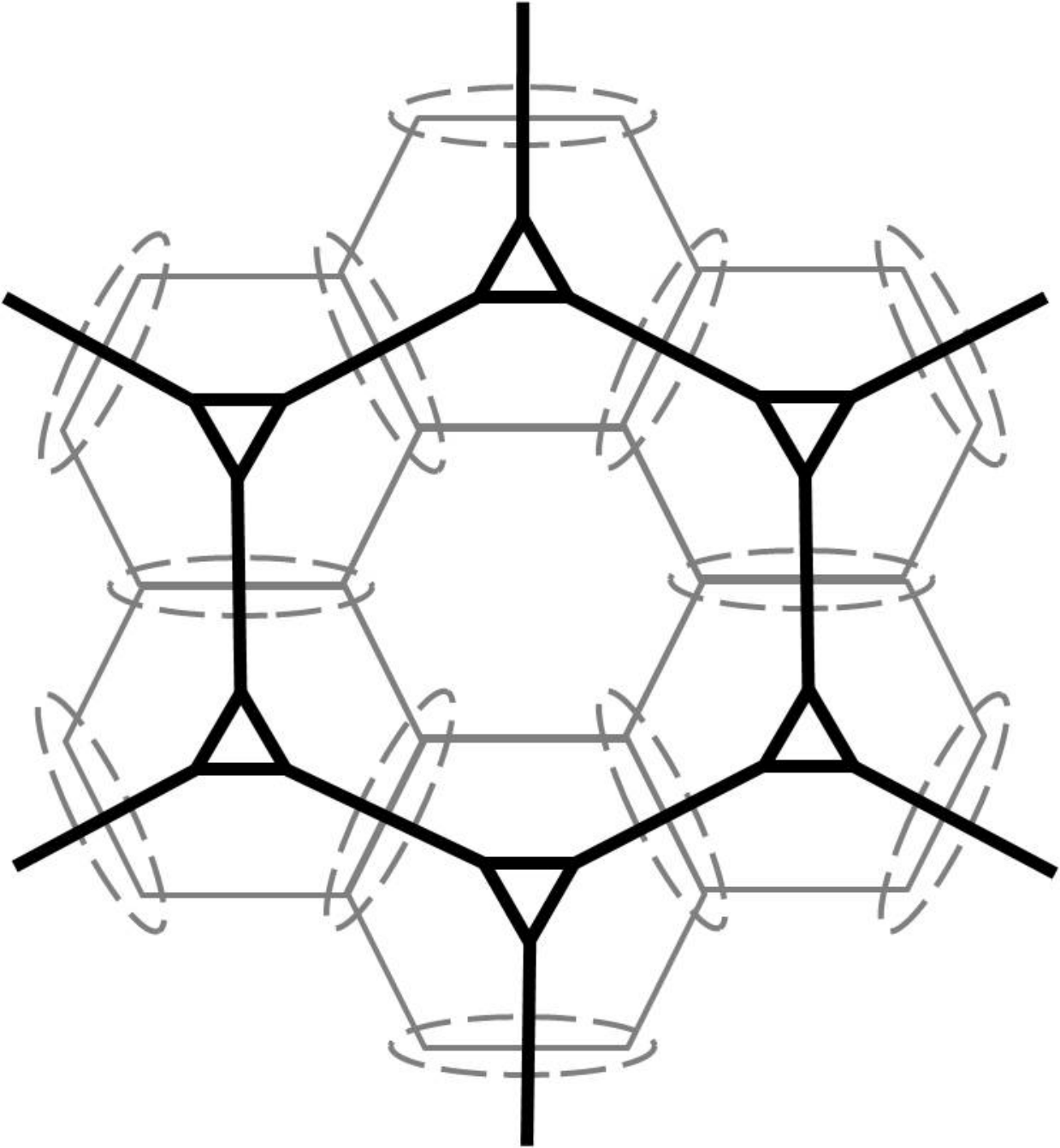}
\caption{Original hexagonal lattice (gray line) and renormalized lattice (black line) after F-move has been applied to the neighboring pairs of sites circled by dash line.} \label{fig:RG_hex}
\end{figure}

On a hexagonal lattice, we do F-move on the chosen
neighboring pairs of sites (dash-circled in
Fig. \ref{fig:RG_hex}), so that the tensor network is changed
into a configuration shown by thick dark lines in
Fig. \ref{fig:RG_hex}. Physical indices are omitted from this
figure. Now by grouping together the three tensors that meet
at a triangle, we can map the tensor network back into a
hexagonal lattice, with $1/3$ the number of sites in the
original lattice. This is achieved by the P-move introduced
in the next section.

\subsubsection{Step 2: P-move}
\begin{figure}[tb] \centering
\includegraphics[width=3.5in]{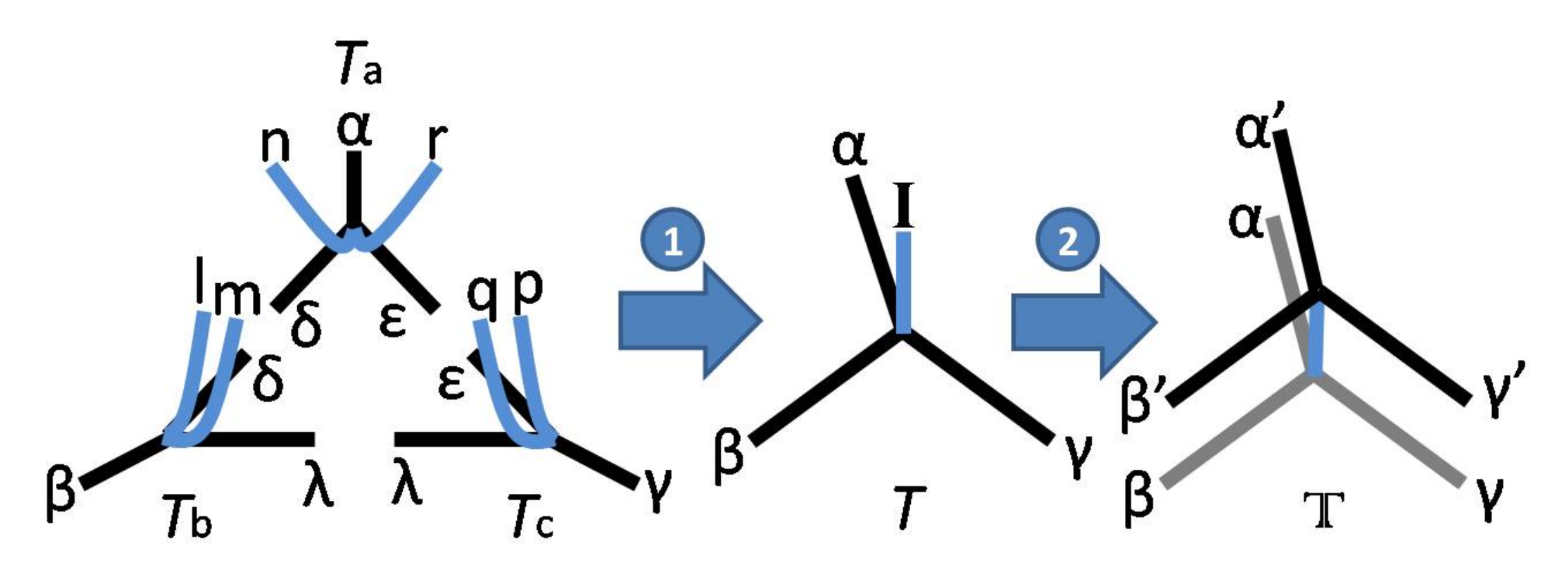}
\caption{
(Color online) P-move in the renormalization procedure: (1) contracting three tensors that meet at a triangle $T_a,T_b,T_c$ to form a new tensor $T^{(1)}$ on one site of the renormalized hexagonal lattice. (2)constructing the double tensor $\mathbb{T}^{(1)}$ from $T^{(1)}$ so that we can start to do F-moves again.} \label{fig:P-move}
\end{figure}

Now we contract the three tensors that meet at a triangle together to form a new tensor in the renormalized lattice as shown in the first step in Fig. \ref{fig:P-move}
\begin{equation}
T^{I}_{\alpha\beta\gamma}=\sum_{\delta\epsilon\lambda} T^{nr}_{a,\alpha\delta\epsilon} \times T^{lm}_{b,\beta\lambda\delta} \times T^{pq}_{c,\gamma\epsilon\lambda}
\label{P-move}
\end{equation}
where $I$ is the physical index of the new tensor which
includes all the physical indices of
$T_a,T_b,T_c$: $l,m,n,r,p,q$. Note that in the contraction,
only inner indices are contracted and the physical indices
are simply group together.

Constructing the double tensor $\mathbb{T}$ from $T$, we get
the renormalized double tensor on the new hexagonal lattice
which is in the same form as $\mathbb{T}_1$,$\mathbb{T}_2$
and we can go back again and do the F-move.

\subsubsection{Complications: corner double line}

\begin{figure}[tb] \centering
\includegraphics[width=1.2in]{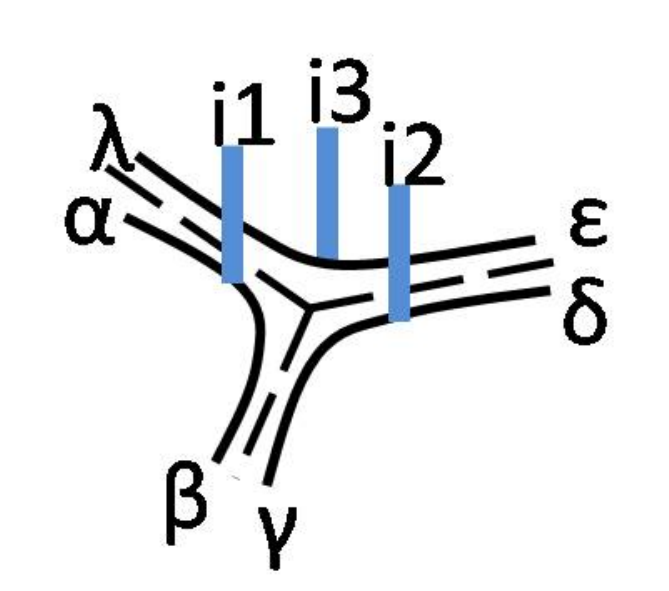}
\caption{
(Color online) The corner
double line tensor which is a fixed point of the
renormalization algorithm. The three groups of indices
$\{\alpha,\beta,i_1\},\{\gamma,\delta,i_2\},\{\epsilon,\lambda,i_3\},$
are entangled within each group but not between the groups.}
\label{fig:CDL}
\end{figure}

One problem with the above renormalization algorithm is that, instead of having one isolated fixed-point tensor for each phase, the algorithm has a continuous family of fixed points which all correspond to the same phase. Consider a tensor with structure shown in Fig. \ref{fig:CDL}.
The tensor is a tensor product of three parts which include
indices
$\{\alpha,\beta,i_1\},\{\gamma,\delta,i_2\},\{\epsilon,\lambda,i_3\}$
respectively. It can be shown that this structure remains
invariant under the renormalization flow. Therefore, any tensor of this structure is a fixed point of our renormalization flow. However, it is easy to see that the state it represents is a tensor product of loops around each
plaquette, which can be disentangled locally into a trivial
product state. Therefore, the states all have only short-range entanglement and correspond to the topologically trivial phase. The trivial phase has then a continuous family of fixed-point tensors. This situation is very similar to that discussed in \Ref{GW0931,LN0701}. We will keep the terminology and call such a tensor a corner double line tensor. Not only does corner double line tensor complicate the situation in the trivial phase, it leads to a continuous family of fixed points in every phase. It can be checked that the tensor
product of a corner double line with any other fixed-point
tensor is still a fixed-point tensor. The states they correspond to differ only by small loops around each plaquette and represent the same topological order. Therefore any single
fixed-point tensor gets complicated into a continuous class
of fixed-point tensors. In practical application of the renormalization algorithm, in order to identify the topological order of the fixed-point tensor, we need to get rid of such corner double line structures. Due to their simple structure, this can always be done, as discussed in the next section.

\section{Applications of the renormalization for tensor
product states}

Now we present some examples where our algorithm is used to
determine the phase of a tensor product state. The algorithm
can be applied both to symmetry breaking phases and
topological ordered phases. In the study of symmetry
breaking/topological ordered phases, suppose that we have
obtained some tensor product description of the ground state
of the system Hamiltonian. We can then apply our algorithm
to the tensors, flow them to the fixed point, and see
whether they represent a state in the symmetry breaking
phase/topological ordered phase or a trivial phase.

For system with symmetry/topological order related to gauge
symmetry, it is very important to keep the symmetry/gauge
symmetry in the variational approach to ground state and
search within the set of tensors that have this
symmetry/gauge symmetry\cite{CZG1003}. The resulting tensor will be
invariant under such symmetries/gauge symmetries, but the
state they correspond to may have different orders. In the
symmetry breaking case, the state could have this symmetry
or could spontaneous break it. In the topological ordered
phase, the state could have nontrivial topological order or
be just trivial. Our algorithm can then be applied to decide
which is the case. In order to correctly determine the phase
for such symmetric tensors, it is crucial that we maintain
the symmetry/gauge symmetry of the tensor throughout our
renormalization process. We will discuss in detail two
cases: the Ising symmetry breaking phase and the
$\mathbb{Z}_2$ topological ordered phase. For simplicity of discussion and to demonstrate the generality of our renormalization scheme, we will first introduce the square lattice version of the algorithm. All subsequent applications are carried out on square lattice. (Algorithm on a hexagonal lattice would give qualitatively similar result, though quantitatively they might differ, e.g. on the position of critical point.)

\subsection{Renormalization on square lattice}

Tensor product states on a square lattice are represented
with one tensor $T^{i}_{\alpha\beta\gamma\delta}$ on each
vertex, where $i$ is the physical index and
$\alpha\beta\gamma\delta$ are the four inner indices in the
up,down,left,right direction respectively. We will assume
translational invariance and require the tensor to be the
same on every vertex. The renormalization procedure is be a
local unitary transformation on the state which flows the
form of the tensor until it reaches the fixed point. It is
implemented in the following steps. First, we form the
double tensor $\mathbb{T}$ from tensor $T$
\begin{equation*}
\mathbb{T}_{\alpha'\beta'\gamma'\delta',\alpha\beta\gamma\delta}
=\sum_i \left(T^{i}_{\alpha'\beta'\gamma'\delta'}\right)^*\times
T^{i}_{\alpha\beta\gamma\delta}
\end{equation*}
\begin{figure}[tb] \centering
\includegraphics[width=2.5in]{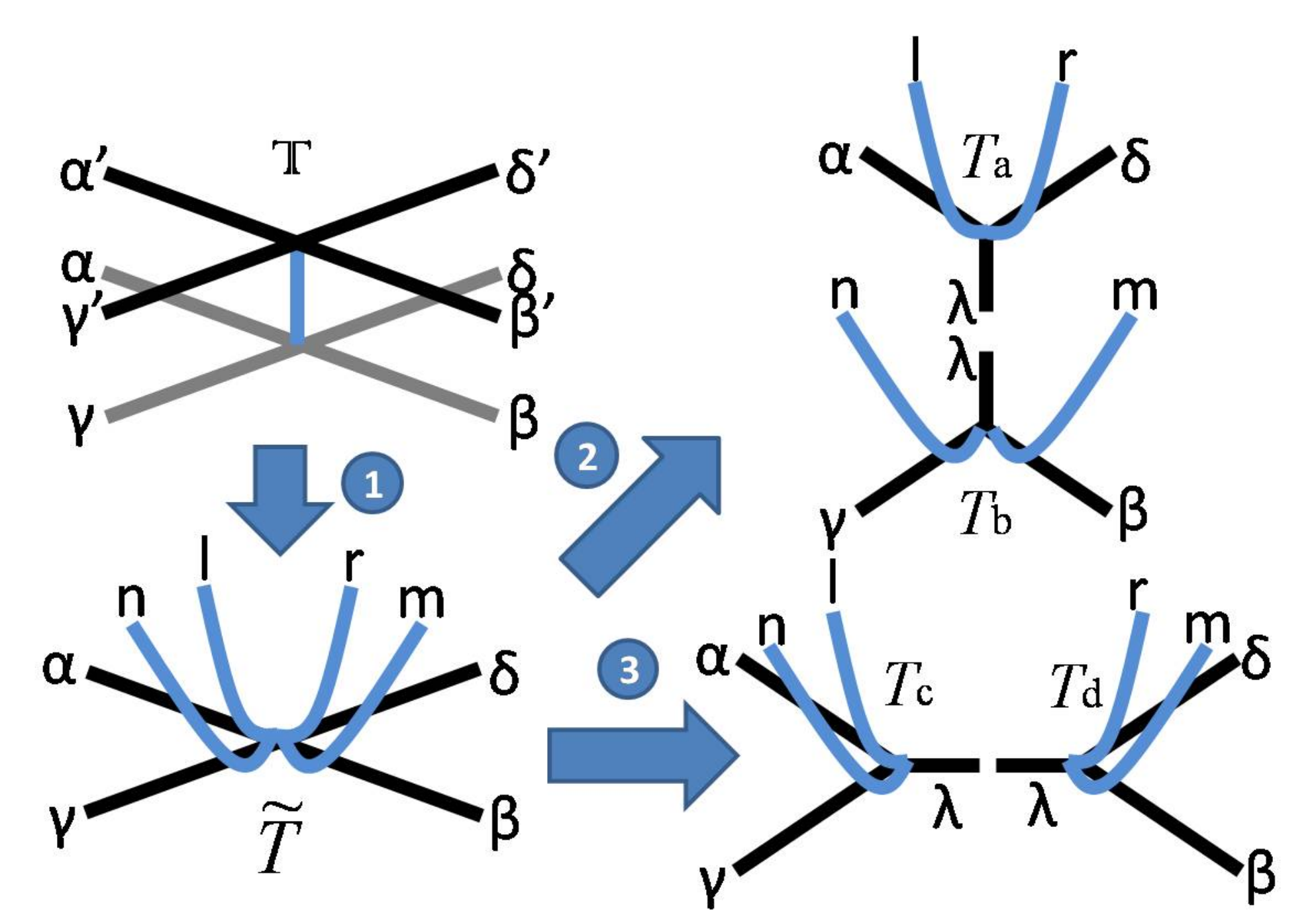}
\caption{
(Color online) Renormalization procedure on square lattice
part 1: (1) decomposing double tensor $\mathbb{T}$ into
tensor $\t{T}$ (2) SVD decomposition of $\t{T}$ in two
different directions, resulting in tensors $T_a$, $T_b$ and
$T_c$, $T_d$ respectively.}
\label{fig:RG_SQ_1}
\end{figure}
Then do the spectral decomposition of positive operator $\mathbb{T}$ into
\begin{equation*}
\mathbb{T}_{\alpha'\beta'\gamma'\delta',\alpha\beta\gamma\delta}
=\sum_j \lambda_j \left(\hat{T}^{j}_{\alpha'\beta'\gamma'\delta'}\right)^*
\times \hat{T}^{j}_{\alpha\beta\gamma\delta}
\end{equation*}
and form a new tensor $\t{T}^{lmnr}_{\alpha\beta\gamma\delta}$
\begin{equation*}
\t{T}^{lmnr}_{\alpha\beta\gamma\delta}=\sum_j \sqrt{\lambda_j}
\left(\hat{T}^{j}_{lmnr}\right)^*\times \hat{T}^{j}_{\alpha\beta\gamma\delta}
\end{equation*}
$l$,$m$,$n$,$r$ are physical indices in the up, down, left,
right directions respectively. This is illustrated in step 1
of Fig. \ref{fig:RG_SQ_1}. This step is very similar to the
second step in the F-move on hexagonal lattice.  Next we do
SVD decomposition of tensor $\t{T}$. For vertices in
sublattice $A$ we decompose between the up-right and
down-left direction as shown in step 2 of Fig.
\ref{fig:RG_SQ_1}. For vertices in sublattice $B$ we
decompose between the up-left and down-right direction as
shown in step 3 of Fig. \ref{fig:RG_SQ_1}.
\begin{eqnarray*}
\t{T}^{lmnr}_{\alpha\beta\gamma\delta}
=\sum_{\lambda} T^{lr}_{a,\alpha\delta\lambda}\times
T^{mn}_{b,\beta\gamma\lambda} \\
\t{T}^{lmnr}_{\alpha\beta\gamma\delta}
=\sum_{\lambda} T^{lr}_{c,\alpha\delta\lambda}\times
T^{mn}_{b,\beta\gamma\lambda}
\end{eqnarray*}

\begin{figure}[tb] \centering
\includegraphics[width=1.2in]{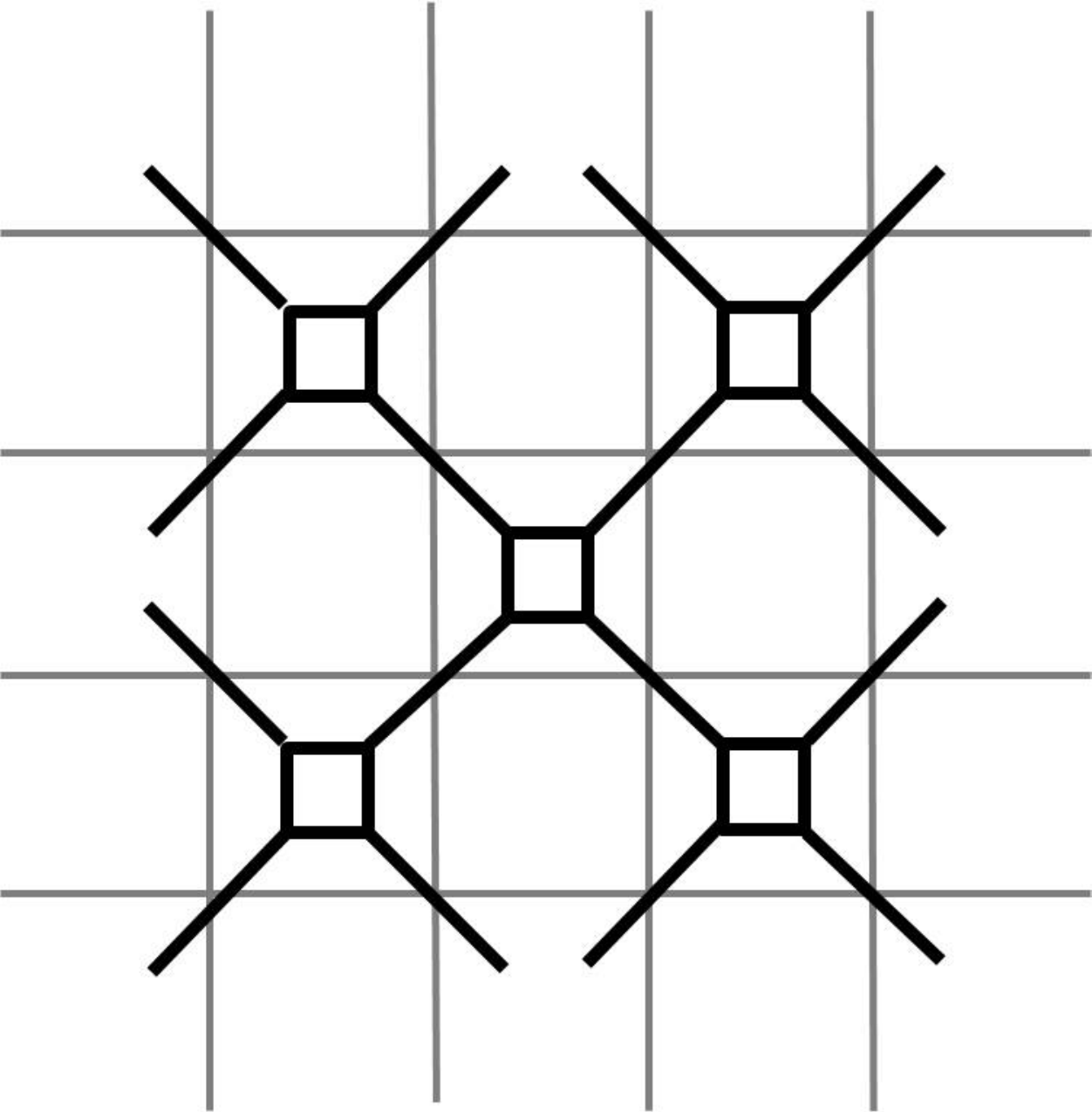}
\caption{
Original square lattice (gray line) and renormalized
lattice (black line) after the operations in Fig.
\ref{fig:RG_SQ_1} has been applied. Physical indices of the
tensors are not drawn here.}
\label{fig:RG_SQ_2}
\end{figure}

\begin{figure}[tb] \centering
\includegraphics[width=2.5in]{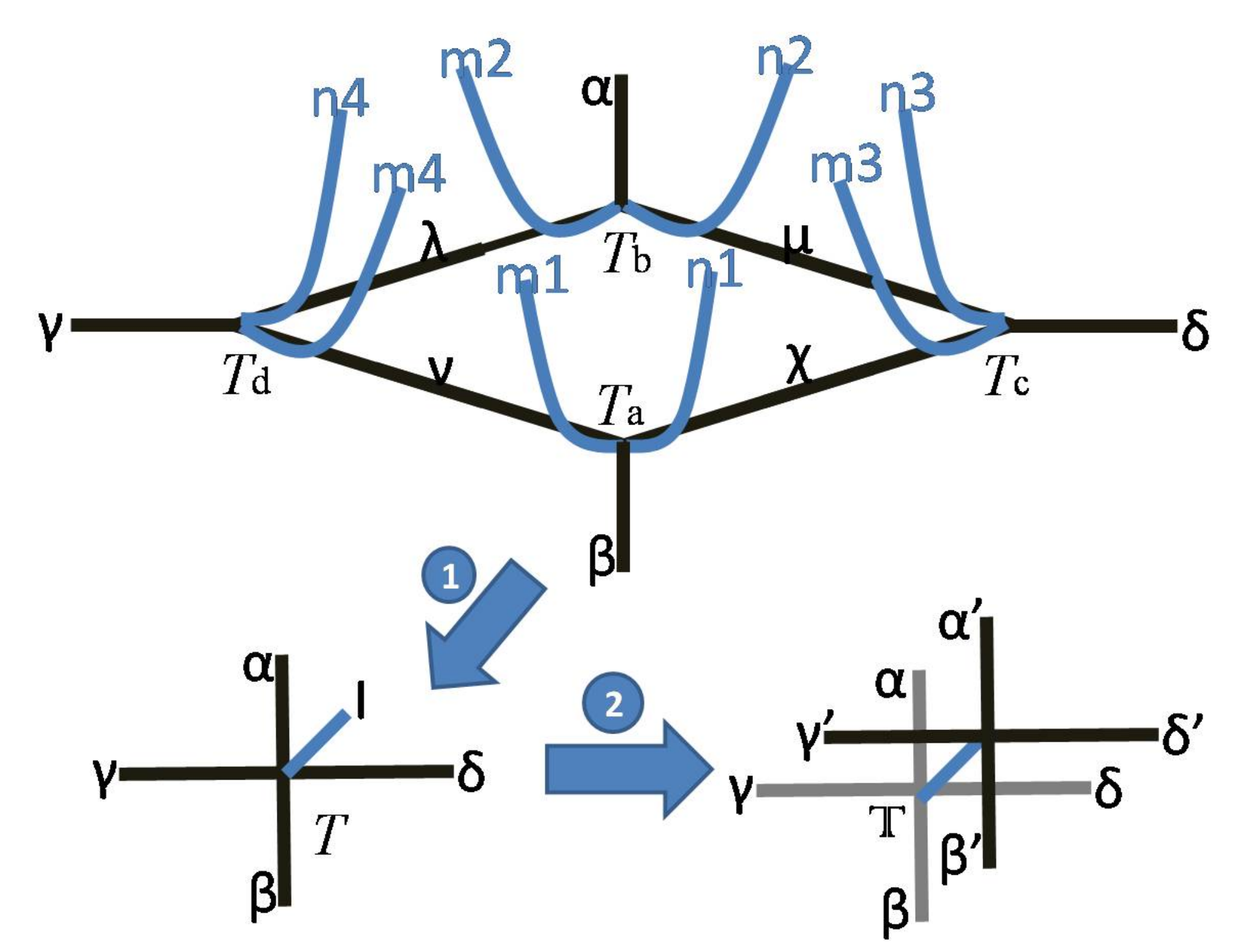}
\caption{
(Color online) Renormalization procedure on square lattice
part 2: (1) combining four tensors that meet at a square
into a single tensor (2) constructing a double tensor from
it.}
\label{fig:RG_SQ_3}
\end{figure}

After the decomposition, the original lattice (gray lines in
Fig. \ref{fig:RG_SQ_2}) is transformed into the
configuration shown by thick dark lines in Fig.
\ref{fig:RG_SQ_2}. Physical indices are omitted from this
figure. If we now shrink the small squares, we get a tensor
product state on a renormalized square lattice. Fig.
\ref{fig:RG_SQ_3} shows how this is done.

In step 1, we contract the four tensors that meet at a small square
\begin{equation*}
T^{I}_{\alpha\beta\gamma\delta}=
\sum_{\lambda\mu\nu\chi} T^{m_1n_1}_{a,\beta\nu\chi}\times
T^{m_2n_2}_{b,\alpha\lambda\mu}\times T^{m_3n_3}_{c,\delta\mu\chi}
\times T^{m_4n_4}_{d,\gamma\lambda\nu}
\end{equation*}
where $I$ stands for the combination of all physical indices
$m_i,n_i$, $i=1...4$. In step 2, we construct a double
tensor $\mathbb{T}$ form tensor $T$ and completes one round
of renormalization. Now we can go back to step 1 in Fig.
\ref{fig:RG_SQ_1} and flow the tensor further.

Now we are ready to discuss two particular examples, the
Ising symmetry breaking phase and the $\mathbb{Z}_2$
topological ordered phase, to demonstrate how our algorithm
can be used to determine the phase of a tensor product
state.

\subsection{Ising symmetry breaking phase}

A typical example for symmetry breaking phase transition is
the transverse field Ising model.  Consider a square
lattice with one spin $1/2$ on each site. The transverse field
Ising model is
\begin{equation*}
H_{Ising}=\sum_{ij} Z_iZ_j +
\epsilon \sum_k X_k
\end{equation*}
where $\{ij\}$ are nearest neighbor sites. The hamiltonian
is invariant under spin flip transformation $\prod_{k} X_k$
for any $\epsilon$.

When $\epsilon=0$, the ground state spontaneously breaks
this symmetry into either the all spin up state
$|00...0\rangle$ or the all spin down state
$|11...1\rangle$. In this case any global superposition $\al
|00...0\rangle + \bt |11...1\rangle$ represents a degenerate
ground state. When $\epsilon=\infty$, the ground state has
all spin polarized in the $X$ direction ($|++...+\rangle$)
and does not break this symmetry.

In the variational study of this system, we can
require that the variational ground state always have this
symmetry, regardless if the system is in the symmetry breaking
phase or not. Then we will find for $\epsilon=0$ the ground
state to be $|00...0\rangle + |11...1\rangle$. Such a
global superposition represents the
spontaneous symmetry breaking.
For $\epsilon=\infty$, we
will find the ground state to be $|++...+\rangle$ and does
not break the symmetry. For $0<\epsilon<\infty$, we will
need to decide which of the previous two cases it belongs
to. We can first find a tensor product representation of an
approximate ground state which is symmetric under the spin
flip transformation, then apply the renormalization
algorithm to find the fixed point and decide which phase the
state belongs to. Below we will assume a simple form of
tensor and demonstrate how the algorithm works.

Suppose that the tensors obtained from the variational
study $T^{i}_{\alpha\beta\gamma\delta}$, where
$i$,$\alpha$,$\beta$,$\gamma$,$\delta$ can be $0$ or $1$, takes the
following form \begin{equation} \begin{array}{l}
T^{0}_{\alpha\beta\gamma\delta}=\lambda^{\alpha+\beta+\gamma+\delta}\\
T^{1}_{\alpha\beta\gamma\delta}=\lambda^{4-(\alpha+\beta+\gamma+\delta)}
\end{array}
\label{T_Ising}
\end{equation}
$\lambda$ is a parameter between $0$ and $1$.  Under an $X$
operation to the physical index, the tensor is changed to
$\tilde{T}$
\begin{equation*}
\begin{array}{l}
\tilde{T}^{0}_{\alpha\beta\gamma\delta}=\lambda^{4-(\alpha+\beta+\gamma+\delta)}
\nonumber\\
\tilde{T}^{1}_{\alpha\beta\gamma\delta}=\lambda^{\alpha+\beta+\gamma+\delta}
\end{array}
\end{equation*}
$\tilde{T}$ can be mapped back to $T$ by switching the
$0$,$1$ label for the four inner indices
$\alpha\beta\gamma\delta$. Such a change of basis for the inner
indices does not change the contraction result of the tensor
and hence the state that is represented. Therefore, the
state is invariant under the spin flip transformation
$\prod_k X_{k}$ and we will say that the tensor has this
symmetry also.

When $\lambda=0$, the tensor represents state
$|00...0\rangle + |11...1\rangle$, which corresponds to the
spontaneous symmetry breaking phase.  We note that the
$\lambda=0$ tensor is a direct sum of dimension-1 tensors.
Such a direct-sum structure corresponds to spontaneous
symmetry breaking, as discussed in detail in \Ref{GW0931}.
When $\lambda=1$, the tensor
represents state $|++...+\rangle$ which corresponds to the
symmetric phase. When $0<\lambda<1$, there must be a phase
transition between the two phases. However, as $\lambda$
goes from $0$ to $1$, the tensor varies smoothly with well
defined symmetry. It is hard to identify the phase
transition point. Now we can apply our algorithm to the
tensor. First, we notice that at $\lambda=0$ or $1$, the
tensor is a fixed point for our algorithm. Next, we find
that for $\lambda<0.358$, the tensor flows to the form with
$\lambda=0$, while for $\lambda>0.359$, it flows to the form
with $\lambda=1$. Therefore, we can clearly identify the
phase a state belongs to using this algorithm and find the
phase transition point.

Note that in our algorithm, we explicitly keep the spin flip
symmetry in the tensor. That is, after each renormalization
step, we make sure that the renormalized tensor is invariant
under spin flip operations up to change of basis for the
inner indices. If the symmetry is not carefully preserved,
we will not be able to tell the two phases apart.

\begin{figure}[t] \centering
\includegraphics[width=2.5in]{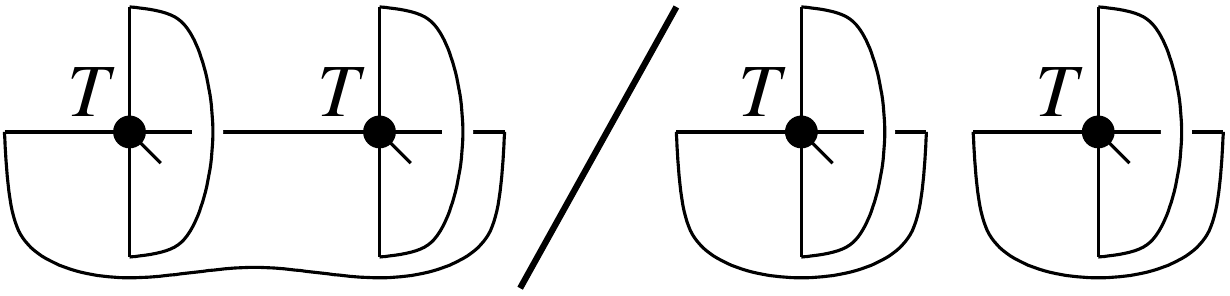}
\caption{Quantity $X_2/X_1$ obtained by taking the ratio of
the contraction value of the double tensor in two different
ways.  $X_2/X_1$ is invariant under change of scale, basis
transformation and corner double line structures of the
double tensor and can be used to distinguish different fixed
point tensors. For clarity, only one layer of the double
tensor is shown. The other layer connects in exactly the
same way.}
\label{fig:X2X1}
\end{figure}

We also need to mention that for arbitrary $\lambda$, the
fixed point that the tensor flows to can be different from the
tensor at $\lambda=0$ or $1$ by a corner double line
structure. We need to get rid of the corner double line
structure in the result to identify the real fixed point.
This is possible by carefully examining the fixed point
structure. Another way to distinguish the different fixed
points without worrying about corner double lines is to
calculate some quantities from the fixed-point tensors that
are invariant with the addition of corner double lines. We
also want the quantity to be invariant under some trivial
changes to the fixed point, such as a change in scale $T
\to \eta T$ or the change of basis for physical and inner
indices. One such quantity is given by the ratio of $X_2$
and $X_1$ defined as
$X_1=\left(\sum_{\alpha'\gamma',\alpha\gamma}
\mathbb{T}_{\alpha'\alpha'\gamma'\gamma',\alpha\alpha\gamma\gamma}\right)^2
$, and $
X_2
=\sum_{\alpha'\beta'\gamma'\delta',\alpha\beta\gamma\delta}
\mathbb{T}_{\alpha'\alpha'\gamma'\delta',\alpha\alpha\gamma\delta}
\times
\mathbb{T}_{\beta'\beta'\delta'\gamma',\beta\beta\delta\gamma}
$.
Fig. \ref{fig:X2X1} gives a graphical representation of these
two quantities. In this figure, only one layer of the double
tensor is shown. The other layer connects in the exactly the
same way. It is easy to verify that $X_2/X_1$ is invariant
under the change of scale, basis transformation and corner
double lines.

\begin{figure}[tb] \centering
\includegraphics[width=3.7in]{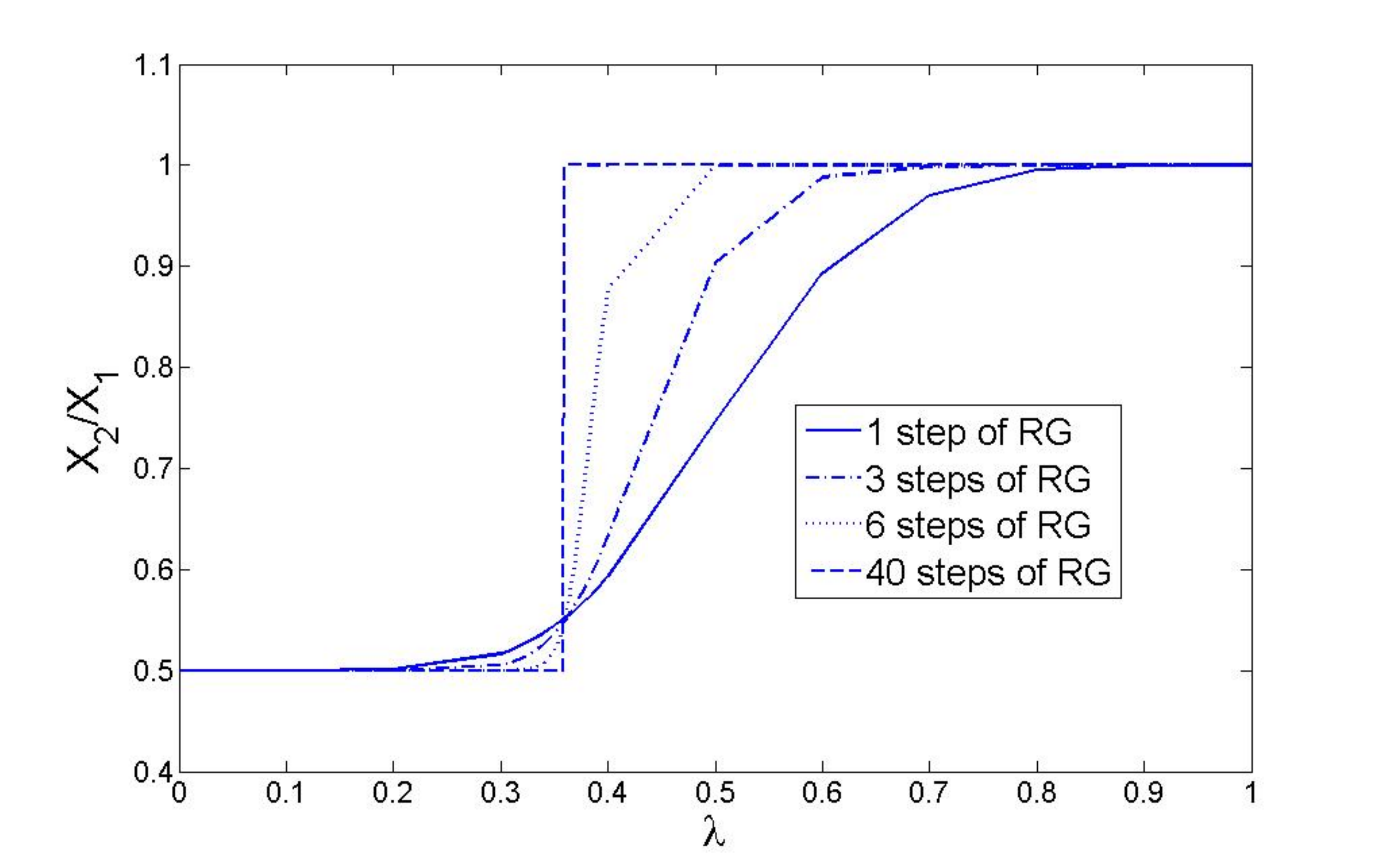}
\caption{(Color online)
$X_2/X_1$ for tensors \eq{T_Ising} under the renormalization
flow. As the number of RG steps increases, the transition in
$X_2/X_1$ becomes sharper and finally approaches a step
function at fixed point. The critical point is at
$\lambda_c=0.358$.}
\label{fig:X1_Ising}
\end{figure}

We calculate $X_2/X_1$ along the renormalization flow. The
result is shown in Fig. \ref{fig:X1_Ising}. At the
$\lambda=0$ fixed point, $X_2/X_1=0.5$ while at $\lambda=1$,
$X_2/X_1=1$.  As we increase the number of renormalization
steps, the transition between the two fixed points becomes
sharper and finally approaches a step function with critical
point at $\lambda_c=0.358$. Tensors with $\lambda<\lambda_c$
belongs to the symmetry breaking phase  while tensors with
$\lambda>\lambda_c$ belongs to the symmetric phase.

\subsection{$\mathbb{Z}_2$ topological ordered phase}

The algorithm can also be used to study topological order of
quantum states. In this section, we will demonstrate how the
algorithm works with $\mathbb{Z}_2$ topological order.

Consider again a square lattice but now with one spin $1/2$
per each link. A simple Hamiltonian on this lattice with
$\mathbb{Z}_2$ topological order can be defined as
\begin{equation*} H_{\mathbb{Z}_2}=\sum_p \prod_{i \in p} X_i
+\sum_v \prod_{j \in v} Z_j \label{H_Z2}
\end{equation*}
where $p$ means plaquettes and $i \in p$ is all the spin $1/2$s
around the plaquette and $v$ means vertices and $j \in v$ is
all the spin $1/2$s connected to the vertex.  The ground state
wave function of this Hamiltonian is a fixed-point wave
function and corresponds to the $N=1$ loop state with
$\eta=1$ as discussed in the previous section.

The ground state wave function has a simple tensor product
representation. For simplicity of discussion we split
every spin $1/2$ into two and associate every vertex with four
spins. The tensor
$T^{ijkl}_{\alpha\beta\gamma\delta,\mathbb{Z}_2}$ has four
physical indices $i,j,k,l=0,1$ and three inner indices
$\alpha,\beta,\gamma,\delta=0,1$.
\begin{equation*}
\begin{array}{l}
T^{ijkl}_{ijkl,\mathbb{Z}_2}=1, \text{\ if\ } \text{mod}(i+j+k+l,2)=0\\
\text{all other terms being $0$}
\label{T_Z2}
\end{array}
\end{equation*}
It can be checked that $T_{\mathbb{Z}_2}$ is a fixed-point
tensor of our algorithm. This tensor has a $\mathbb{Z}_2$
gauge symmetry. If we apply $Z$ operation to all the inner
indices, where $Z$ maps $0$ to $0$ and $1$ to $-1$, the
tensor remains invariant as only even configurations of the
inner indices are nonzero in the tensor.

Consider then the following set of tensor parameterized by $g$
\begin{equation}
\begin{array}{l}
T^{ijkl}_{ijkl}=g^{i+j+k+l}, \text{\ if\ } \text{mod}(i+j+k+l,2)=0\\
\text{all other terms being $0$}
\label{T_Z2_g}
\end{array}
\end{equation}
At $g=1$, this is exactly $T_{\mathbb{Z}_2}$ and the
corresponding state has topological order. At $g=0$, the
tensor represents a product state of all $0$ and we denote
the tensor as $T_0$. At some critical point in $g$, the
state must go through a phase transition. This set of
tensors are all invariant under gauge transformation $ZZZZ$
on their inner indices and the tensor seems to vary smoothly
with $g$. One way to detect the phase transition is to apply
our algorithm. We find that, at $g>g_c$, the tensors flow to
$T_{\mathbb{Z}_2}$, while at $g<g_c$, the tensors flow to
$T_0$. We determine $g_c$ to be between $0.804 \sim 0.805$.
As this model is mathematically equivalent to two
dimensional classical Ising model where the transition point
is known to great accuracy, we compare our result to that
result and find our result to be within $1\%$ accuracy
($g_c=0.8022$). Again in the renormalization algorithm, we
need to carefully preserve the $\mathbb{Z}_2$ gauge symmetry
of the tensor so that we can correctly determine the phase
of the states.

\begin{figure}[tb] \centering
\includegraphics[width=3.7in]{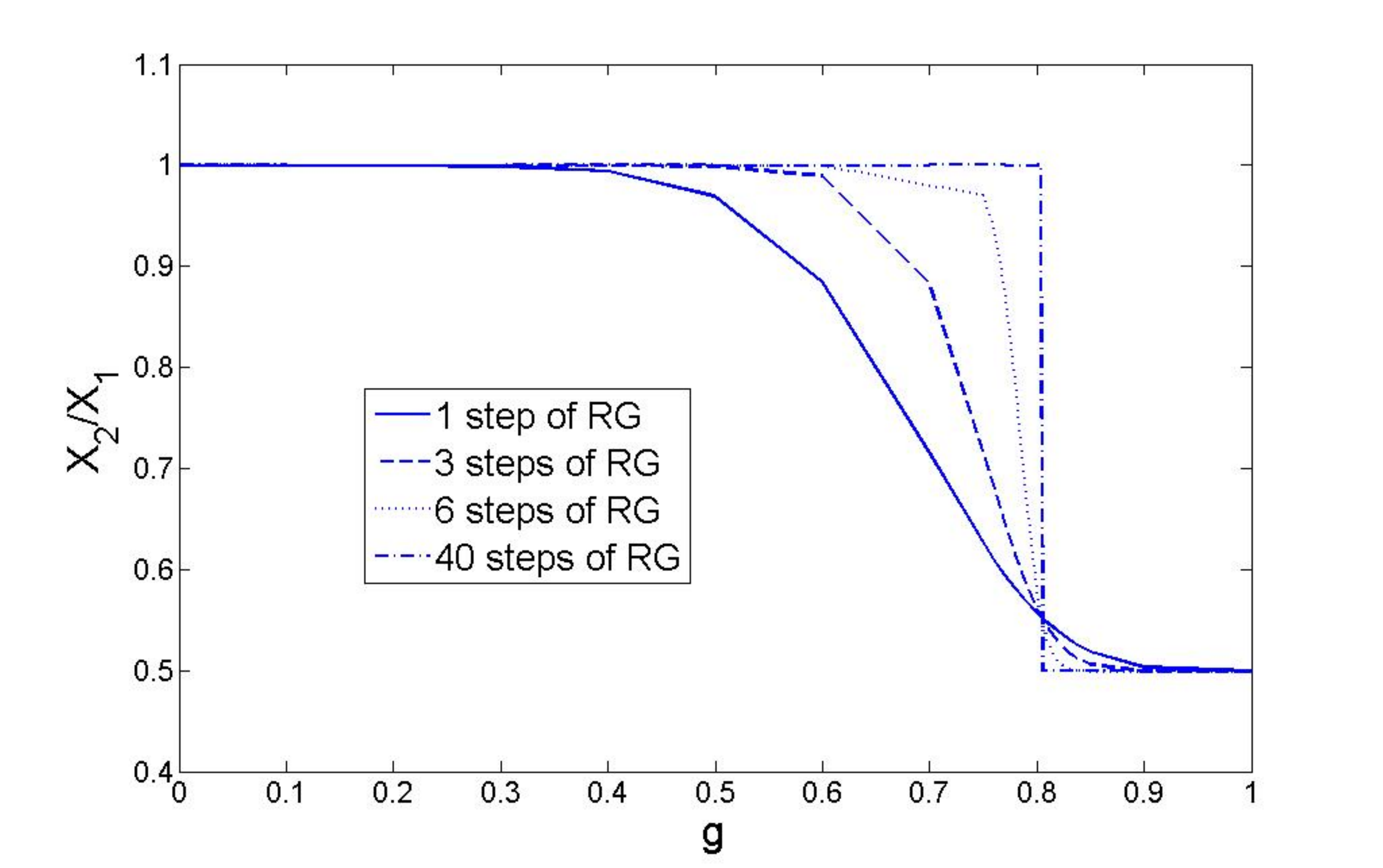}
\caption{(Color online)
$X_2/X_1$ for tensors \eq{T_Z2_g} under the renormalization
flow. As the number of RG steps increases, the transition in
$X_2/X_1$ becomes sharper and finally approaches a step
function at fixed point. The critical point is at
$g_c=0.804$.}
\label{fig:X1_Z2}
\end{figure}

The fixed-point tensor structure might also be complicated
by corner double line structures, but it is always possible
to identify and get rid of them. Similarly, we can calculate
the invariance quantity $X_2/X_1$ to distinguish
the two fixed points. $X_2/X_1=1$ for $T_{\mathbb{Z}_2}$
while $X_2/X_1=0.5$ for $T_0$. The result is plotted in Fig.
\ref{fig:X1_Z2} and we can see that the transition in
$X_2/X_1$ approaches a step function after a large number of
steps of RG, i.e. at the fixed point. The critical point is
at $g_c=0.804$. For $g<g_c$, the tensor belongs to the
trivial phase, while for $g>g_c$, the tensor belongs to the
$\mathbb{Z}_2$ topological ordered phase.

Our algorithm can also be used to demonstrate the stability
of topological order against local perturbation. As is shown
in Ref.  \onlinecite{CZG1003}, local perturbations to the
$\mathbb{Z}_2$ Hamiltonian correspond to variations in
tensor that do not break the $\mathbb{Z}_2$ gauge symmetry.
We picked tensors in the neighborhood of $T_{\mathbb{Z}_2}$
which preserve this gauge symmetry randomly and applied our
renormalization algorithm (gauge symmetry is kept throughout
the renormalization process). We find that as long as the
variation is small enough, the tensor flows back to
$T_{\mathbb{Z}_2}$, up to a corner double line structure.
This result shows that the $\mathbb{Z}_2$ topological
ordered phase is stable against local perturbations.

\section{Summary}

In this paper, we discuss a defining relation between local
unitary transformation and quantum phases.  We argue that
two gapped states are related by a local unitary
transformation if and only if the two states belong to the same
quantum phase.

We can use the equivalent classes of local unitary
transformations to define ``patterns of long range
entanglement''.  So the patterns of long range entanglement
correspond to universality classes of quantum phases, and
are the essence of topological orders.\cite{W0275}

As an application of this point of view of quantum phases
and topological order, we use the generalized local unitary
transformations to generate a wave function renormalization,
where the wave functions flow within the same universality
class of a quantum phase (or the same equivalent class of
the local unitary transformations).  In other words, the
renormalization flow of a wave function does not change its
topological order.

Such a wave function renormalization allows us to classify
topological orders, by classifying the fixed-point wave
functions and the associated fixed-point local unitary
transformations.

First, we find that the fixed-point local unitary
transformations are described by the data
$(N_{ijk}, F^{ijm,\al\bt}_{kln,\chi\del},P_i^{kj,\al\bt})$
that satisfy
\begin{align}
\label{Ncond}
& N_{ijk}=N_{jki} =N_{k^*j^*i^*} \geq 0,\ \ \ \
 \sum_{j,k=0}^N N_{ii^*k} N_{jk^*j^*} \geq 1 ,
\nonumber\\
& \sum_{m=0}^N N_{jim^*} N_{kml^*}
=\sum_{n=0}^N N_{kjn^*} N_{l^*ni} ,
\end{align}
\begin{align}
\label{gLUcondF}
& (F^{ijm,\al\bt}_{kln,\chi\del})^*
= F^{jkn,\chi\del}_{l^*i^*m^*,\bt\al} ,
\nonumber\\
&
\sum_{n,\chi,\del}
F^{ijm',\al'\bt'}_{kln,\chi\del}
(F^{ijm,\al\bt}_{kln,\chi\del})^*
=\del_{m\al\bt,m'\al'\bt'},
\nonumber\\
&
\sum_{t}
\sum_{\eta=1}^{N_{kjt^*}}
\sum_{\vphi=1}^{N_{tin^*}}
\sum_{\ka=1}^{N_{lts^*}}
F^{ijm,\al\bt}_{knt,\eta\vphi}
F^{itn,\vphi\chi}_{lps,\ka\ga}
F^{jkt,\eta\ka}_{lsq,\del\phi}
\nonumber\\
&\ \ \ \ \ =
\e^{\imth \th_F}
\sum_{\eps=1}^{N_{qmp^*}}
F^{mkn,\bt\chi}_{lpq,\del\eps}
F^{ijm,\al\eps}_{qps,\phi\ga} ,
\end{align}
\begin{align}
\label{gLUcondFP}
&
\e^{\imth \th_{P1}}
P_i^{kj,\al\bt}=
\sum_{m,\la,\ga,l,\nu,\mu}
F^{jj^*k,\bt\al}_{i^*i^*m^*,\la\ga}
F^{i^*mj,\la\ga}_{m^*i^*l,\nu\mu}
P_{i^*}^{lm,\mu\nu} ,
\nonumber\\
&
\e^{\imth \th_{P2}}
P^{jp,\al\eta}_i \del_{im} \del_{\bt\del}
= \sum_{\chi=1}^{N_{kjk^*}} F^{ijm,\al\bt}_{klk,\chi\del} P^{jp,\chi\eta}_{k^*}
\nonumber\\
&\ \ \ \ \ \ \ \ \ \ \ \text{for all } k,i,l \text{ satisfying }
N_{kil^*}>0 .
\end{align}
[see eqns. (\ref{NNNN}, \ref{Di}, \ref{NNstar}
\ref{2FFstar}, \ref{1FstarF}, \ref{penid}, \ref{PFFP},
\ref{PFP})]. From the data $(N_{ijk},
F^{ijm,\al\bt}_{kln,\chi\del},P_i^{kj,\al\bt})$ we can
further find out the fixed-point wave function by solving
the following equations for $A^i, i=0,...,N$:
\begin{align}
\label{wavcond}
& A^i=\e^{\imth \th_A} A^{i^*}\neq 0,\ \ \ \ \sum_i A^i (A^i)^*=1 ,
\nonumber\\
& P^{mj,\ga\la}_i A^i =\e^{\imth \th_{A1}} P^{m^*i^*,\la\ga}_{j^*} A^j ,
\nonumber\\
& \Phi^\th_{ikj,\al\bt} = \e^{\imth \th'}
\sum_{m,\la,\ga}
F^{ijk^*,\al\bt}_{j^*im,\la\ga} P^{mj,\ga\la}_i A^i ,
\nonumber\\
&  \Phi^\th_{ikj,\al\bt}
=  \e^{\imth \th_{A1}} \Phi^\th_{kji,\al\bt},
\nonumber\\
& \Phi^\th_{ikj,\al\bt}=0, \text{ if } N_{ikj}=0,
\nonumber\\
& \det(\Phi^\th_{ikj,\al\bt})\neq 0.
\end{align}
[see eqns. (\ref{Anonz}, \ref{AiAistar}, \ref{PAPA},
\ref{PhiPA}, \ref{Phiikj}, \ref{Phinonz})].

The combined data $(N_{ijk},
F^{ijm,\al\bt}_{kln,\chi\del},P_i^{kj,\al\bt},A^i)$ that
satisfy the conditions \eq{Ncond}, \eq{gLUcondF},
\eq{gLUcondFP} and \eq{wavcond} classify a large class of
topological orders.  We see that the problem of classifying
a large class of topological orders becomes the problem of
solving a set of non-linear algebraic equations \eq{Ncond},
\eq{gLUcondF},  \eq{gLUcondFP}, and \eq{wavcond}.

The combined data $(N_{ijk},
F^{ijm,\al\bt}_{kln,\chi\del},P_i^{kj,\al\bt},A^i)$ that
satisfy the conditions \eq{Ncond}, \eq{gLUcondF},  \eq{gLUcondFP},
and \eq{wavcond} also classify a large class of time
reversal invariant topological orders, if we restrict
ourselves to real solutions.  The solutions related by local
orthogonal transformations all belong to the same phase,
since the local orthogonal transformations always connect to
identity if we enlarge the Hilbert space.

We like to point out that we cannot claim that the solutions
$(N_{ijk}, F^{ijm,\al\bt}_{kln,\chi\del},P_i^{kj,\al\bt},A^i)$
classify all topological orders since we have assumed that
the fixed-point local unitary transformations are described
by tensors of finite dimensions. It appears that chiral
topological orders, such as quantum Hall states, are
described by tensors of infinite dimensions.

We note that the data $(N_{ijk},
F^{ijm,\al\bt}_{kln,\chi\del},P_i^{kj,\al\bt},A^i)$ just
characterize different fixed-point wave functions.  It is
not guaranteed that the different data will represent
different topological orders.  However, for the simple
solutions discussed in this paper, they all coincide with
string-net states, where the topological properties, such as
the ground state degeneracy, number of quasiparticle types,
the quasiparticle statistics {\it etc}, were calculated from
the data. From those topological properties, we know that
those different simple solutions represent different
topological orders.  Also $(
F^{ijm,\al\bt}_{kln,\chi\del},P_i^{kj,\al\bt},A^i)$ are real
for the simple solutions discussed here. Thus they also
represent topological orders with time reversal symmetry.
(Certainly, at the same time, they represent stable
topological orders even without time reversal symmetry.)

We also like to point out that our description of
fixed-point local unitary transformations is very similar to
the description of string-net states. However, the
conditions \eq{Ncond}, \eq{gLUcondF},  \eq{gLUcondFP}, and
\eq{wavcond} on the data appear to be weaker than (or
equivalent to) those\cite{LWstrnet} on the string-net data.
So the fixed-point wave functions discussed in this paper
may include all the string-net states (in 2D).

Last, we present a wave function renormalization scheme,
based on the gLU transformations for generic TPS.  Such a
wave function renormalization always flows within the same
phase (or within the same equivalence class of LU
transformations). It allows us to determine which phase a
generic TPS belongs to by studying the resulting fixed-point
wave functions.  We demonstrated the effectiveness of our
method for both symmetry breaking phases and topological
ordered phases.  We find that we can even use tensors that
do not break symmetry to describe spontaneous symmetry
breaking states: if a state described by a symmetric tensor
has a spontaneous symmetry breaking, the symmetric tensor
will flow to a fixed-point tensor that has a form of direct
sum.

We would like to thank I. Chuang, M. Hastings, M. Levin, F.
Verstraete, Z.-H. Wang, Y.-S. Wu, S. Bravyi for some very helpful
discussions.  XGW is supported by  NSF Grant No.
DMR-0706078. ZCG is supported in part by the NSF Grant No.
NSFPHY05-51164.

\section*{Appendix: Equivalence relation between quantum states in the same phase}

In section \ref{QphUcl} and section \ref{TOLREn}, we argued about the equivalence relation between gapped quantum ground states in the same phase. We concluded that two states are in the same phase if and only if they can be connected by local unitary evolution or constant depth quantum circuit:
\begin{eqnarray*}
\label{LUtrans}
 |\Phi(1)\> \sim |\Phi(0)\> &\text{\ iff\ }&
 |\Phi(1)\> =  \cT[e^{-i\int_0^1 dg\, \t H(g)}] |\Phi(0)\>\\
 |\Phi(1)\> \sim |\Phi(0)\> &\text{\ iff\ }&
 |\Phi(1)\> = U^M_{circ} |\Phi(0)\>
\end{eqnarray*}

Now we want to make these arguments more rigorous, by
stating clearly what is proved and what is conjectured, and
by giving precise definition of two states being the same,
the locality of operators, etc. We will show the equivalence
in the following steps:

1. if two gapped ground states are in the same phase, then they are connected by local unitary evolution

2. a gapped ground state remains in the same phase under local unitary evolution

3. a local unitary evolution can be simulated by a constant depth quantum circuit and vice-verse.

(All these discussions can be generalized to the case where the system has certain symmetries. $\t H(g)$ and $U_{circ}$ used in the equivalence relation will then have the same symmetry as the system Hamiltonian $H$.)

First, according to the definition in section \ref{QphUcl},
two states $|\Phi(0)\>$ and $|\Phi(1)\>$ are in the same
phase if we can find a family of local Hamiltonians $H(g)$,
$g \in [0,1]$ with $|\Phi(g)\>$ being its ground state such
that the ground state average of any local operator $O$,
$\<\Phi(g)|O|\Phi(g)\>$ changes smoothly from $g=0$ to
$g=1$. Here we allow a more general notion of locality for
the Hamiltonian \cite{BHM1044} and require $H(g)$ to be a
sum of local operators $H_Z(g)$:
\begin{align}
 H(g)=\sum_{Z\in \cZ} H_Z(g)
\end{align}
where $H_Z(g)$ is a Hermitian operator defined on a compact
region $Z$.  $\sum_{Z\in \cZ}$ sums over a set $\cZ$ of
regions.  The set $\cZ$ contains regions that differ by
translations.  The set $\cZ$ also contains regions with
different sizes.  However, $H_Z(g)$ approaches zero
exponentially as the size of the region $Z$ approaches
infinity.  Or more precisely, for all sites $u$ in the
lattice
\begin{equation}
\sum_{Z\in \cZ, Z \ni u} \left \Vert H_Z(g) \right \| \left | Z \right \vert exp(\mu \text{diam}(Z))=O(1)
\label{local_H}
\end{equation}
for some positive constant $\mu$. Here $\sum_{Z\in \cZ, Z
\ni u}$ sums over all regions in the set $\cZ$ that cover the site
$u$, $\left \Vert ...  \right \|$ denotes operator norm,
$\left | Z \right \vert$ is the cardinality of $Z$,  and
$\text{diam}(Z)$ is the diameter of $Z$. Therefore, instead
of being exactly zero outside of a finite region, the
interaction terms can have an exponentially decaying tail.

If $|\Phi(0)\>$ and $|\Phi(1)\>$ are gapped ground states of
$H(0)$ and $H(1)$, then we assume that for all
$\<\Phi(g)|O|\Phi(g)\>$ to be smooth, $H(g)$ must remain
gapped for all $g$. If $H(g)$ closes gap for some $g_c$,
then there must exist a local operator $O$ such that
$\<\Phi(g)|O|\Phi(g)\>$ has a singularity at $g_c$.  We call
the gapped $H(g)$ an adiabatic connection between two states
in the same phase.

The existence of an adiabatic connection gives rise to a
local unitary evolution between $|\Phi(0)\>$ and
$|\Phi(1)\>$. A slight modification of Lemma 7.1 in
\Ref{BHM1044} gives that
\begin{theorem}
Let $H(g)$ be a differentiable family of local Hamiltonians and  $|\Phi(g)\>$ be its ground state. If the excitation gap above $|\Phi(g)\>$ is larger than some finite value $\Delta$ for all $g$, then we can define $\t H(g)=i\int dtF(t)exp(iH(g)t)\left(\partial_g H(g) \right) exp(-iH(g)t)$, such that $|\Phi(1)\> =  \cT[e^{-i\int_0^1 dg\, \t H(g)}] |\Phi(0)\>$.
\end{theorem}
where $F(t)$ is a function which has the following
properties: 1. The fourier transform of $F(t)$, $\t
F(\omega)$ is equal to $-1/\omega$ for $|\omega|>\Delta$.
2. $\t F(\omega)$ is infinitely differentiable. 3.
$F(t)=-F(-t)$. Under this definition, $\t H(g)$ is
local (almost) and satisfies
\begin{equation}
\left \Vert \left[ \t H_Z(g), O_B \right] \right \| \leq h'(\text{dist}(Z,B)) \left | Z \right \vert \left \Vert \t H_Z(g) \right \| \left \Vert O_B \right \|
\label{local_tH}
\end{equation}
where $O_B$ is any operator supported on site $B$,
$\text{dist}(Z,B)$ is the smallest distance between $B$ and
any site in $Z$, and $h'(l)$ is a function which decays
faster than any negative power of $l$. $[...]$ denotes
commutator of two operators. This is called the
quasi-adiabatic continuation of states.

Therefore, we can show that if $|\Phi(0)\>$ and $|\Phi(1)\>$
are in the same phase, then we can find a local (as defined
in \eq{local_tH}) Hamiltonian $\t H_Z(g)$ such that
$|\Phi(1)\> =  \cT[e^{-i\int_0^1 dg\, \t H(g)}] |\Phi(0)\>$.
In other words, states in the same phase are equivalent
under local unitary evolution.

Note that here we map $|\Phi(0)\>$ exactly to $|\Phi(1)\>$.
There is another version of quasi-adiabatic continuation
\cite{HW0541}, where the mapping is approximate. In that
case, $\t H_Z(g)$ can be defined to have only exponentially
small tail outside of a finite region instead of a tail
which decays faster than any negative power.
$\cT[e^{-i\int_0^1 dg\, \t H(g)}] |\Phi(0)\>$ will not be
exactly the same as $|\Phi(1)\>$, but any local measurement
on them will give approximately the same result.

Next we want to show that the reverse is also true. Suppose
that $|\Phi(0)\>$ is the gapped ground state of a local
Hamiltonian $H(0)$, $H(0)=\sum_Z H_Z(0)$ and each $H_Z(0)$
is supported on a finite region $Z$. Apply a local unitary
evolution generated by $\t H(s)$ to $|\Phi(0)\>$ and take it
to $|\Phi(g)\>$, $g=0\sim 1$, $|\Phi(g)\> =  U_g |\Phi(0)\>$,
where $U_g=\cT[e^{-i\int_0^g ds\, \t H(s)}] $. $|\Phi(g)\>$
is then ground state of $H(g)=U_gH(0)U_g^{\dagger}=\sum_Z
U_gH_Z(0)U_g^{\dagger} =\sum_Z H_Z(g)$. Under unitary
transformation $U_g$ the spectrum of $H(0)$ doesn't change,
therefore $H(g)$ remains gapped. To show that $H(g)$ also
remains local, we use the Lieb-Robinson bound derived in
\Ref{BHM1044}, which gives
\begin{eqnarray*}
\left \Vert \left[H_Z(g), O_B\right] \right \| &=& \left \Vert \left[U_gH_Z(0)U_g^{\dagger}, O_B\right] \right \|\\
& \leq& h(\text{dist}(Z,B)) \left |Z\right \vert \left \Vert H_Z(0) \right \| \left \Vert O_B \right \|
\end{eqnarray*}
where $h(l)$ decays faster than any negative power of $l$.
Therefore, $H_Z(g)$ remains local up to a tail which decays
faster than any negative power. $H(g)$ then forms a local
gapped adiabatic connection between $|\Phi(0)\>$ and
$|\Phi(1)\>$.

To detect for phase transition, we must check whether the
ground state average value of any local operator $O$,
$\<\Phi(g)|O|\Phi(g)\>$, has a singularity or not.
$\<\Phi(g)|O|\Phi(g)\>=\<\Phi(0)|U_g^{\dagger}OU_g|\Phi(0)\>$.
Using the Lieb-Robinson bound given in \Ref{BHM1044}, we
find  $U_g^{\dagger}OU_g$ remains local (in the sense of
\eq{local_tH}) and evolves smoothly with $g$. Therefore,
$\<\Phi(g)|O|\Phi(g)\>$ changes smoothly. In fact, because
$H(g)$ is differentiable, the derivative of
$\<\Phi(g)|O|\Phi(g)\>$ to any order always exists.
Therefore, there is no singularity in the ground state
average value of any local operator $O$ and $|\Phi(0)\>$ and
$|\Phi(1)\>$ are in the same phase. We have hence shown that
states connected with local unitary evolution are in the
same phase.

This completes the equivalence relation stated in
\eq{LUdef}. The definition of locality is slightly different
in different cases, so there is still some gap in the
equivalence relation. However, we believe that the
equivalence relation should be valid with a slight
generalization of the definition of locality.

Lastly, we want to show that the equivalence relation is
still valid if we use constant depth quantum circuit instead
of local unitary evolution. This is true because we can
always simulate a local unitary evolution using a constant
depth quantum circuit and vice-verse.

To simulate a local unitary evolution of the form
$\cT[e^{-i\int_0^1 dg\, \t H(g)}]$, first divide the total
time into small segments $\delta t$ such that
$\cT[e^{-i\int_{m\delta t}^{(m+1)\delta t} dg\, \t H(g)}]
\simeq e^{-i\delta t H(m\delta t)}$. $H(m\delta t)=\sum_Z
H_Z(m\delta t)$. The set of local operators $\{H_Z(m\delta
t)\}$ can always be divided into a finite number of subsets
$\{H_{Z_1^i}(m\delta t)\}$, $\{H_{Z_2^i}(m\delta t)\}$...
such that elements in the same subset commute with each
other. Then we can do Trotter expansion and approximate
$e^{-i\delta t H(m\delta t)}$ as $e^{-i\delta t H(m\delta t)}\simeq
\left(U^{(1)}U^{(2)}...\right)^n$, where $U^{(1)}=\prod_i
e^{-i\delta tH_{Z_1^i}/n}$, $U^{(2)}=\prod_i e^{-i\delta
tH_{Z_2^i}/n}$... Each term in $U^{(1)}$, $U^{(2)}$ commute,
therefore they can be implement as a piece-wise local
unitary operator. Putting these piece-wise local unitary
operators together, we have a quantum circuit which
simulates the local unitary evolution. The depth of circuit
is proportional to $n\times (1/\delta t)$. It can be shown
that to achieve a simulation with constant error, a constant
depth circuit would suffice\cite{Lloyd96}.

On the other hand, to simulate a constant depth quantum
circuit
$U_{circ}=U^{(1)}_{pwl}U^{(2)}_{pwl}...U^{(2)}_{pwl}$, where
$U^{(k)}_{pwl}=\prod_i U^{(k)}_i$ with a local unitary
evolution, we can define the time dependent Hamiltonian as
$H(t)=\sum_i H^{(k)}_i$, such that for $(k-1)\delta t< t <
k\delta t$, $e^{iH^{(k)}_i\delta t}=U^{(k)}_i$. It is easy
to check that $\cT[e^{-i\int_0^{M\delta t} dg\, \t
H(g)}]=U_{circ}$. The simulation time needed is $M\delta t$.
We can always choose a finite $\delta t$ such that $\left
\Vert H^{(k)}_i \right \|$ is finite and the evolution time
is finite.

We have shown that constant depth quantum circuit and local
unitary evolution can simulate each other. The equivalence
relation \eq{LUdef} can therefore  also be stated in terms of
constant depth circuit as in \eq{PhiUcPhi}.

\end{document}